\documentclass[preprint,superscriptaddress,amsfonts,amssymb,amsmath,
  preprintnumbers,floatfix,aps,nofootinbib ]{revtex4}
\makeatletter

\usepackage[dvipdfm]{graphicx}

\providecommand{\abs}[1]{\left|#1\right|}
\newcommand{\bra}[1]{\langle#1|}
\newcommand{\ket}[1]{|#1\rangle}
\newcommand{\n}{\notag \\}
\newcommand{\del}{\partial}
\newcommand{\diag}{\text{diag}\,}

\makeatother

\begin{document}

\title{
Tau and muon lepton flavor violations\\
in the littlest Higgs model with T parity
}
\date{\today}
\author{Toru Goto}
\email[E-mail address:]{tgoto@post.kek.jp}
\affiliation{
  KEK Theory Center, Institute of Particle and Nuclear Studies, 
  KEK, Tsukuba, Ibaraki 305-0801, Japan
}
\author{Yasuhiro Okada}
\email[E-mail address:]{yasuhiro.okada@kek.jp}
\affiliation{
  KEK Theory Center, Institute of Particle and Nuclear Studies, 
  KEK, Tsukuba, Ibaraki 305-0801, Japan
}
\affiliation{
	Department of Particle and Nuclear Physics, 
	The Graduate University for Advanced Studies (Sokendai), 
	Tsukuba, Ibaraki 305-0801, Japan
}
\author{Yasuhiro Yamamoto}
\email[E-mail address:]{yamayasu@post.kek.jp}
\affiliation{
  KEK Theory Center, Institute of Particle and Nuclear Studies, 
  KEK, Tsukuba, Ibaraki 305-0801, Japan
}
\affiliation{
	Department of Particle and Nuclear Physics, 
	The Graduate University for Advanced Studies (Sokendai), 
	Tsukuba, Ibaraki 305-0801, Japan
}
\begin{abstract} 

Lepton flavor violation in $\tau$ and $\mu$ processes is studied in
the littlest Higgs model with T parity.
We consider various asymmetries defined in polarized $\tau$ and $\mu$
decays.
Correlations among branching ratios and asymmetries
are shown in the following lepton flavor violation processes:
$\mu^+   \to e^+ \gamma$, 
$\mu^+   \to e^+ e^+ e^-$,
$\mu^- A \to e^- A$ ($A$ = Al, Ti, Au and Pb),
$\tau^+ \to \mu^+ \gamma$,  
$\tau^+ \to \mu^+ \mu^+ \mu^-$,
$\tau^+ \to \mu^+ e^+   e^-$,
$\tau^+ \to \mu^+ P$ ($P$ = $\pi^0 $, $\eta  $ and $\eta'$),
$\tau^+ \to \mu^+ V$ ($V$ = $\rho^0$, $\omega$ and $\phi $),
$\tau^+ \to e^+ \gamma$, 
$\tau^+ \to e^+ e^+   e^-$,
$\tau^+ \to e^+ \mu^+ \mu^-$,
$\tau^+ \to e^+ P$,
$\tau^+ \to e^+ V$,
$\tau^+ \to \mu^+ \mu^+ e^-$ and
$\tau^+ \to e^+   e^+   \mu^-$.
It is shown that large parity asymmetries and
time-reversal asymmetries are allowed in $\mu^+ \to e^+ e^+ e^-$.
For $\tau$ lepton flavor violation processes, sizable asymmetries are 
possible reflecting
characteristic chirality structure of lepton flavor violating interactions in 
this model.
\end{abstract}

\preprint{KEK-TH-1426}

\maketitle
\section{Introduction} 
Although the standard model (SM) is successful in describing 
almost all experimental results of high energy physics in
terms of gauge theory, the dynamics responsible for the 
elerctroweak symmetry breaking is unknown.
In the minimal version of the Higgs sector, one Higgs doublet
field is introduced as an effective description of the dynamics
below some cutoff scale $\Lambda$ . Constraints from electroweak
precision measurements indicate that  $\Lambda$ is at least beyond 3 TeV. 
On the other hand, this scale already introduces
a fine-tuning problem of the Higgs mass term 
at the $O(1)$ \% level. This is called a little hierarchy 
problem~\cite{Barbieri:1999tm}.

Little Higgs models~\cite{ArkaniHamed:2001nc} were proposed as 
a solution to the little hierarchy problem.
In these models, the Higgs doublet field appears as pseudo
Nambu-Goldstone (NG) bosons of new strong dynamics at the cutoff scale.
A remarkable property is that the one-loop quadratic divergence to
the renormalization of the Higgs mass term is cancelled by a proper
choice of global and local symmetries, so that the cutoff scale can be 
pushed to $O(10)$TeV without a severe fine-tuning. 
A simple case is called the littlest Higgs model~\cite{ArkaniHamed:2002qy} 
where the Higgs field is realized as a part of the NG bosons
associated with the global symmetry breaking of SU(5) to SO(5) and 
the gauge symmetry is [SU(2)$\times$U(1)]$^2$.
Then the quadratic divergence of the Higgs mass term in the SM is 
cancelled by the extra gauge bosons and the heavy partner of 
the top quark.
Subsequent studies~\cite{Csaki:2002qg}, however, showed that 
these heavy partners need to be heavier than several TeV to 
satisfy severe constraints imposed by precise electroweak measurements, 
hence reintroducing some degree of the fine-tuning problem.
In the littlest Higgs model with T parity (LHT)~\cite{Cheng:2003ju}, 
the model is extended to have a $Z_2$ parity so that the heavy
gauge bosons assigned to be T-odd particles do not directly 
couple with a pair of the SM fermions,
and the phenomenological constraints are somewhat relaxed.

Flavor physics provides an interesting possibility to explore the LHT. 
In order to assign T parity, we need to introduce extra fermions 
(T-odd fermions) which are left-handed SU(2) doublets.
Then, the flavor transition can arise in the vertices of heavy gauge bosons,
T-odd fermions and ordinary SM fermions.
There are two 3$\times$3 unitary matrices describing these flavor 
transitions associated with quark and lepton sectors 
besides the ordinary flavor mixing matrices in the 
quark and lepton sectors i.e. the Cabibbo-Kobayashi-Maskawa (CKM)~\cite{CKM} 
and Pontecorvo-Maki-Nakagawa-Sakata (PMNS)~\cite{ref:PMNS} matrices.
These matrices are new sources of flavor changing neutral current (FCNC)
processes in the quark sector and lepton flavor violating (LFV) processes 
in the charged lepton sector.
In the early literature on this subject various observable quantities of 
FCNC and LFV processes were calculated in
the LHT and showed that the effects of T-odd 
partners can be sizable and the correlations among various observable
quantities can be different from other new physics model such as 
supersymmetric (SUSY) models~\cite{RefHLP,Blanke:2006sb,Blanke:2006eb,Choudhury:2006sq,Blanke:2007db}.
It was pointed out later that calculations missed a type of diagrams and 
the logarithmic dependences on the cutoff scale in the
FCNC and LFV amplitudes disappear thanks to new 
contributions~\cite{Goto:2008fj,delAguila:2008zu},
thus reducing theoretical ambiguity associated with the physics 
at the ultraviolet cutoff scale.
Branching ratios of various FCNC and LFV processes were reevaluated 
including the new contributions~\cite{Blanke:2009am,delAguila:2010nv}.
The qualitative feature turned out to be similar to the previous calculation 
though there were sizable changes at the quantitative level.

In this paper, we present the results of further studies in tau and
muon LFV processes.
In addition to branching fractions, we also study observables defined 
with the help of polarizations of the initial muon and tau 
lepton~\cite{Okada:1999zk,Kitano:2000fg}. 
Polarized muon experiments can be done using surface muon beams
that are 100\% polarized in the opposite direction of the $\mu^+$ momentum.
In fact, a $\mu^+ \to e^+ \gamma$ experiment with initial muon polarization
is under consideration \cite{RefMihara}.
In the tau pair production at $e^+\ e^-$ colliders, 
the polarization information 
can be obtained by taking an angular correlation with tau decays 
on the opposite side \cite{Kitano:2000fg}.
The study of polarized tau decays is also being considered 
for the $\tau^+ \to \mu^+ \mu^+ \mu^-$
process at the CERN Large Hadron Collider (LHC)
where we can utilize the tau polarization from $W$ 
decays~\cite{Giffels:2008ar}.
In the following, we consider the processes 
$\mu^+   \to e^+ \gamma$, $\mu^+   \to e^+ e^+ e^-$,
$\mu^- A \to e^- A$ ($A =$ Al, Ti, Au and Pb),
$\tau^+ \to \mu^+ \gamma$,  
$\tau^+ \to \mu^+ \mu^+ \mu^-$,
$\tau^+ \to \mu^+ e^+   e^-$,
$\tau^+ \to \mu^+ P$ ($P =\pi^0 $, $\eta  $ and $\eta'$),
$\tau^+ \to \mu^+ V$ ($V =\rho^0$, $\omega$ and $\phi $),
$\tau^+ \to e^+ \gamma$, 
$\tau^+ \to e^+ e^+   e^-$,
$\tau^+ \to e^+ \mu^+ \mu^-$,
$\tau^+ \to e^+ P$,
$\tau^+ \to e^+ V$,
$\tau^+ \to \mu^+ \mu^+ e^-$ and,
$\tau^+ \to e^+   e^+   \mu^-$.
We define a parity asymmetry in two-body decays and two
parity and one time-reversal asymmetries in three-body decays.
In addition, we can define forward-backward asymmetry and
forward-backward-angular asymmetries in the cases of 
$\tau^+ \to \mu^+ e^+ e^-$ and $\tau^+ \to e^+ \mu^+ \mu^-$.
We show that the parity asymmetries of two-body decays reflect
the characteristic chirality structure of the LFV interactions. 
For three-body decays, we find that there are useful relations 
among various asymmetries.
We calculate the rates of $\mu$-$e$ conversions for different muonic atoms
and show that the ratios of the conversion rates can vary within 
1 order of magnitude 
over most of the LHT model parameter space.
These features, as well as correlations of various branching ratios,
can be useful in discriminating 
different new physics models.

This paper is organized as follows:
Section~II is a brief review of the LHT.
LFV processes are classified and various asymmetries are defined 
in Sec.~III. 
Numerical results on the various observable quantities are shown in Sec.~IV.
Section~V is the conclusion.
Appendix~A shows the general formulae 
of branching ratios and asymmetries for the processes studied.
Functions for the Wilson coefficients are given in Appendix~B.
Useful formulae to perform the consistency test of the LHT are presented 
in Appendix~C.
\section{The littlest Higgs model with T parity}

In this section we review the LHT in order to fix the notations 
we use in this paper.
We use the Lagrangian given in Ref.~\cite{Blanke:2006eb} with 
corrections discussed in Refs.~\cite{Goto:2008fj,delAguila:2008zu}.

\subsection{Gauge and Higgs sectors} 
Gauge and Higgs sectors of the littlest Higgs model are described as
a nonlinear $\sigma$ model with the spontaneous global symmetry breaking
from SU(5) to SO(5) with scalar fields, $\Sigma$, 
which is transformed as $\mathbf{15}$ representation of the SU(5).
The vacuum expectation value of the scalar fields, $\Sigma_0$,
which breaks the global SU(5) symmetry is
\begin{align}
  \Sigma_0 =
  \left(
    \begin{array}{ccccc}
      0 & 0 & 0 & 1 & 0 \\
      0 & 0 & 0 & 0 & 1 \\
      0 & 0 & 1 & 0 & 0 \\
      1 & 0 & 0 & 0 & 0 \\
      0 & 1 & 0 & 0 & 0
    \end{array}
  \right).
\end{align}
Generators of the unbroken SO(5) symmetry, $T^a$, and 
the broken ones of SU(5)/SO(5), $X^a$, satisfy the following relations,
\begin{align}
	\Sigma_0 T^a \Sigma_0 = -(T^a)^T, \qquad
	\Sigma_0 X^a \Sigma_0 =  (X^a)^T.
\label{EqTgen}
\end{align}

The Lagrangian of the NG bosons and gauge fields is given by
\begin{align}
  \mathcal{L}_{\text{NG-gauge}} =&
  \frac{f^2}{8} \text{tr}[ (D^{\mu} \Sigma)^{\dagger} D_{\mu} \Sigma ] +
	\sum_{i=1,2} \left( 
	  -\frac{1}{2} \text{tr} [ W_i{}^{\mu\nu} W_{i\mu \nu} ]
	  -\frac{1}{4} B_i{}^{\mu\nu} B_{i\mu \nu}
	\right),
\label{EqLagnggauge}
\end{align}
where $f$ is the decay constant of the nonlinear $\sigma$ model and
the NG boson field is written as follows,
\begin{align}
	\Sigma &= \xi \Sigma_0 \xi^T = \xi^2 \Sigma_0, \quad
	\xi = e^{i\Pi/f}, 
\end{align}
\begin{align}
	\Pi &= \pi^a X^a =
		\begin{pmatrix}
			-\frac{\omega^0}{2}-\frac{\eta}{2\sqrt{5}} &
			-\frac{\omega^+}{\sqrt{2}} & -i\frac{\pi^+}{\sqrt{2}} &
			-i\phi^{++} & -i\frac{\phi^+}{\sqrt{2}} \\ 
			-\frac{\omega^+}{\sqrt{2}} & 
			\frac{\omega^0}{2}-\frac{\eta}{2\sqrt{5}} &
			\frac{v+h+i\pi^0}{2} & -i\frac{\phi^+}{\sqrt{2}} &
			\frac{ -i\phi^0+\phi^{\text{P}} }{\sqrt{2}} \\ 
			i\frac{\pi^-}{\sqrt{2}} & \frac{v+h-i\pi^0}{2} & 
			\frac{2\eta}{\sqrt{5}} & -i\frac{\pi^+}{\sqrt{2}} &
			\frac{v+h+i\pi^0}{2} \\ 
			i\phi^{--} & i\frac{\phi^-}{\sqrt{2}} & i\frac{\pi^-}{\sqrt{2}} &
			-\frac{\omega^0}{2}-\frac{\eta}{2\sqrt{5}} & 
			-\frac{\omega^-}{\sqrt{2}} \\ 
			i\frac{\phi^-}{\sqrt{2}} 
			&\frac{ i\phi^0 +\phi^{\text{P}} }{\sqrt{2}} &
			\frac{v+h-i\pi^0}{2} & -\frac{\omega^+}{\sqrt{2}} &
			\frac{\omega^0}{2}-\frac{\eta}{2\sqrt{5}} 
		\end{pmatrix} .
\label{EqNgboson}
\end{align}
Fourteen NG bosons are denoted by $h$, $\pi^{\pm,0}$, $\omega^{\pm,0}$, 
$\eta$, $\phi^{\pm,0}$, $\phi^P$, and $\phi^{\pm \pm}$.
As explained later, $H = ( -i\pi^+ /\sqrt{2}, (v+h+i\pi^0)/2 )^T$ is
identified as the SM Higgs doublet. 
$v$ is the electroweak symmetry breaking vacuum expectation value, 
$v = 246$ GeV.
$W_i^{a\mu}$ and $B_i^{\mu}$ ($i = 1,2$) are gauge fields of the
SU(2)$_i \times$U(1)$_i$, which are subgroups of the SU(5).
The generators of the gauged subgroups are 
\begin{align}
	Q_1^a =& \frac{1}{2} \begin{pmatrix} \sigma^a & 0 & 0 \\ 0 & 0 & 0 \\ 
				0 & 0 & 0 \end{pmatrix},&
	Y_1 =& \frac{1}{10} \diag (3,3,-2,-2,-2) ,\\ 
	Q_2^a =& \frac{1}{2} \begin{pmatrix} 0 & 0 & 0 \\ 0 & 0 & 0 \\ 
				0 & 0 & -\sigma^{aT} \end{pmatrix},&
	Y_2 =& \frac{1}{10} \diag (2,2,2,-3,-3),
\end{align}
where $\sigma^a$ ($a=$1, 2, 3) are the Pauli matrices.
The covariant derivative of $\Sigma$ is defined as 
\begin{align}
  D_{\mu} \Sigma = 
   \partial_{\mu} \Sigma 
  -\sqrt{2} i\sum_{j=1,2} \Bigl(
    g  W_{j\mu}^a ( Q_j^a \Sigma +\Sigma Q_j^{aT}) 
	+g' B_{j\mu}   ( Y_J   \Sigma +\Sigma Y_j     )
  \Bigr),
\end{align}
where $g$ and $g'$ are gauge coupling constants of SU(2)$_L \times$U(1)$_Y$.

The Lagrangian \eqref{EqLagnggauge} has a discrete symmetry under the
following $Z_2$ transformation, which is called T parity~\cite{Cheng:2003ju}.
\begin{align}
	W_1^{a\mu} &\leftrightarrow W_2^{a\mu}, \qquad
	B_1^{ \mu} \leftrightarrow B_2^{\mu},\\
	\Pi &\leftrightarrow -\Omega \Pi \Omega, \qquad
	\Omega = \diag(1,1,-1,1,1).
\label{EqTparing}
\end{align}
The T parity eigenstates of the gauge bosons are 
\begin{align}
  & W_L^{\pm ,3\mu} =
     \frac{W_1^{\pm ,3\mu} +W_2^{\pm ,3\mu}}{\sqrt{2}},\quad
    B_L^{\mu} =
	  \frac{B_1^{\mu} +B_2^{\mu}}{\sqrt{2}} \qquad (\text{T-even}),\\
  & W_H^{\pm ,3\mu} =
     \frac{W_1^{\pm ,3\mu} -W_2^{\pm ,3\mu}}{\sqrt{2}}, \quad 
    B_H^{\mu} =
	  \frac{B_1^{\mu} -B_2^{\mu}}{\sqrt{2}} \qquad (\text{T-odd}),
\end{align}
where $W_j^{\pm \mu} = ( W_j^{1\mu} \mp iW_j^{2\mu} )/\sqrt{2}$ ($j = 1,2$).
As for the NG boson fields, we can show from 
Eqs.~\eqref{EqNgboson} and \eqref{EqTparing} that 
$h$, $\pi^{\pm,0}$ are T-even particles and 
the others are T-odd particles.
$W_H^{\pm,3\mu}$ and $B_H^{\mu}$ receive mass of the order of $f$ 
from the gauge symmetry breaking 
[SU(2)$\times$U(1)]$^2\to$ SU(2)$_L \times$U(1)$_Y$ by absorbing
$\omega^{\pm,0}$ and $\eta$.
After the electroweak symmetry breaking,
$\pi^{\pm,0}$ are absorbed by the SM gauge fields,
$W_L^{\pm,3\mu}$ and $B_L^{\mu}$.
Mass eigenstates of the neutral gauge bosons are defined by
\begin{eqnarray}
  Z_L^{\mu} &=& W_L^{3\mu} \cos\theta_W - B_L^{\mu} \sin\theta_W,
\qquad
  A_L^{\mu}  =  W_L^{3\mu} \sin\theta_W + B_L^{\mu} \cos\theta_W,
\\
  Z_H^{\mu} &=& W_H^{3\mu} \cos\theta_H - B_H^{\mu} \sin\theta_H,
\qquad
  A_H^{\mu}  =  W_H^{3\mu} \sin\theta_H + B_H^{\mu} \cos\theta_H,
\end{eqnarray}
where $Z_L^{\mu}$ and $A_L^{\mu}$ are the SM $Z$ boson and photon.
$\theta_W$ is the weak mixing angle determined as
$\sin\theta_W=g'/\sqrt{g^2 + g^{\prime 2}}$.
The mixing angle of the T-odd gauge bosons is given by
\begin{align}
  \tan2\theta_H &=
  -\frac{g g' c_v^2 s_v^2}{ g^2 - \frac{1}{5} g^{\prime 2}
    - \frac{1}{2}(g^2 - g^{\prime 2})c_v^2 s_v^2},
\end{align}
where $c_{v} = \cos(v/(\sqrt{2}f))$ and $s_{v} = \sin(v/(\sqrt{2}f))$.
Expanding in terms of $v^2/f^2$, we obtain
\begin{align}
	\sin \theta_H &= 
	 -\frac{gg'}{g^2 -\frac{1}{5}g'^2} 
	  \frac{v^2}{4f^2} + O(\frac{v^4}{f^4}) ,\\
	\cos \theta_H &= 1 + O(\frac{v^4}{f^4}) .
\end{align}
Gauge boson masses are given as
\begin{subequations}
\begin{eqnarray}
  m_{W_L}^2 &=& \frac{g^2 f^2}{2} s_v^2,
\qquad
  m_{Z_L}^2 = \frac{m_{W_L}^2}{\cos^2\theta_W},
\\
  m_{W_H}^2 &=& g^2 f^2 \left(1 - \frac{s_v^2}{2} \right),
\\
   m_{Z_H}^2 &=&
   \frac{g^2 f^2}{c_H^2-s_H^2}
   \left[
     \left( 1 - \frac{c_v^2 s_v^2}{2} \right) c_H^2
     -
     \frac{g^{\prime 2}}{5g^2}
     \left( 1 - \frac{5}{2} c_v^2 s_v^2 \right) s_H^2
   \right]
\nonumber\\&=&
  m_{W_H}^2 + O(\frac{v^4}{f^2}),
\\
   m_{A_H}^2 &=&
   \frac{g^{\prime 2} f^2}{5(c_H^2-s_H^2)}
   \left[
     \left( 1 - \frac{5}{2} c_v^2 s_v^2 \right) c_H^2
     -
     \frac{5g^2}{g^{\prime 2}}
     \left( 1 - \frac{c_v^2 s_v^2}{2} \right) s_H^2
   \right]
\nonumber\\&=&
 \frac{g^{\prime 2}f^2}{5}
  \left( 1 - \frac{5v^2}{4f^2} +  O(\frac{v^4}{f^4}) \right),
\end{eqnarray}
\end{subequations}
where $s_H=\sin\theta_H$ and $c_H=\cos\theta_H$.
We consider terms up to $O(v^4/f^4)$ so that 
we neglect the difference between $m_{W_H}$ and $m_{Z_H}$ in our analysis.

\subsection{Fermion sector} 

The fermion sector of the LHT consists of
three families of quark and lepton fields $q_{1,2}$, $\ell_{1,2}$,
$u_R$, $d_R$, $\nu_R$, $e_R$, $q_{HR}$ and $\ell_{HR}$ for
each generation, and a set of the ``top-partner''
fermions, $t'_{1,2}$ and $t'_{1R,2R}$.
Following the procedure given in Ref.~\cite{Goto:2008fj}, we assign gauge
charges of the fermion fields as summarized 
in Table~\ref{TabCharge}.
The kinetic terms of left-handed fermions are given by 
\begin{align}
	\mathcal{L}_{\text{left}} = &
	\bar{\Psi}_1 \gamma^{\mu} 
	\left(i\del_{\mu} -\sqrt{2}gQ_1^a W_{1\mu}^a 
	 -\sqrt{2} g' B_{1\mu} Y_1^{\Psi_1} 
	 -\sqrt{2} g' B_{2\mu} Y_2^{\Psi_1} \right) P_L \Psi_1 \n
	& + \bar{\Psi}_2 \gamma^{\mu} 
	\left(i\del_{\mu} +\sqrt{2}gQ_2^{aT} W_{2\mu}^a 
	 -\sqrt{2} g' B_{1\mu} Y_1^{\Psi_2} 
	 -\sqrt{2} g' B_{2\mu} Y_2^{\Psi_2} \right) P_L \Psi_2 \,,
\label{EqGaugeint}
\end{align}
where these fermions are introduced as incomplete multiplets of the SU(5),
\begin{align}
	\Psi_1 =& \begin{pmatrix} -i\sigma^2 \ell_1\\ 0 \\ 0 \end{pmatrix}, \quad
	\Psi_2 = \begin{pmatrix} 0 \\ 0 \\ -i\sigma^2 \ell_2\end{pmatrix} 
\end{align}
and $P_L = \frac{1-\gamma_5}{2}$.
Under the T parity, they transform as
\begin{align}
	\Psi_1 \leftrightarrow -\Sigma_0 \Psi_2 .
\end{align}
Thus the SM leptons and the T-odd leptons are given as 
the T parity eigenstates,
\begin{align}
	\ell_L = \frac{\ell_1 -\ell_2}{\sqrt{2}}\quad (\text{T-even}), \\
	\ell_H = \frac{\ell_1 +\ell_2}{\sqrt{2}}\quad (\text{T-odd }),
\end{align}
respectively. 
The right-handed SM fermions are introduced in the same manner as in the SM.

The right-handed T-odd leptons are introduced as 
a nonlinear representation of the SU(5),
\begin{align}
	\Psi_{R} = 
	 \begin{pmatrix} \tilde{\psi}_{R} \\ \chi_{R} \\ 
	                -i\sigma^2 \ell_{HR} \end{pmatrix},
\end{align}
where upper three components, $\tilde{\psi}_R$ and $\chi_R$, are heavy
enough to be neglected in low energy dynamics.
$\Psi_R$ changes the sign with the T parity.
In order to construct the gauge invariant kinetic term,
we define $\Psi_{R}^{\xi}$ as
\begin{align}
  \Psi_{R}^{\xi} = \xi\Psi_{R} 
\end{align}
which transform linearly under the SU(5) transformation.
Thus the kinetic term is given by
\begin{align}
	\mathcal{L}_{\text{HR}} 
	&= \frac{1}{2} \overline{\Psi_{R}^{\xi}} \gamma^{\mu}
		\left( i\del_{\mu} -\sqrt{2}g W_{1\mu}^a Q_1^a 
		+\sqrt{2}g W_{2\mu}^a Q_2^{aT} \right.
		\left. -\sqrt{2}g' B_{1\mu} Y_1^{\Psi_R} 
			-\sqrt{2}g' B_{2\mu} Y_2^{\Psi_R} \right)
			P_R \Psi_{R}^{\xi} \n 
	&+(\text{T parity conjuation}),
\end{align}
where 
$Y_1^{\Psi_R}= \frac{1}{10} \text{diag}(3,3,-2,-2,-2)$ and 
$Y_2^{\Psi_R}= \frac{1}{10} \text{diag}(2,2, 2,-3,-3)$.
The quark sector is constructed in the same procedure.
For the quark sector,
$Y_1^{\Psi_R}= \frac{1}{30} \text{diag}(19,19,4,4,4)$ and 
$Y_2^{\Psi_R}= \frac{1}{30} \text{diag}(16,16,16,1,1)$.

\begin{table}[tbp] 
\begin{displaymath}
\begin{array}[t]{|c|cccc|}
\hline
  & \text{SU(2)}_1 & \text{SU(2)}_2 & Y_1 & Y_2 
\\
\hline
  q_1 = \left(\begin{array}{c} u_1 \\ d_1 \end{array}\right)
  & \mathbf{2} & \mathbf{1} & \frac{1}{30} & \frac{4}{30}
\\
  q_2 = \left(\begin{array}{c} u_2 \\ d_2 \end{array}\right)
  & \mathbf{1} & \mathbf{2} & \frac{4}{30} & \frac{1}{30}
\\
\hline
  t'_1
  & \mathbf{1} & \mathbf{1} & \frac{16}{30} & \frac{4}{30}
\\
  t'_2
  & \mathbf{1} & \mathbf{1} & \frac{4}{30} & \frac{16}{30}
\\
\hline
  \ell_1 = \left(\begin{array}{c} \nu_1 \\ e_1 \end{array}\right)
  & \mathbf{2} & \mathbf{1} & -\frac{3}{10} & -\frac{2}{10}
\\
  \ell_2 = \left(\begin{array}{c} \nu_2 \\ e_2 \end{array}\right)
  & \mathbf{1} & \mathbf{2} & -\frac{2}{10} & -\frac{3}{10}
\\
\hline
\end{array}
\qquad 
\begin{array}[t]{|c|cccc|}
\hline
  & \text{SU(2)}_1 & \text{SU(2)}_2 & Y_1 & Y_2 
\\
\hline
  u_R
  & \mathbf{1} & \mathbf{1} 
  & \frac{1}{3} & \frac{1}{3}
\\
  d_R
  & \mathbf{1} & \mathbf{1} 
  & -\frac{1}{6} & -\frac{1}{6}
\\
\hline
  t'_{1R}
  & \mathbf{1} & \mathbf{1} 
  & \frac{16}{30} & \frac{4}{30}
\\
  t'_{2R}
  & \mathbf{1} & \mathbf{1} 
  & \frac{4}{30} & \frac{16}{30}
\\
\hline
  \nu_R
  & \mathbf{1} & \mathbf{1} 
  & 0 & 0
\\
  e_R
  & \mathbf{1} & \mathbf{1} 
  & -\frac{1}{2} & -\frac{1}{2}
\\
\hline
\end{array}
\end{displaymath}
\caption{
  Quantum numbers of fermion fields.
  $Y_{1,2}$ are charges of U(1)$_{1,2}$.
  Generation indices are suppressed.
  For $q_{HR}$ and $\ell_{HR}$, see the text.}
\label{TabCharge}
\end{table}
%
\subsection{The Yukawa sector} 

The Yukawa couplings of the T-odd fermions are 
\begin{align}
  \mathcal{L}_{\text{T-odd-Yukawa}} =
	-\kappa^{ij} f (\bar{\Psi}_2^i \xi 
	+\bar{\Psi}_1^i \Sigma_0 \Omega \xi^{\dagger}\Omega )\Psi_{R}^j 
	+\text{H.c.},
\end{align}
with generation indices, $i$ and $j$.
Thus their masses are given by
\begin{align}
	m_{H\ell}^i &= \sqrt{2} \kappa^i f , \\
	m_{H\nu}^i &= \frac{\kappa^i f}{\sqrt{2}}( 1+c_{v} )
				= m_{H\ell}^i \left( 1 +O\left(\frac{v^2}{f^2}\right) \right),
\end{align}
where $i$ is the generation index and $\kappa^i$ are eigenvalues of 
the Yukawa coupling matrix, $\kappa^{ij}$.
We neglect $O(v^2/f^2)$ differences between $m_{H\ell}^i$ and $m_{H\nu}^i$
in the calculation of LFV processes because
these differences only contribute to the higher order corrections
in the $v^2/f^2$ expansion.
For the SM charged leptons,
the gauge invariant Yukawa coupling terms are written as
\begin{align}
	\mathcal{L}_{\text{T-even-Yukawa}} =
	-\frac{ \lambda_{\text{d}}^{ij} }{2\sqrt{2}} f \epsilon_{ab} \epsilon_{xyz}
		\left[ (\bar{\Psi}_2^{Xi})_x (\Sigma )_{ay}
		(\Sigma )_{bz} -(\bar{\Psi}_1^{Xi} \Sigma_0)_x (\tilde{\Sigma})_{ay}
		(\tilde{\Sigma})_{bz} \right] e_{R}^j + \text{H.c.},
\label{EqDowny}
\end{align}
where $a,\,b = (1,2)$ and $x,\,y,\,z = (3,4,5)$ are SU(5) indices and
$i$ and $j$ are generation indices.
The left-handed SM lepton doublets are embedded in SU(5) multiplets 
$\Psi_1^{Xi}$ and $\Psi_1^{Xi}$ as
\begin{align}
 \Psi_1^{Xi} &=
  \begin{pmatrix} i\tilde{X} \ell_1^i \\ 0 \\ 0 \end{pmatrix},\quad
 \Psi_2^{Xi}  =
  \begin{pmatrix} 0 \\ 0 \\ iX \ell_2^i \end{pmatrix},
\end{align}
where $X$ and $\tilde{X}$ are SU(2)$_i$ singlet scalar fields
whose U(1) gauge charges are
$(Y_1,\,Y_2)= (\frac{1}{10},\, -\frac{1}{10})$ and 
$(-\frac{1}{10},\, \frac{1}{10})$, respectively.
Following Ref.~\cite{Chen:2006cs}, we use
$(\Sigma_{33})^{-1/4}$ and its T parity conjugate 
as $X$ and $\tilde{X}$, respectively.
\subsection{New mixing matrices} 

As explained in Ref.~\cite{RefHLP}, 
the mass matrices of the SM fermions and the T-odd fermions are not 
simultaneously diagonalized in general.
Consequently, new mixing matrices are
introduced in the interaction among the T-odd gauge bosons and fermions.
We extract the interaction terms of the T-odd $W$ boson and leptons 
from Eq.~(\ref{EqGaugeint}) as
\begin{align}
  \mathcal{L}_{\text{left}} \supset &
  \frac{ig}{\sqrt{2}} (V_{H\ell})_{ij} 
    \bar{\nu}_H^i \gamma_{\mu} P_L \ell^j W_H^{+\mu}
  +\frac{ig}{\sqrt{2}} (V_{H\nu})_{ij} 
    \bar{\ell}_H^j \gamma_{\mu} P_L \nu^j W_H^{-\mu} +\text{H.c.},
\end{align}
where $V_{H\ell}$ and $V_{H\nu}$ are new mixing matrices.
These two mixing matrices are related to each other by 
\begin{align}
	V_{PMNS} = V_{H\ell}^{\dagger} V_{H\nu},
\end{align}
where $V_{PMNS}$ is the PMNS matrix \cite{ref:PMNS},
because the mass matrices of $\ell_H$ and $\nu_H$ are 
determined by the same Yukawa coupling matrix $\kappa^{ij}$.
$V_{H\ell}$ and $V_{H\nu}$ also appear in the interaction of
the neutral T-odd gauge bosons and leptons.
We parameterize $V_{H\ell}$ as
\begin{align}
	V_{H\ell} &=
	\begin{pmatrix}
		1 & 0 & 0 \\ 
		0 & c_{23}^{\ell} & s_{23}^{\ell} 
		e^{-i\delta_{23}^{\ell} } \\
		0 & -s_{23}^{\ell} e^{i\delta_{23}^{\ell} } & c_{23}^{\ell} 
	\end{pmatrix}
	\begin{pmatrix}
		c_{13}^{\ell} & 0 & s_{13}^{\ell} 
		e^{-i\delta_{13}^{\ell} } \\
		0 & 1 & 0 \\ 
		-s_{13}^{\ell} e^{i\delta_{13}^{\ell} } & 0 & c_{13}^{\ell} 
	\end{pmatrix}
	\begin{pmatrix}
		c_{12}^{\ell} & s_{12}^{\ell} 
		e^{-i\delta_{12}^{\ell}} & 0 \\
		-s_{12}^{\ell} e^{i\delta_{12}^{\ell}} & 
		c_{12}^{\ell} & 0 \\
		0 & 0 & 1 
	\end{pmatrix},
\label{EqMix}
\end{align}
where $s_{ij}^{\ell} = \sin \theta_{ij}^{\ell}$ and 
$c_{ij}^{\ell} = \cos \theta_{ij}^{\ell}$.
We take $\theta_{ij}^{\ell}$ and $\delta_{ij}^{\ell}$ as input parameters.
Notice that there remain three unremovable phases in $V_{H\ell}$ 
after the phase convention of the PMNS matrix is fixed, 
as pointed out in Ref.~\cite{Blanke:2006eb}.
For the quark sector, we define new mixing matrices $V_{Hd}$ and 
$V_{Hu}$ in the same way as $V_{H\ell}$ and $V_{H\nu}$.
We take $V_{Hd}$ as a matrix parametrized like Eq.~\eqref{EqMix} and
the other mixing matrix, $V_{Hu}$, is expressed by 
$V_{Hd}$ and the CKM matrix \cite{CKM} in the following way:
\begin{align}
  V_{Hu} = V_{Hd} V_{CKM}^{\dagger}.
\end{align}

\section{Lepton flavor violation in the littlest Higgs model with T parity}
\label{SecLhtlfv} 

In this section, we present formulae for observables in the LFV 
processes in the LHT.
In contrast to previous works \cite{Choudhury:2006sq,Blanke:2007db,%
delAguila:2008zu,Blanke:2009am,delAguila:2010nv}, we consider various
angular distributions in polarized $\mu$ and $\tau$ decays in addition
to the branching ratios and the $\mu-e$ conversion rates 
in various muonic atoms.
Studying the angular distributions is useful to extract information
about the chirality structure of the low energy effective Lagrangian.

The LFV processes that we consider in this paper are the following.
\begin{itemize}
\item
Radiative two-body decays: $\mu^+\to e^+ \gamma$,
$\tau^+\to \mu^+ \gamma$ and $\tau^+\to e^+ \gamma$.
\item
Leptonic three-body decays of $\mu$ and $\tau$.
We classify these decay modes into three types according to the lepton
flavor combination in the final state.
\begin{itemize}
\item Type I: 
$\mu^+\to e^+ e^+ e^-$, $\tau^+\to \mu^+ \mu^+ \mu^-$ and
$\tau^+\to e^+ e^+ e^-$.
All the three leptons in the final state have the same flavor.
\item Type II:
$\tau^+\to \mu^+ e^+ e^-$ and $\tau^+\to e^+ \mu^+ \mu^-$.
In these processes, the final state consists of a pair of same flavor
leptons with opposite charges and another lepton with different flavor.
There are no identical particles in the final state.
\item Type III:
$\tau^+\to \mu^+ \mu^+ e^-$ and $\tau^+\to e^+ e^+ \mu^-$.
Lepton flavor changes by more than one.
\end{itemize}
\item Semileptonic two-body decay of $\tau$.
We consider $\tau^+\to \mu^+ P$ and $\tau^+\to \mu^+ V$ modes, where $P$
and $V$ stand for a pseudoscalar meson ($\pi^0$, $\eta$ or $\eta'$) and a
vector meson ($\rho^0$, $\omega$ or $\phi$), respectively, as well as
corresponding $\tau^+\to e^+$ modes.
We do not consider $\tau^+\to \mu^+(e^+) K^{(*)}$ modes, where both quark
and lepton flavors are violated.
\item $\mu-e$ conversion in a muonic atom.
In these processes, $\mu^- A \to e^- A$, we discuss aluminum (Al), 
titanium (Ti), gold (Au) and lead (Pb) as target atom, $A$, 
because the conversion rates depend on target nuclei~\cite{Kitano:2002mt}.
\end{itemize}

In the LHT, LFV processes are induced by loop diagrams with the 
T-odd gauge bosons
and leptons in the internal lines.
Contributions of the T-even particles are negligible because they are
suppressed by neutrino masses.
Since only the left-handed SM (T-even) leptons couple to the T-odd gauge
bosons, the low energy effective Lagrangian for LFV has a restricted
chirality structure: the lepton flavor changing occurs in the left-handed
lepton sector only.
In the following subsections, we present the formulae for the effective
Lagrangians, branching ratios, and various asymmetries, which can be defined 
based on angular distributions.
Except for the $\mu-e$ conversion, we explicitly show formulae for
$\tau$ to $\mu$ transitions.
Those for $\mu$ to $e$ and $\tau$ to $e$ are obtained by appropriate
replacements of flavor indices.
For simplicity, we neglect lepton masses in the final states.

In the following, formulae of the low energy effective Lagrangians and 
the observable quantities are shown
based on the LFV interactions relevant to the LHT.
Those in the case of the most general low energy effective 
Lagrangians are given in Appendix~\ref{AppFormulae}.
\subsection{$\mu^+ \to e^+ \gamma$, $\tau^+ \to \mu^+ \gamma$ and, 
            $\tau^+ \to e^+ \gamma$}

The effective Lagrangian for $\tau^+ \to \mu^+ \gamma$ in the LHT
consists of the following dipole moment type operator.
\begin{align}
	\mathcal{L}_{\gamma} =&
	-\frac{4G_F}{\sqrt{2}}
	  \left[
            m_{\tau} A_R^{\text{LHT}} 
	    \bar{\tau}_R \sigma^{\mu \nu} \mu_L F_{\mu \nu}
	+ \text{H.c.}
      \right],
\label{eq:llg}
\end{align}
where
\begin{align}
  A_R^{\text{LHT}} =& 
    \frac{ e }{ ( 4 \pi )^2 } \frac{ v^2 }{ 8 f^2 }
    \sum_l ( V_{H\ell} )^{\ast}_{l3} ( V_{H\ell} )_{l2}
	  \left( 
	    N_{CM}(x_l) + \frac{1}{2} N_{NM}(x_l) + \frac{1}{10} N_{NM}(x'_l) 
	  \right),
\label{EqDipole}
\end{align}
with
\begin{align}
  x_l  = \frac{m_{H\ell}^l{}^2}{m_{W_H}^2}, \quad 
  x'_l = \frac{m_{H\ell}^l{}^2}{m_{A_H}^2} = \frac{5}{\tan^2 \theta_{W}} x_l.
\end{align}
The functions $N_{CM}$ and $N_{NM}$ are given in Appendix~\ref{AppWilson}.
The coefficient of the operator with the opposite chirality,
$\bar{\tau}_L\sigma^{\mu\nu}\mu_R F_{\mu\nu}$, is suppressed by
$m_\mu/m_\tau$ due to the chirality properties in the interaction of the
T-odd gauge bosons.

\begin{figure}[htb]
  \includegraphics[width=14em,clip]{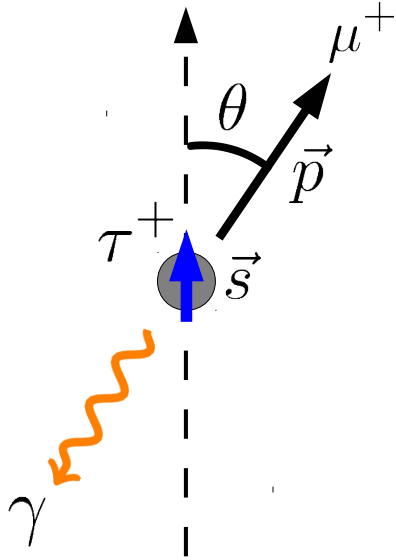}
\caption{
  Configuration of $\tau^+ \to \mu^+ \gamma$ decay.
  The dashed arrow denotes the direction of the $\tau^+$ spin.
  The solid and the wavy arrows show the directions of 
  momenta of the $\mu^+$ and the photon, respectively.
  $\theta$ is defined as the angle between the directions of 
  the $\tau^+$ spin and the $\mu^+$ momentum.
}
\label{fig:pllg}
\end{figure}

For a $\tau^+\to \mu^+\gamma$ decay of a polarized $\tau^+$, we define a
decay angle $\theta$ as the angle between the spin of $\tau^+$ and the
momentum of $\mu^+$ in the rest frame of $\tau^+$.
The configuration of spin and momenta are shown in Fig.~\ref{fig:pllg}.

The differential branching ratio with respect to $\cos\theta$ is given
by
\begin{align}
  \frac{d}{d\cos\theta}
  \text{Br}(\tau^+ \to \mu^+ \gamma)_{\text{LHT}} &=
  \tau_{\tau} \frac{G_F^2 m_{\tau}^5 }{\pi}
  \abs{ A_R^{\text{LHT}} }^2
  (1 - \cos\theta).
\label{eq:dBr-tmg}
\end{align}
We define the polarization asymmetry, which characterizes the chirality
structure of the effective interaction in (\ref{eq:llg}), as
\begin{align}
  A_{\gamma}(\tau^+ \to \mu^+ \gamma)
&=
 \frac{1}{\text{Br}(\tau^+ \to \mu^+ \gamma)}
 \left(
  \int_{0}^{1}
  -
  \int_{-1}^{0}
\right)d\cos\theta
  \frac{d\text{Br}(\tau^+ \to \mu^+ \gamma)}{d\cos\theta},
\label{EqAsymgamma}
\end{align}
where the integrated branching ratio is
\begin{align}
  \text{Br}(\tau^+ \to \mu^+ \gamma)
&=
  \int_{-1}^{1}d\cos\theta
  \frac{d\text{Br}(\tau^+ \to \mu^+ \gamma)}{d\cos\theta}.
\end{align}
$A_{\gamma}$ is a parity odd quantity since $\cos\theta$ is
proportional to the inner product $\vec{s}\cdot\vec{p}$, where $\vec{s}$
and $\vec{p}$ are spin of initial $\tau^+$ and spatial momentum of final
$\mu^+$, respectively.
By substituting (\ref{eq:dBr-tmg}), we obtain
\begin{align}
  \text{Br}(\tau^+ \to \mu^+ \gamma)_{\text{LHT}} &=
  \tau_{\tau} \frac{ 2 G_F^2 m_{\tau}^5 }{\pi}
  \abs{ A_R^{\text{LHT}} }^2,
\\
  A_{\gamma}(\tau^+ \to \mu^+ \gamma)_{\text{LHT}}
&=
  -\frac{1}{2}.
\end{align}
The parameter independent result of $A_{\gamma}$ is a consequence of
the special chirality structure of the LHT.

\subsection{$\mu^+  \to e^+ e^+ e^-$, $\tau^+ \to \mu^+ \mu^+ \mu^-$ and  
            $\tau^+ \to e^+ e^+ e^-$}
\label{sec:3leptontypeI}

The effective Lagrangian for type I leptonic three-body decay,
$\tau^+\to \mu^+\mu^+\mu^-$, in the LHT is given as
\begin{align}
	\mathcal{L}_{\text{I}} =
	-\frac{4G_F}{\sqrt{2}} & \bigl[
	m_{\tau} A_R^{\text{LHT}} 
	  \bar{\tau}_{R} \sigma^{\mu \nu} \mu_L F_{\mu \nu}
	+ g_{Ll}^{\text{I,LHT} } ( \bar{\tau}_L \gamma^{\mu} \mu_L )
	         ( \bar{\mu}_L \gamma_{\mu} \mu_L ) \n &
	+ g_{Lr}^{\text{I,LHT} } ( \bar{\tau}_L \gamma^{\mu} \mu_L )
				( \bar{\mu}_R \gamma_{\mu} \mu_R )
	+ \text{H.c.}
	\bigr],
\label{eq:Leff-tmmm}
\end{align}
where
\begin{align}
  g_{Ll}^{\text{I,LHT} } =& 
	 \frac{ g^2 }{ ( 4 \pi )^2 } \frac{ v^2 }{ 8 f^2 }
    \sum_l ( V_{H\ell} )^{\ast}_{l3} ( V_{H\ell} )_{l2} \n &
	 \quad \times
    \left( 
	 - \sin^2 \theta_W P_{\gamma}(x_l) 
	 + \left( -\frac{1}{2} + \sin^2 \theta_W \right) P_Z(x_l) 
	   + \sum_m ( V_{H\ell} )^{\ast}_{m2} ( V_{H\ell} )_{m2}
	  B_{(e)}(x_l,x_m) 
	 \right), \label{EqVeclr1} \\
  g_{Lr}^{\text{I,LHT} } =& 
	 \frac{ g^2 }{ ( 4 \pi )^2 } \frac{ v^2 }{ 8 f^2 }
    \sum_l ( V_{H\ell} )^{\ast}_{l3} ( V_{H\ell} )_{l2}\,
    ( - \sin^2 \theta_W ) \bigl( P_{\gamma}(x_l) - P_Z (x_l) \bigr).
\label{EqVeclr2}
\end{align}
The functions $P_{\gamma}$, $P_{Z}$ and $B_{(e)}$ represent
contributions of photon penguin, $Z$ penguin and box diagrams,
respectively, and are given in Appendix~\ref{AppWilson}.
$A_R^{\text{LHT}}$ is shown in (\ref{EqDipole}).

\begin{figure}[htb]
  \includegraphics[scale=0.8,clip]{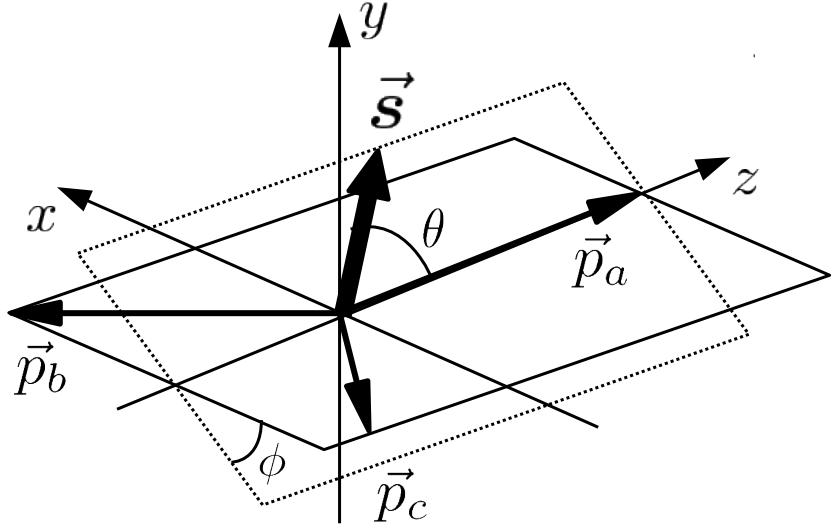}
\caption{
The spin and momentum configuration of a three-body decay in the rest
frame of the initial lepton.
The decay plane is identified as the $z$-$x$ plane where
the $z$ axis is taken to be the direction of particle $a$'s momentum.
the $x$ axis is defined so that the $x$ component of particle $b$'s momentum
is positive.
The direction of the initial lepton polarization vector $\vec{s}$ is
parametrized by the polar and the azimuthal angles $\theta$ and $\phi$,
respectively.
}
\label{FigPl3l}
\end{figure}

The spin and momentum configuration of a three-body decay of a polarized
$\tau^+$ at rest is determined by two energy variables and two angular
variables, as depicted in Fig.~\ref{FigPl3l}.
The three spatial momenta of the particles in the final state are denoted as
$\vec{p}_a$, $\vec{p}_b$, and $\vec{p}_c$.
We take $z$ axis in the direction of $\vec{p}_a$ and the decay plane is
identified as the $z$-$x$ plane.
The direction of $x$ axis is chosen in such a way that the $x$ component
of $\vec{p}_b$ has a positive value.
We take the $y$ axis along $\vec{p}_a\times \vec{p}_b$.
The direction of the polarization vector of the initial particle is
parametrized by the polar and the azimuthal angles $\theta$ and $\phi$,
respectively.
For the $\tau^+ \to \mu^+ \mu^+ \mu^-$ decay, we identify the momentum of
$\mu^-$ as $\vec{p}_a$.
Two $\mu^+$'s are distinguished by their energies.
The momentum of the $\mu^+$ with larger (smaller) energy is identified as
$\vec{p}_b$ ($\vec{p}_c$).

As for the energy variables, we define
\begin{align}
  x_{b,c} = \frac{2 E_{b,c}}{m_{\tau}},
\label{EqDefx}
\end{align}
where $E_{b,c}$ are energies of $\mu^+$'s with momenta $\vec{p}_{b,c}$.
$E_{b}$ is larger than $E_{c}$ by definition.
Possible ranges of $x_{b,c}$ in an approximation of massless final
particles are
\begin{align}
  \frac{1}{2} \leq x_{b} \leq 1,
\qquad
  1 - x_{b} \leq x_{c} \leq x_{b}.
\end{align}
However, the $\tau^+\to\mu^+\mu^+\mu^-$ decay amplitude is singular for
$x_{b}\to 1$, where the invariant mass of the pair of $\mu^+$ (with
$\vec{p}_c$) and $\mu^-$ is vanishing within the approximation of 
neglecting the muon mass.
The singularity comes from the pole of the photon propagator in the
contribution from the dipole moment operator.
We introduce a cutoff parameter $0<\delta \ll 1$ and take the integration
interval of $x_{b}$ as
\begin{align}
   \frac{1}{2} \leq x_{b} \leq 1 - \delta,
\end{align}
in order to avoid this collinear singularity.

The $\delta$-dependent branching ratio is calculated by integration over
$x_{b,c}$, $\cos\theta$ and $\phi$ as
\begin{align}
  \text{Br} ( \delta ) &=
  \int_{-1}^{1} d\cos\theta
  \int_{0}^{2\pi} d\phi
  \int_{1/2}^{1-\delta} dx_b
  \int_{1-x_b}^{x_b} dx_c
  \frac{d^4\text{Br}}{dx_b dx_c \,d\phi \,d\!\cos \theta}.
\end{align}
The general formulae of differential branching ratios are shown 
in Appendix \ref{AppFormulae}.
We also define the following angular asymmetries:
\begin{subequations}
\begin{align}
  A_{Z}(\delta) &=
  \frac{1}{\text{Br}(\delta)}
  \left(
    \int_{0}^{1}
    -
    \int_{-1}^{0}
  \right) d\cos\theta
  \int_{0}^{2\pi}  d\phi
  \int_{1/2}^{1-\delta}  dx_b
  \int_{1-x_b}^{x_b}  dx_c
  \frac{d^4\text{Br}}{dx_b dx_c \,d\phi \,d\!\cos \theta},
\label{eq:AZ-tmmm}
\\
  A_{X}(\delta) &=
  \frac{1}{\text{Br}(\delta)}
  \int_{-1}^{1}  d\cos\theta
  \left(
    \int_{-\pi/2}^{\pi/2}
    -
    \int_{\pi/2}^{3\pi/2}
  \right) d\phi
  \int_{1/2}^{1-\delta}  dx_b
  \int_{1-x_b}^{x_b}  dx_c
  \frac{d^4\text{Br}}{dx_b dx_c \,d\phi \,d\!\cos \theta},
\label{eq:AX-tmmm}
\\
  A_{Y}(\delta) &=
  \frac{1}{\text{Br}(\delta)}
  \int_{-1}^{1}  d\cos\theta
  \left(
    \int_{0}^{\pi}
    -
    \int_{-\pi}^{0}
  \right) d\phi
  \int_{1/2}^{1-\delta}  dx_b
  \int_{1-x_b}^{x_b}  dx_c
  \frac{d^4\text{Br}}{dx_b dx_c \,d\phi \,d\!\cos \theta}.
\label{eq:AY-tmmm}
\end{align}
\end{subequations}

The branching ratio and asymmetries in the LHT are expressed as
\begin{align}
  \text{Br}( \tau^+ \to \mu^+ \mu^+ \mu^- )_{\text{LHT}}( \delta ) =&
  \text{Br}( \tau^+ \to \bar{\nu}_{\tau} e^+ \nu_e )
  B^{\text{I,LHT} }(\delta),
\end{align}
\begin{align}
  B^{\text{I,LHT} }(\delta) =&
	  C_{R1}^{\text{I,LHT} } A_1 (\delta) 
	+ C_{L2}^{\text{I,LHT} } A_2 (\delta)
	+ C_{L3}^{\text{I,LHT} } A_3 (\delta)
	+ C_{J1}^{\text{I,LHT} } A_4 (\delta) 
	+ C_{J3}^{\text{I,LHT} } A_5 (\delta),
\end{align}
\begin{subequations}
\begin{align}
	A_{Z}^{\text{I,LHT} } (\delta) =& \frac{1}{2B^{\text{I,LHT} }(\delta)} 
	\bigl(
	  C_{R1}^{\text{I,LHT} } B_1 (\delta) +C_{L2}^{\text{I,LHT} } B_2 (\delta) 
	 -C_{L3}^{\text{I,LHT} } A_3 (\delta) \n & \qquad \qquad
	 -C_{J1}^{\text{I,LHT} } A_4 (\delta) +C_{J3}^{\text{I,LHT} } A_5 (\delta)
	\bigr),\\
	A_{X}^{\text{I,LHT} } (\delta) =& \frac{1}{2 B^{\text{I,LHT} }(\delta)} 
	\bigl(
	 -C_{R1}^{\text{I,LHT} } C_1 (\delta) +C_{L2}^{\text{I,LHT} } C_2 (\delta) 
	 -C_{J1}^{\text{I,LHT} } C_3 (\delta) +C_{J3}^{\text{I,LHT} } C_4 (\delta)
	\bigr),\\
	A_{Y}^{\text{I,LHT} } (\delta) =& \frac{1}{2B^{\text{I,LHT} }(\delta)} 
	\bigl(
	 -C_{J5}^{\text{I,LHT} }  C_3 (\delta) +C_{J6}^{\text{I,LHT} } C_4 (\delta)
	\bigr).
\end{align}
\end{subequations}
The functions $A_i(\delta)$, $B_i(\delta)$, and $C_i(\delta)$ are given
in Appendix~\ref{AppFormulaeT1}.
$C_{R1}^{\text{I,LHT} }$, {\it etc.} are written in terms of the Wilson
coefficients in (\ref{eq:Leff-tmmm}) as
\begin{subequations}
\begin{align}
  C_{R1}^{\text{I,LHT} }&= 
    \abs{ eA_R^{\text{LHT}} }^2 ,\quad
  C_{L2}^{\text{I,LHT} } = 
    \abs{ g_{Lr}^{\text{I,LHT} } }^2 ,\quad
  C_{L3}^{\text{I,LHT} } = 
    \abs{ g_{Ll}^{\text{I,LHT} } }^2 , \\
  C_{J1}^{\text{I,LHT} }&= 
    \text{Re}[ eA_R^{\text{LHT}} g_{Ll}^{\text{I,LHT} \ast} ] ,\quad
  C_{J3}^{\text{I,LHT} } = 
    \text{Re}[ eA_R^{\text{LHT}} g_{Lr}^{\text{I,LHT} \ast} ] ,\\
  C_{J5}^{\text{I,LHT} }&= 
    \text{Im}[ eA_R^{\text{LHT}} g_{Ll}^{\text{I,LHT} \ast} ] ,\quad
  C_{J6}^{\text{I,LHT} } = 
    \text{Im}[ eA_R^{\text{LHT}} g_{Lr}^{\text{I,LHT} \ast} ] .
\end{align}
\end{subequations}
As explained in Appendix~\ref{AppFormulaeT1}, 
$A_Z$ and $A_X$ are parity odd asymmetries and $A_Y$ is
a time-reversal asymmetry.

The time-reversal asymmetry $A_Y$ is induced by the CP violation in the
effective Lagrangian, since we are considering a theory with CPT
invariance.
In fact, $C_{J5}^{\text{I,LHT}}$ and $C_{J6}^{\text{I,LHT}}$ are
proportional to the Jarlskog invariant \cite{Jarlskog:1985ht} defined
for $V_{H\ell}$:
\begin{align}
  C_{J5,J6}^{\text{I,LHT}} \propto
  J(V_{H\ell}) &=
  \text{Im} \left[ (V_{H\ell})_{11}^* (V_{H\ell})_{22}^*
    (V_{H\ell})_{12} (V_{H\ell})_{21} \right]
\nonumber\\
  &= c_{12}^{\ell} c_{13}^{\ell}{}^2 c_{23}^{\ell}
    s_{12}^{\ell} s_{13}^{\ell} s_{23}^{\ell} 
    \sin( \delta_{12}^{\ell} -\delta_{13}^{\ell} +\delta_{23}^{\ell} ).
\end{align}
Therefore, a three-generation mixing is necessary for nonvanishing
$A_Y$.
Notice that two of three complex phases in $V_{H\ell}$ are removable
when T-even neutrino masses are neglected.
That is why $J(V_{H\ell})$ depends on only one combination of the
phases.
\subsection{$\tau^+ \to \mu^+ e^+ e^-$ and $\tau^+ \to e^+ \mu^+ \mu^-$} 
\label{sec:tmeeII}

The effective Lagrangian for a type II trilepton decay
$\tau^+ \to \mu^+ e^+ e^-$ has a similar structure to
(\ref{eq:Leff-tmmm}):
\begin{align}
	\mathcal{L}_{\text{II}} =
	-\frac{4G_F}{\sqrt{2}} \bigl[
	& m_{\tau} A_R^{\text{LHT}} 
	  \bar{\tau}_{R} \sigma^{\mu \nu} \mu_L F_{\mu \nu}
	+ g_{Ll}^{\text{II,LHT}} 
	  ( \bar{\tau}_{ L} \gamma^{\mu} \mu_L )
	  ( \bar{e}_{L} \gamma_{\mu} e_{L} ) \n &
	+ g_{Lr}^{\text{II,LHT}} 
	  ( \bar{\tau}_{ L} \gamma^{\mu} \mu_L )
	  ( \bar{e}_{R} \gamma_{\mu} e_{R} )
	+ \text{H.c.}
	\bigr],
\end{align}
where
\begin{align}
  g_{Ll}^{\text{II,LHT}} = 
    \frac{ g^2 }{ ( 4 \pi )^2 } \frac{ v^2 }{ 8 f^2 } &\biggl(
    \sum_l ( V_{H\ell} )^{\ast}_{l3} ( V_{H\ell} )_{l2}
    \Bigl[ 
	   - \sin^2 \theta_W P_{\gamma}(x_l) 
		+ \Bigl( -\frac{1}{2} +\sin^2 \theta_W \Bigr) P_Z(x_l) 
	 \Bigr] \n &
   +\sum_{l,m} \Bigl[
      ( V_{H\ell} )^{\ast}_{l3} ( V_{H\ell} )_{l2}
	   ( V_{H\ell} )^{\ast}_{m1} ( V_{H\ell} )_{m1} \n & \qquad
    + ( V_{H\ell} )^{\ast}_{l3} ( V_{H\ell} )_{l1}
	   ( V_{H\ell} )^{\ast}_{m1} ( V_{H\ell} )_{m2}
	 \Bigr]
	 B_{(e)}(x_l,x_m) 
	 \biggr), 
\label{EqWilsontypeII}
\end{align}
and $g_{Lr}^{\text{II,LHT}}=g_{Lr}^{\text{I,LHT}}$.
$A_R^{\text{LHT}}$ is the same as (\ref{EqDipole}).

Kinematical variables are defined in a similar way to the type I case.
For the momentum assignment, we identify particles $a$, $b$ and $c$ in
Fig.~\ref{FigPl3l} as $\mu^+$, $e^+$ and $e^-$, respectively
\footnote{The convention of particle assignment is different from 
that in Ref.~\cite{Kitano:2000fg}}.
A difference from the type I case is that both $x_b>x_c$ and $x_b<x_c$
are possible in type II modes because there are no identical particles
in the final state.
Therefore, the range of $x_b$ and $x_c$ are
\begin{align}
  \delta \leq x_b \leq 1,
\qquad
  1 + \delta - x_b \leq x_c \leq 1,
\label{eq:xbxcrange-tmee1}
\end{align}
where the cutoff parameter $0<\delta\ll 1$ is introduced to avoid the
collinear singularity in $x_b+x_c\to 1$.

Formulae for the branching ratio and the angular asymmetries $A_{Z,X,Y}$ in
$\tau^+\to \mu^+ e^+ e^-$ decay in the LHT are obtained as
\begin{align}
  \text{Br}(\tau^+ \to \mu^+ e^+ e^-)_{\text{LHT}}(\delta) =&
  \text{Br}(\tau^+ \to \bar{\nu}_{\tau} e^+ \nu_e)
   B^{\text{II,LHT}} (\delta),
\end{align}
\begin{align}
  B^{\text{II,LHT}} (\delta) =&
	 C_{R1}^{\text{II,LHT}} D_1(\delta)
	+( C_{L3}^{\text{II,LHT}} +C_{L4}^{\text{II,LHT}}) D_3(\delta) \n &
	+( C_{J1}^{\text{II,LHT}} +C_{J3}^{\text{II,LHT}}) D_4(\delta),
\end{align}
\begin{subequations}
\begin{align}
 A_{Z}^{\text{II,LHT}} (\delta) = \frac{1}{2 B^{\text{II,LHT} }} 
 \Bigl(&
   -C_{R1}^{\text{II,LHT}} D_5(\delta)
	+( C_{L3}^{\text{II,LHT}} +C_{L4}^{\text{II,LHT}}) D_6(\delta) \n &
	-\frac{1}{3}( C_{J1}^{\text{II,LHT}} +C_{J3}^{\text{II,LHT}}) D_4(\delta)
  \Bigr),
\label{eq:AZ-II}
\\
  A_{X}^{\text{II,LHT}} (\delta) = \frac{\pi}{2 B^{\text{II,LHT} }}
  \bigl(&
     ( -C_{L3}^{\text{II,LHT}} +C_{L4}^{\text{II,LHT}} ) E_1(\delta)
   + (  C_{J1}^{\text{II,LHT}} -C_{J3}^{\text{II,LHT}} ) E_2(\delta)
  \bigr),
\\
  A_{Y}^{\text{II,LHT}} (\delta) = \frac{\pi}{2 B^{\text{II,LHT}} }
  \bigl(&
   -( C_{J7}^{\text{II,LHT}} - C_{J8}^{\text{II,LHT}}) E_3(\delta)
  \bigr),
\end{align}
\end{subequations}
where
\begin{subequations}
\begin{align}
  C_{R1}^{\text{II,LHT}} &
  = \abs{ eA_R^{\text{LHT}} }^2 ,\quad
  C_{L3}^{\text{II,LHT}} 
  = \abs{ g_{Lr}^{\text{II,LHT}} }^2 ,\quad
  C_{L4}^{\text{II,LHT}} 
  = \abs{ g_{Ll}^{\text{II,LHT}} }^2 , \\
  C_{J1}^{\text{II,LHT}} &
  = \text{Re}[ eA_R^{\text{LHT}} g_{Ll}^{\text{II,LHT}\ast} ] ,\quad
  C_{J3}^{\text{II,LHT}} 
  = \text{Re}[ eA_R^{\text{LHT}} g_{Lr}^{\text{II,LHT}\ast} ] ,\\
  C_{J7}^{\text{II,LHT}}&
  = \text{Im}[ eA_R^{\text{LHT}} g_{Ll}^{\text{II,LHT}\ast} ] ,\quad
  C_{J8}^{\text{II,LHT}} 
  = \text{Im}[ eA_R^{\text{LHT}} g_{Lr}^{\text{II,LHT}\ast} ] .
\end{align}
\end{subequations}
The functions $D_i(\delta)$, $E_i(\delta)$ and $F_i(\delta)$ are given
in Appendix~\ref{AppFormulaeT2}.

In the type II decay, another class of asymmetries can be defined.
We define the energy asymmetry in $\tau^+\to\mu^+ e^+ e^-$ as the
asymmetry between the partial widths with $E_{e^+} > E_{e^-}$ and
$E_{e^+} < E_{e^-}$ in the rest frame of initial $\tau^+$:
\begin{align}
  A_{FB} &=
  \frac{1}{\text{Br}}
  \left(
    \int_{x_b>x_c} - \int_{x_b<x_c}
  \right) dx_b dx_c
  \frac{d^2\text{Br}}{dx_b dx_c}.
\label{EqDeffbasym}
\end{align}
This asymmetry is also called as the forward-backward asymmetry, because
$x_b > x_c$ ($x_b < x_c$) corresponds to
$\vec{p}_{\mu^+}\cdot \vec{p}_{e^-}>0$
($\vec{p}_{\mu^+}\cdot \vec{p}_{e^-}<0$) in the rest frame of the
$e^+\,e^-$ pair.
Furthermore, we define the following asymmetries combining asymmetric
integrations over $(x_b,\,x_c)$ and $(\cos\theta,\,\phi)$:
\begin{subequations}
\begin{align}
	A_{ZFB} &=
  \frac{1}{\text{Br}(\delta)}
  \left(
    \int_{0}^{1}
    -
    \int_{-1}^{0}
  \right) d\cos\theta
  \int_{0}^{2\pi}  d\phi
  \left(
    \int_{x_b > x_c} 
    -
    \int_{x_b < x_c}
  \right) dx_b dx_c
  \frac{d^4\text{Br}}{dx_b dx_c \,d\phi \,d\!\cos \theta}, \\
	A_{XFB} &=
  \frac{1}{\text{Br}(\delta)}
  \int_{-1}^{1}  d\cos\theta
  \left(
    \int_{-\pi/2}^{\pi/2}
    -
    \int_{\pi/2}^{3\pi/2}
  \right) d\phi
  \left(
    \int_{x_b > x_c}
	 -
    \int_{x_b < x_c}
  \right) dx_b dx_c
  \frac{d^4\text{Br}}{dx_b dx_c \,d\phi \,d\!\cos \theta}, \\
	A_{YFB} &=
  \frac{1}{\text{Br}(\delta)}
  \int_{-1}^{1}  d\cos\theta
  \left(
    \int_{0}^{\pi}
    -
    \int_{-\pi}^{0}
  \right) d\phi
  \left(
    \int_{x_b > x_c}
	 -
    \int_{x_b < x_c}
  \right) dx_b dx_c
  \frac{d^4\text{Br}}{dx_b dx_c \,d\phi \,d\!\cos \theta}.
\label{EqDeffbasymY}
\end{align}
\end{subequations}
In the LHT, we obtain
\begin{subequations}
\begin{align}
  A_{FB}^{\text{II,LHT}} (\delta) = \frac{1}{B^{\text{II,LHT} }} 
  \biggl(&
	-\frac{1}{4}( C_{L3}^{\text{II,LHT}} -C_{L4}^{\text{II,LHT}}) D_2(\delta)
	+\frac{1}{2}( C_{J1}^{\text{II,LHT}} -C_{J3}^{\text{II,LHT}}) D_4(\delta)
  \biggr),
\\
  A_{ZFB}^{\text{II,LHT}} (\delta) = \frac{1}{2 B^{\text{II,LHT} }} 
  \biggl(&
     \frac{1}{4}( C_{L3}^{\text{II,LHT}} -C_{L4}^{\text{II,LHT}}) D_2(\delta) 
	 -\frac{1}{2}( C_{J1}^{\text{II,LHT}} -C_{J3}^{\text{II,LHT}}) D_4(\delta)
  \biggr),
\\
  A_{XFB}^{\text{II,LHT}} (\delta) = \frac{1}{2 B^{\text{II,LHT} }} 
  \biggl(&
    C_{R1}^{\text{II,LHT}} E_4(\delta)
   +\frac{4}{3}( C_{L3}^{\text{II,LHT}} 
	             +C_{L4}^{\text{II,LHT}} ) E_1 (\delta) \n &
   +\frac{4}{3}( C_{J1}^{\text{II,LHT}} 
	             +C_{J3}^{\text{II,LHT}} ) E_2 (\delta)
  \biggr),
\\
  A_{YFB}^{\text{II,LHT}} (\delta) = \frac{1}{2 B^{\text{II,LHT} }} 
  \biggl(&
   -\frac{4}{3}( C_{J7}^{II} + C_{J8}^{II}) E_3 (\delta) 
  \biggr),
\end{align}
\end{subequations}
where the functions $G_i(\delta)$ and $H_i(\delta)$ are given in
Appendix~\ref{AppFormulaeT2}.

\subsection{$\tau^+ \to \mu^+ \mu^+ e^-$ and $\tau^+ \to e^+ e^+ \mu^-$} 

The type III leptonic decay modes, such as $\tau^+ \to \mu^+ \mu^+ e^-$,
are those in which the lepton flavor changes by two.
At the one-loop level, only the box diagrams contribute to these decay
modes.
The effective Lagrangian for $\tau^+ \to \mu^+ \mu^+ e^-$ is
\begin{align}
	\mathcal{L}_{\text{III}} =
	-\frac{4G_F}{\sqrt{2}} \bigl[ &
	 g_{Ll}^{\text{III,LHT} } 
	   ( \bar{\tau}_L \gamma^{\mu} \mu_L )
	   ( \bar{e}_L    \gamma_{\mu} \mu_L )
	+ \text{H.c.}
	\bigr],
\end{align}
where
\begin{align}
  g_{Ll}^{\text{III,LHT} } =&
    \frac{ g^2 }{ ( 4 \pi )^2 } \frac{ v^2 }{ 8 f^2 }
   \sum_{l,m}
      ( V_{H\ell} )^{\ast}_{l3} ( V_{H\ell} )_{l2}
	   ( V_{H\ell} )^{\ast}_{m1} ( V_{H\ell} )_{m2} 
	 B_{(e)}(x_l,x_m).
\end{align}

The kinematics of the type III decay is treated in a way  similar to the type I
case.
For $\tau^+ \to \mu^+ \mu^+ e^-$ mode, $e^-$ and one of the $\mu^+$'s,
which has a larger energy in the rest frame of $\tau^+$, is identified
as particles $a$ and $b$ in Fig.~\ref{FigPl3l}, respectively.
We define the angular asymmetries $A_Z$, $A_X$ and $A_Y$ by
(\ref{eq:AZ-tmmm}), (\ref{eq:AX-tmmm}) and (\ref{eq:AY-tmmm}),
respectively.
The forward-backward type asymmetries are not defined because there are
identical particles in the final state as in the type I case.

The branching ratio in the LHT is written as
\begin{align}
	\frac{\text{Br}(\tau^+ \to \mu^+ \mu^+ e^- )_{\text{LHT}}(\delta)}
        {\text{Br}(\tau^+ \to \bar{\nu}_{\tau} e^+ \nu_e) }
	=& C_{L3}^{\text{III,LHT} } A_3 (\delta),
\end{align}
where
\begin{align}
   C_{L3}^{\text{III,LHT} } &=
    \abs{ g_{Ll}^{\text{III,LHT} } }^2 
\end{align}
and $A_3(\delta)$ is given in Appendix~\ref{AppFormulaeT1}.
Since the differential width is written in terms of one combination of
the Wilson coefficient, $C_{L3}^{\text{III,LHT} }$, the angular
asymmetries are determined independently of the input parameters:
\begin{align}
	A_{Z}^{\text{III,LHT} } =& -\frac{1}{2},
\qquad
	A_{X}^{\text{III,LHT} } = 0,
\qquad
	A_{Y}^{\text{III,LHT} } = 0,
\end{align}
\subsection{$\tau^+ \to \mu^+ \pi, \eta, \eta' $ and 
            $\tau^+ \to e^+ \pi, \eta, \eta' $ }  

The effective Lagrangian for semileptonic $\tau\to\mu$ decay processes
are written as
\begin{align}
	\mathcal{L}_{\text{had}} =
	-\frac{4G_F}{\sqrt{2}} & \Bigl[
	m_{\tau} A_R^{\text{LHT}}
	  \bar{\tau}_{R} \sigma^{\mu \nu} \mu_L F_{\mu \nu} \n &
	+ \sum_{q = u,d,s} \left(
	    g_{Ll(q)}^{\text{LHT}} 
	    ( \bar{\tau}_L \gamma^{\mu} \mu_L ) ( \bar{q}_L \gamma_{\mu} q_L )
	  + g_{Lr(q)}^{\text{LHT}} 
	    ( \bar{\tau}_L \gamma^{\mu} \mu_L ) ( \bar{q}_R \gamma_{\mu} q_R ) 
	  \right)
	+ \text{H.c.}
	\Bigl].
\label{EqHlfv}
\end{align}
The Wilson coefficients are given as
\begin{align}
 g_{Ll(q)}^{\text{LHT}} &
 = \frac{ g^2 }{ ( 4 \pi )^2 } \frac{ v^2 }{ 8 f^2 }
   \sum_l ( V_{H\ell} )^{\ast}_{l3} ( V_{H\ell} )_{l2} \n & \quad \times
   \left( 
    Q_q \sin^2 \theta_W P_{\gamma}(x_l) 
	 + ( T^3_q -Q_q \sin^2 \theta_W ) P_Z(x_l) 
    + \sum_m ( V_{Hq} )^{\ast}_{m1} ( V_{Hq} )_{m1} B_{(q)}(x_l,x_m) 
   \right),
\label{eq:gLlq-tm}
\\
 g_{Lr(q)}^{\text{LHT}} & 
 = \frac{ g^2 }{ ( 4 \pi )^2 } \frac{ v^2 }{ 8 f^2 }
   \sum_l ( V_{H\ell} )^{\ast}_{l3} ( V_{H\ell} )_{l2}\,
   Q_q \sin^2 \theta_W \bigl( P_{\gamma}(x_l) - P_Z (x_l) \bigr),
\label{eq:gLrq-tm}
\end{align}
where $T_q^3$ and $Q_q$ are the weak isospin and 
the electromagnetic charge for quarks, respectively.
$A_R^{\text{LHT}}$ is given in (\ref{EqDipole}).

For the two-body decay process $\tau^+\to\mu^+ P$, where $P$ stands for a
neutral pseudoscalar meson $\pi^0$, $\eta$ or $\eta'$, the differential
branching ratio is written as
\begin{align}
  \frac{d\text{Br}(\tau^+ \to \mu^+ P)_{ \text{LHT}} }{d\cos\theta} &= 
  \tau_{\tau} \frac{G_F^2 m_{\tau}^3}{8\pi}
  \left( 1 - \frac{m_P^2}{m_{\tau}^2} \right)^2 \abs{ G_{RP}^{\text{LHT}} }^2
  (1 + \cos\theta),
\label{eq:dBdtheta-tmP}
\end{align}
where the decay angle $\theta$ is defined in the same way as in the
$\tau^+\to\mu^+\gamma$ case.
The effective coupling $G_{RP}^{\text{LHT}}$ for $P=\pi^0$ and $P=\eta$
are written in terms of the Wilson coefficients in (\ref{EqHlfv}) as
\begin{align}
  G_{R \pi}^{\text{LHT}} &=
    \frac{f_{\pi} }{2\sqrt{2}}
	 \left( g_{Lr(u)}^{\text{LHT}} -g_{Lr(d)}^{\text{LHT}} 
	  -g_{Ll(u)}^{\text{LHT}} +g_{Ll(d)}^{\text{LHT}} \right), \\
  G_{R \eta}^{\text{LHT}} &=
    \frac{1}{2} \left(
	 \frac{f_{\eta}^q}{\sqrt{2}}
	   ( g_{Lr(u)}^{\text{LHT}} +g_{Lr(d)}^{\text{LHT}} 
		 -g_{Ll(u)}^{\text{LHT}} -g_{Ll(d)}^{\text{LHT}} )
   + f_{\eta}^s ( g_{Lr(d)}^{\text{LHT}} -g_{Ll(d)}^{\text{LHT}} )
	 \right). 
\end{align}
$f_\pi$, $f_\eta^q$ and $f_\eta^s$ are the decay constants of the
pseudoscalar mesons that are defined in Appendix~\ref{AppFormulaeMeson}.
The effective coupling for $\tau^+\to\mu^+\eta'$ is obtained by replacing
the subscript $\eta$ with $\eta'$ in the expression of $G_{R \eta}$.
Formulae for the branching ratio and the polarization asymmetry $A_P$ are
derived from (\ref{eq:dBdtheta-tmP}) as
\begin{align}
  \text{Br}(\tau^+ \to \mu^+ P)_{\text{LHT}} &= 
    \tau_{\tau} \frac{G_F^2 m_{\tau}^3 }{4\pi}
      \left( 1 - \frac{m_P^2}{m_{\tau}^2} \right)^2 \abs{G_{RP}}^2 , 
\\
  A_P(\tau^+ \to \mu^+ P)_{\text{LHT}} &= \frac{1}{2}.
\end{align}
$A_P$ is purely determined by the chirality structure of the effective
Lagrangian (\ref{EqHlfv}) and independent of the values of the input
parameters.

\subsection{$\tau^+ \to \mu^+ \rho, \omega, \phi $ and 
            $\tau^+ \to e^+ \rho, \omega, \phi $ }

The semileptonic two-body decay $\tau^+ \to \mu^+ V$, where $V$ denotes
a neutral vector meson $\rho^0$, $\omega$, or $\phi$ are also described
by the effective Lagrangian (\ref{EqHlfv}).
The branching ratio and the polarization asymmetry $A_V$, which is
defined similarly to $A_\gamma$ and $A_P$, are written as
\begin{align}
 \text{Br}(\tau^+ \to \mu^+ V)_{\text{LHT}} &
  = \tau_{\tau} \frac{G_F^2 m_\tau^3}{\pi} 
    \left( 1 - \frac{m_V^2}{m_{\tau}^2} \right)^2
    B_V^{\text{LHT}},
\\
 B_V^{\text{LHT}} &=
     \abs{G_{RAV}}^2 \left( 2 + \frac{m_V^2}{m_{\tau}^2} \right) 
   + \abs{G_{LV }}^2 \frac{m_{\tau}^2 +2 m_V^2}{4 m_V^2} 
   - 3 \text{Re}[ G_{RAV} G_{LV}^{\ast} ],
\\
 A_V(\tau^+ \to \mu^+ V)_{\text{LHT}} &
 = \frac{1}{2B_V^{\text{LHT}} } \n \times &
\left(
 	  \abs{G_{RAV}}^2 \left( 2 - \frac{m_V^2}{m_{\tau}^2} \right) 
 	+ \abs{G_{LV }}^2 \left( \frac{m_{\tau}^2 -2 m_V^2}{4 m_V^2} \right)
 	+ \text{Re}[ G_{RAV} G_{LV}^{\ast} ]
\right).
\end{align}
The effective coupling constants for $V=\rho^0$, $\omega$ and $\phi$ are
\begin{subequations}
\begin{align}
 G_{RA\rho}^{\text{LHT}} &
 = \frac{f_{\rho}}{\sqrt{2}} \frac{m_{\tau}}{m_{\rho}} 
    eA_R^{\text{LHT}} ,\\
 G_{L\rho}^{\text{LHT}} &
 = \frac{f_{\rho} m_{\rho}}{2\sqrt{2} m_\tau}
	\left( g_{Ll(u)}^{\text{LHT}} -g_{Ll(d)}^{\text{LHT}} 
	 +g_{Lr(u)}^{\text{LHT}} -g_{Lr(d)}^{\text{LHT}} \right) ,\\
 G_{RA\omega}^{\text{LHT}} &
 = \frac{f_{\omega}}{3\sqrt{2}} 
    \frac{m_{\tau}}{m_{\omega}} eA_R^{\text{LHT}}  ,\\
 G_{L\omega}^{\text{LHT}} &
 = \frac{f_{\omega} m_{\omega}}{2\sqrt{2} m_\tau }
	\left( g_{Ll(u)}^{\text{LHT}} +g_{Ll(d)}^{\text{LHT}} 
	 +g_{Lr(u)}^{\text{LHT}} +g_{Lr(d)}^{\text{LHT}} \right) ,\\
 G_{RA\phi}^{\text{LHT}} &
 = \frac{f_{\phi}}{3} \frac{m_{\tau}}{m_{\phi}} 
   \, eA_R^{\text{LHT}}  ,\\
 G_{L\phi}^{\text{LHT}} &
 = -\frac{f_{\phi} m_{\phi}}{2 m_\tau}
	\left( g_{Ll(s)}^{\text{LHT}} +g_{Lr(s)}^{\text{LHT}} \right),
\end{align}
\end{subequations}
where the decay constants $f_{\rho,\omega,\phi}$ are defined in
Appendix~\ref{AppFormulaeMeson}.

\subsection{$\mu-e$ conversion} 

The effective Lagrangian for the coherent $\mu-e$ conversion processes
has the same structure as (\ref{EqHlfv}).
\begin{align}
 \mathcal{L}_{\text{had}} 
 = -\frac{4G_F}{\sqrt{2}} & 
 \Bigl[
   m_{\mu} A_{R}^{\text{LHT}} 
    \bar{\mu}_{R} \sigma^{\mu \nu} e F_{\mu \nu}  \n & 
    +\sum_{q=u,d} \Bigl( 
	   g_{Ll(q)}^{\text{LHT}}
      ( \bar{\mu}_{L} \gamma^{\mu} e_{L}) ( \bar{q}_{L} \gamma_{\mu} q_{L} )
     +g_{Lr(q)}^{\text{LHT}}
      ( \bar{\mu}_{L} \gamma^{\mu} e_{L}) ( \bar{q}_{R} \gamma_{\mu} q_{R} )
	 \Bigr)
   +\text{H.c.}
 \Bigl].
\label{eq:mueconvLeffLHT}
\end{align}
The Wilson coefficients are obtained by replacing the lepton flavor
indices 3($\tau$) and 2($\mu$) with 2($\mu$) and 1($e$) in
Eqs.~(\ref{EqDipole}), (\ref{eq:gLlq-tm}) and (\ref{eq:gLrq-tm}).

We calculate the $\mu-e$ conversion rates following the method given
in Ref.~\cite{Kitano:2002mt}.
The $\mu-e$ conversion branching ratio in the LHT is written as
\begin{align}
	\text{R}(\mu^- A\to e^- A)_{\text{LHT}}
	= \frac{2 G_F^2}{\omega_{\text{capt} }} & 
	\Bigl|
	 - A_{R}^{\text{LHT}} D 
	 + 2( 2g_{Ll(u)}^{\text{LHT}} + 2g_{Lr(u)}^{\text{LHT}}  
	 +     g_{Ll(d)}^{\text{LHT}} +  g_{Lr(d)}^{\text{LHT}} ) V^{(p)} \n &
	 + 2(  g_{Ll(u)}^{\text{LHT}} +  g_{Lr(u)}^{\text{LHT}} 
	 +    2g_{Ll(d)}^{\text{LHT}} + 2g_{Lr(d)}^{\text{LHT}} ) V^{(n)}
   \Bigr|^2 ,
\end{align}
where $\omega_{\text{capt}}$ is the muon capture rate.
$D$, $V^{(p)}$ and $V^{(n)}$ are the overlap integrals defined in
Ref.~\cite{Kitano:2002mt}.
For the reader's convenience we quote the values for 
Al, Ti, Au and Pb in Appendix~\ref{AppFormulaeMeconv}.

\section{Numerical results}
In this section we show numerical results for the LFV observables 
given in the previous section.
We have to specify 19 parameters in order to calculate LFV observables
in the LHT:
the decay constant $f$, 
six masses of the T-odd fermions $m_{H\ell}^i$ and $m_{Hq}^i$, six mixing
angles and six phases in the mixing matrices $V_{H\ell}$ and $V_{Hd}$.
The value of $f$ is constrained to $f\gtrsim 500$~GeV by the electroweak
precision measurements \cite{Asano:2006nr}.
Throughout the analysis in this paper, we fix $f$ as $f=500$~GeV.
Since all the LFV amplitudes are proportional to $f^{-2}$ as shown in
Sec.~\ref{SecLhtlfv}, the branching ratios scale as $f^{-4}$.
We also assume that the T-odd quarks (except for the top partner) are
degenerate in mass for simplicity.
Under this assumption, T-odd particle loops do not induce additional
contributions to the quark FCNC observables, and the mixing and the phase
parameters in $V_{Hd}$ are irrelevant.
We fix the T-odd quark mass as 500~GeV.
The top-partner quark mass is irrelevant for the LFV studied here.
There remain nine free parameters in the T-odd lepton sector.
We vary these parameters independently within the ranges
$100~\text{GeV} \leq m_{H\ell}^i \leq 1~\text{TeV}$ ($i=1,\ 2,\ 3$),
$0\leq\theta^{\ell}_{ij}<2\pi$ [$(ij)=(12),\ (23),\ (13)$]
and
$0\leq\delta^{\ell}_{ij}<2\pi$ [$(ij)=(12),\ (23),\ (13)$].
For the type I and type II trilepton decays, we use following 
cutoff parameters:
\begin{align}
 \delta =
 \begin{cases}
	3\times 10^{-4} \sim 3\left( \frac{2m_e}{m_{\mu}} \right)^2 &
	: \mu^+ \to e^+e^+e^-, \\
   0.04 \sim 3\left( \frac{2m_{\mu}}{m_{\tau}} \right)^2 &
	: \tau^+ \to \mu^+\mu^+\mu^-,\ \tau^+ \to e^+\mu^+\mu^- ,\\
	10^{-6} \sim 3\left( \frac{2m_e}{m_{\tau}} \right)^2 &
	: \tau^+ \to e^+e^+e^-,\ \tau^+ \to \mu^+e^+e^- .
 \end{cases}
\end{align}
For the type III modes we take $\delta=0$ since there are no singularity
in the differential widths.
 
We evaluate the observables explained in Sec.~\ref{SecLhtlfv} for each
model parameter set, and check if that set is allowed under current
experimental constraints.
At present, the upper limits of the branching ratios are available for
various LFV processes.
We use the values summarized in Table~\ref{TabBound}.

\begin{table} 
\begin{tabular}[t]{|lll|}
\hline
Mode & Upper limit & Ref. \\
\hline
$\mu^+ \to e^+ \gamma $ & $1.2\times10^{-11}$ & \cite{RefMuegamma} \\
$\mu^+ \to e^+ e^+ e^-$ & $1.0\times10^{-12}$ & \cite{RefMu3e} \\
$\mu^-$Ti$\to e^-$Ti    & $4.3\times10^{-12}$ & \cite{RefMueconvTi} \\
$\mu^-$Au$\to e^-$Au    & $0.7\times10^{-12}$ & \cite{RefMueconvAu} \\
$\mu^-$Pb$\to e^-$Pb    & $4.6\times10^{-11}$ & \cite{RefMueconvPb} \\
\hline
\end{tabular}
\begin{tabular}[t]{|lll|}
\hline
Mode & Upper limit & Ref. \\
\hline
$\tau^+ \to \mu^+ \gamma     $ & $4.4\times10^{-8}$ & \cite{RefAubert10} \\
$\tau^+ \to \mu^+ \mu^+ \mu^-$ & $3.2\times10^{-8}$ & \cite{RefMiyazaki08} \\
$\tau^+ \to \mu^+ e^+ e^-    $ & $2.7\times10^{-8}$ & \cite{RefMiyazaki08} \\
$\tau^+ \to \mu^+ \pi^0      $ & $1.1\times10^{-7}$ & \cite{RefAubert07i} \\
$\tau^+ \to \mu^+ \eta       $ & $6.5\times10^{-8}$ & \cite{RefMiyazaki07} \\
$\tau^+ \to \mu^+ \eta'      $ & $1.3\times10^{-7}$ & \cite{RefMiyazaki07} \\
$\tau^+ \to \mu^+ \rho^0     $ & $2.6\times10^{-8}$ & \cite{RefAubert09} \\
$\tau^+ \to \mu^+ \omega     $ & $8.9\times10^{-8}$ & \cite{RefNishio} \\
$\tau^+ \to \mu^+ \phi       $ & $1.3\times10^{-7}$ & \cite{RefNishio} \\
\hline
\end{tabular}
\begin{tabular}[t]{|lll|}
\hline
Mode & Upper limit & Ref. \\
\hline
$\tau^+ \to e^+ \gamma     $ & $3.3\times10^{-8}$ & \cite{RefAubert10} \\
$\tau^+ \to e^+ e^+ e^-    $ & $3.6\times10^{-8}$ & \cite{RefMiyazaki08} \\
$\tau^+ \to e^+ \mu^+ \mu^-$ & $3.7\times10^{-8}$ & \cite{RefAubert07bk} \\
$\tau^+ \to e^+ \pi^0      $ & $8.0\times10^{-8}$ & \cite{RefMiyazaki07} \\
$\tau^+ \to e^+ \eta       $ & $9.2\times10^{-8}$ & \cite{RefMiyazaki07} \\
$\tau^+ \to e^+ \eta'      $ & $1.6\times10^{-7}$ & \cite{RefMiyazaki07} \\
$\tau^+ \to e^+ \rho^0     $ & $4.6\times10^{-8}$ & \cite{RefAubert09} \\
$\tau^+ \to e^+ \omega     $ & $1.1\times10^{-7}$ & \cite{RefAubert08k} \\
$\tau^+ \to e^+ \phi       $ & $3.1\times10^{-8}$ & \cite{RefAubert09} \\
\hline
\end{tabular}
\begin{tabular}[t]{|lll|}
\hline
Mode & Upper limit & Ref. \\
\hline
$\tau^+ \to \mu^+ \mu^+ e^-$ & $2.3\times10^{-8}$ & \cite{RefMiyazaki08} \\
$\tau^+ \to e^+ e^+ \mu^-  $ & $2.0\times10^{-8}$ & \cite{RefMiyazaki08} \\
\hline
\end{tabular}
\caption{Experimental upper bounds of various LFV branching ratios.}
\label{TabBound}
\end{table} 
\subsection{$\mu$ LFV} 

The current experimental bounds of $\mu^+ \to e^+ \gamma$ and 
$\mu^+ \to e^+ e^+ e^-$ are given by the MEGA \cite{RefMuegamma} 
and SINDRUM I \cite{RefMu3e} collaborations, respectively.
Also, the SINDRUM II collaboration provides us
with upper limits on the $\mu-e$ conversion rates 
for titanium~\cite{RefMueconvTi}, gold~\cite{RefMueconvAu}, and 
lead~\cite{RefMueconvPb}.
The upper bounds given by these experiments are summarized in
Table~\ref{TabBound}.
At present, the MEG experiment is ongoing in search for
$\mu^+ \to e^+ \gamma$, and an upper bound
$\text{Br}(\mu^+ \to e^+ \gamma)<1.5 \times 10^{-11}$ has been reported
recently~\cite{RefMegichep}.
For $\mu-e$ conversions, COMET and Mu2e
experiments are in preparation~\cite{RefComet}.

\begin{figure}[htbp] 
\begin{tabular}{cc}
  \includegraphics[width=20em,clip]{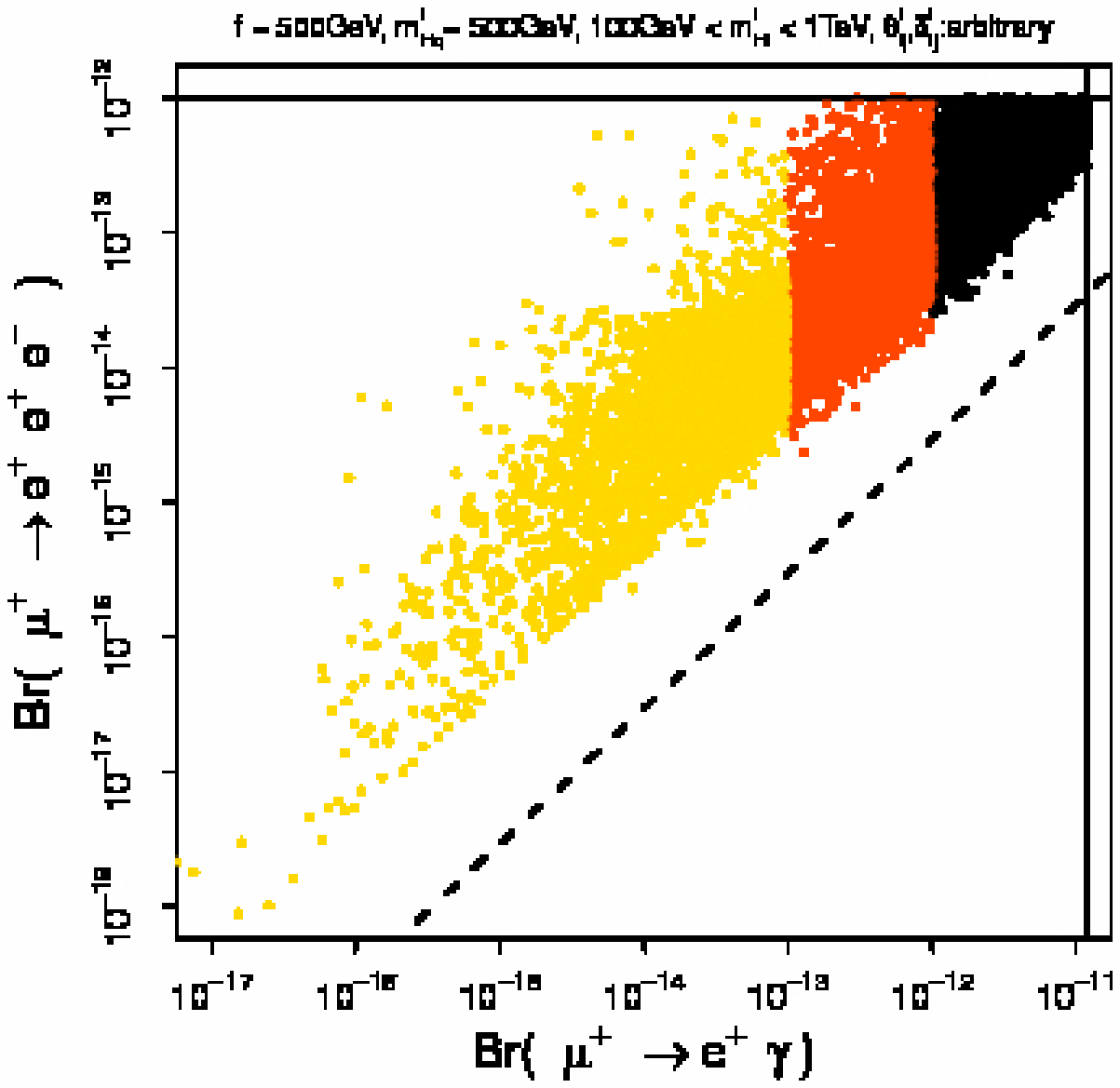} &
  \includegraphics[width=20em,clip]{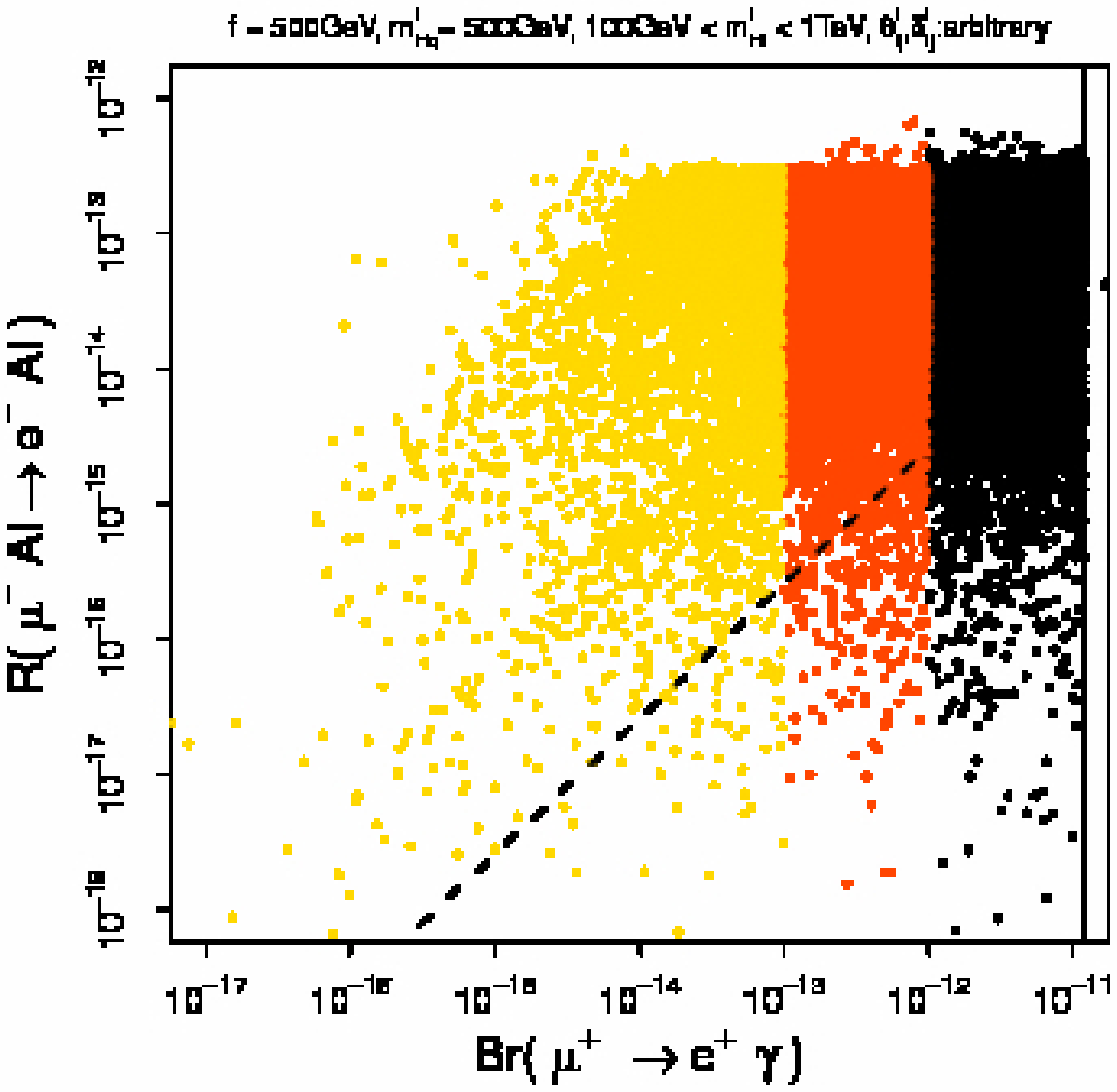} \\
  (a) & (b)
\end{tabular}
\begin{tabular}{c}
  \includegraphics[width=20em,clip]{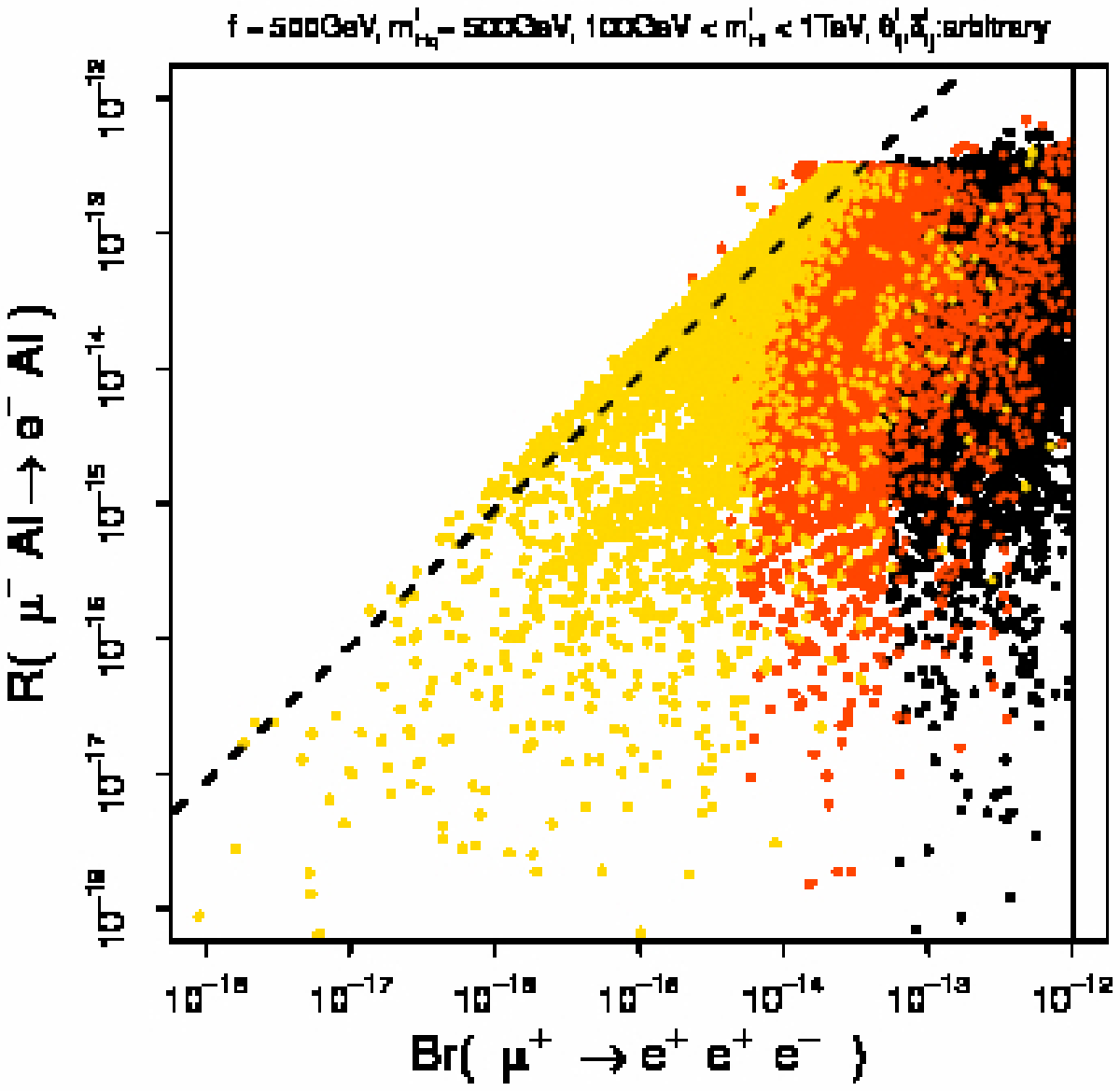} \\
(c)
\end{tabular}
\caption{
Correlations among the branching ratios of 
$\mu^+ \to e^+ \gamma$, $\mu^+ \to e^+ e^+ e^-$ and 
$\mu^- \text{Al} \to e^- \text{Al}$ in the LHT, where the decay constant
$f$ is taken as 500 GeV.
The T-odd lepton masses are varied in the range from 100 GeV to 1 TeV.
The mixing angles and the phases in the mixing matrix $V_{H\ell}$ are also
varied in the whole range.
The T-odd quark masses are fixed as 500 GeV.
The horizontal and vertical lines are experimental upper bounds.
The dashed lines show the branching ratios calculated with the dipole
contributions only.
The color of each dot represents the
value of $\text{Br}(\mu^+\to e^+ \gamma)$: black, red/gray and
yellow/light gray correspond to
$10^{-12}<\text{Br}(\mu^+\to e^+ \gamma)< 1.2\times 10^{-11}$, 
$10^{-13}< \text{Br}(\mu^+\to e^+ \gamma)\leq 10^{-12}$
and
$\text{Br}(\mu^+\to e^+ \gamma)\leq 10^{-13}$, respectively.
}
\label{FigMucor}
\end{figure} 

We present correlations among the branching ratios of
$\mu^+\to e^+ \gamma$, $\mu^+\to e^+ e^+ e^-$
and
$\mu^- \text{Al}\to e^- \text{Al}$ in Fig.~\ref{FigMucor}.
The color of each dot represents the
value of $\text{Br}(\mu^+\to e^+ \gamma)$: black, red/gray and
yellow/light-gray correspond to
$10^{-12}<\text{Br}(\mu^+\to e^+ \gamma)< 1.2\times 10^{-11}$, 
$10^{-13}< \text{Br}(\mu^+\to e^+ \gamma)\leq 10^{-12}$ and,
$\text{Br}(\mu^+\to e^+ \gamma)\leq 10^{-13}$, respectively.
In all the scatter plots hereafter, we use the same color code.
The current experimental upper limits on $\text{Br}(\mu^+\to e^+ \gamma)$
and $\text{Br}(\mu^+\to e^+ e^+ e^-)$ are shown by horizontal and
vertical lines.
The dashed lines show the branching ratios calculated with only the
contributions from the dipole moment type operator
$\bar{\mu}_R \sigma^{\mu\nu} e_L F_{\mu\nu}$.
In a certain class of models, such as SUSY, it is known that the LFV effect 
dominantly appears in the dipole moment operator.
In such a model, the branching ratios are predicted to be on the dashed
lines in the correlation plots.
In contrast with those models, the correlations in the LHT show that
the contributions of the dipole moment operator are less significant
than those from other four-Fermi interaction terms in
$\mu^+\to e^+ e^+ e^-$ and $\mu^- \text{Al}\to e^- \text{Al}$,
as noted in Ref.~\cite{Blanke:2007db}.
We can see that there is rather strong correlation 
between the branching ratios of $\mu^+ \to e^+ \gamma$ and 
$\mu^+ \to e^+ e^+ e^-$
whereas no correlation is observed in the branching ratios of 
$\mu^+ \to e^+ \gamma$ and the conversion rate for Al.

\begin{figure}[htbp] 
\begin{tabular}{cc}
  \includegraphics[width=20em,clip]{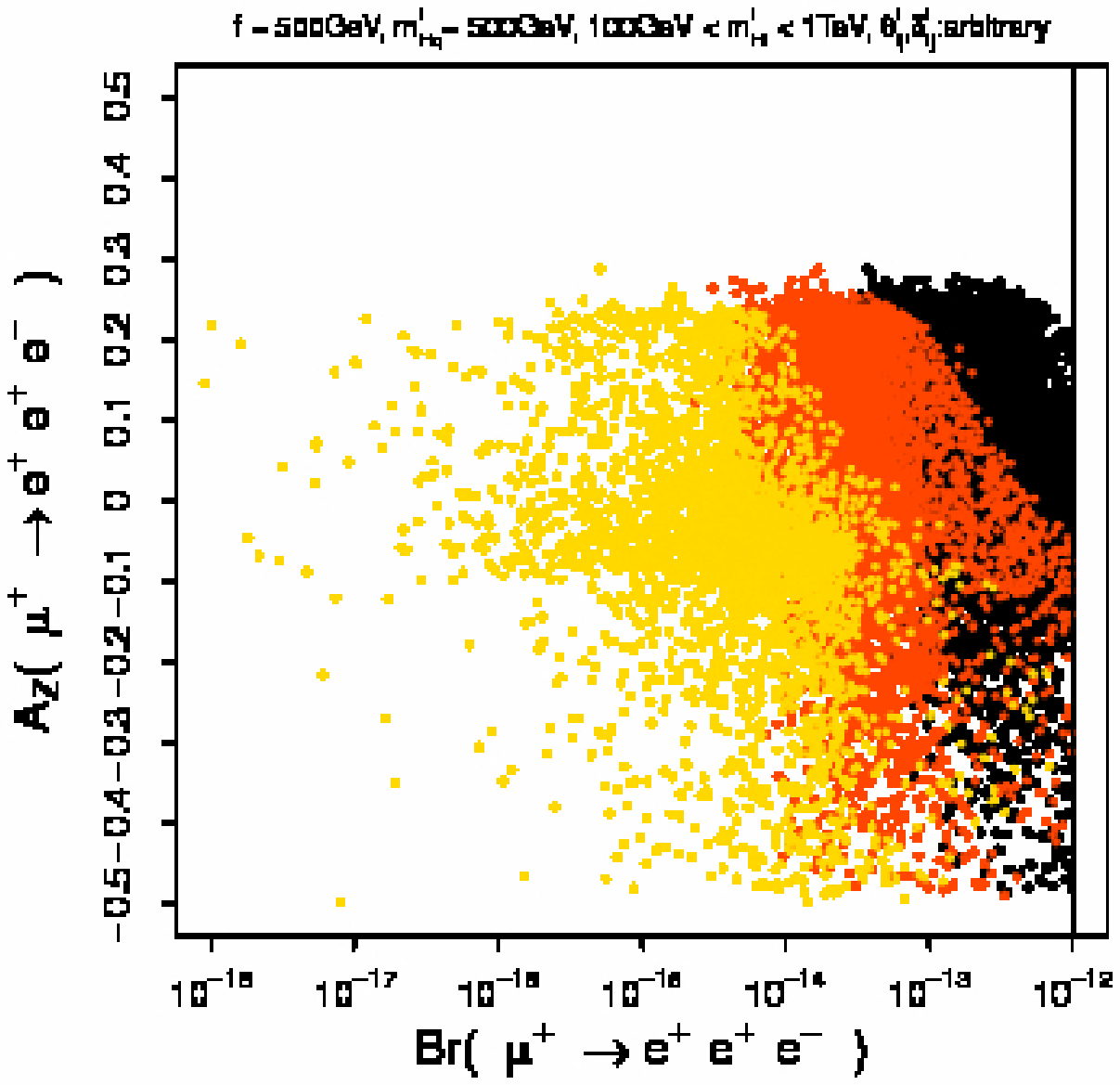} &
  \includegraphics[width=20em,clip]{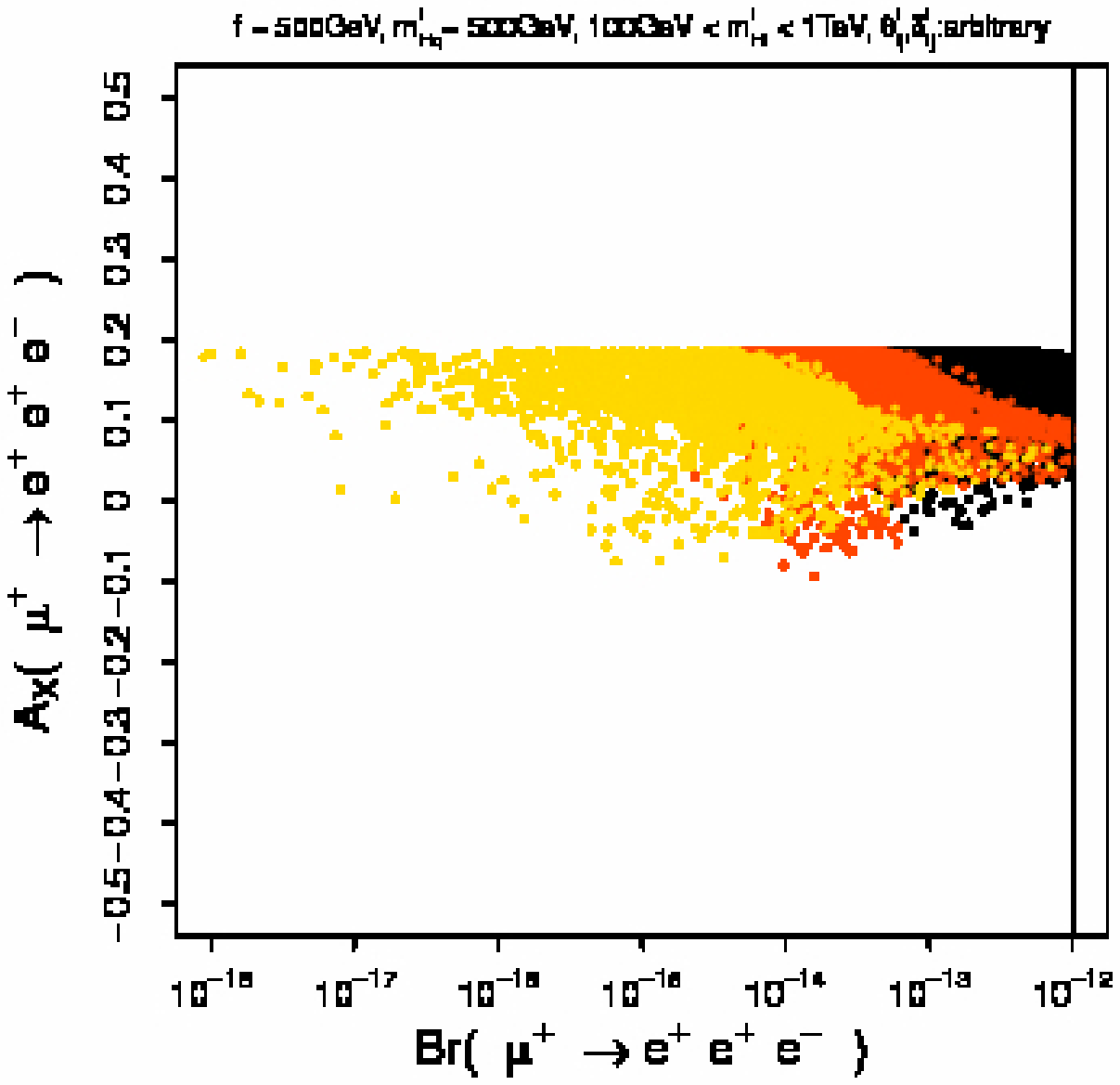} \\
  (a) & (b) \\
  \includegraphics[width=20em,clip]{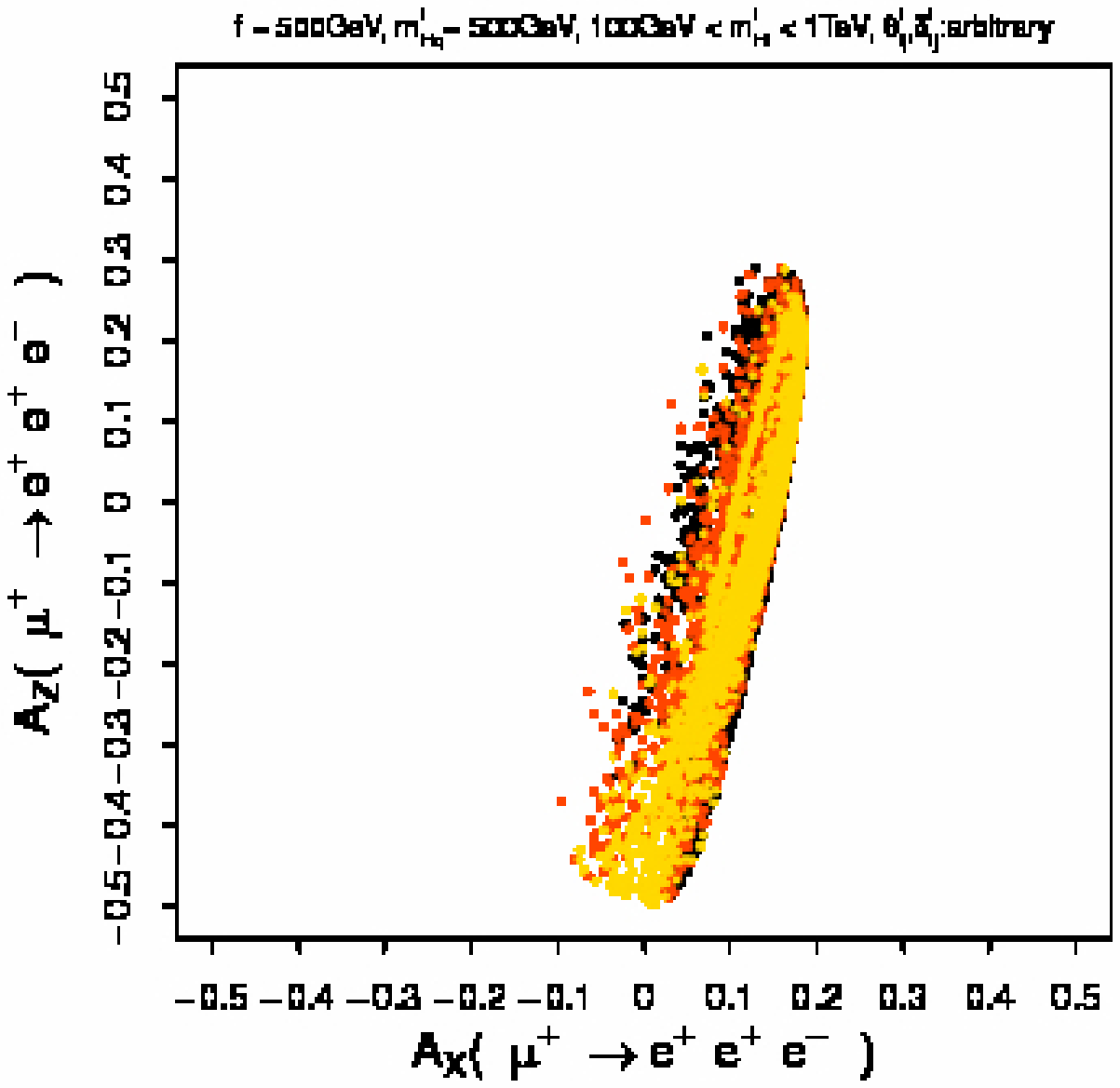} &
  \includegraphics[width=20em,clip]{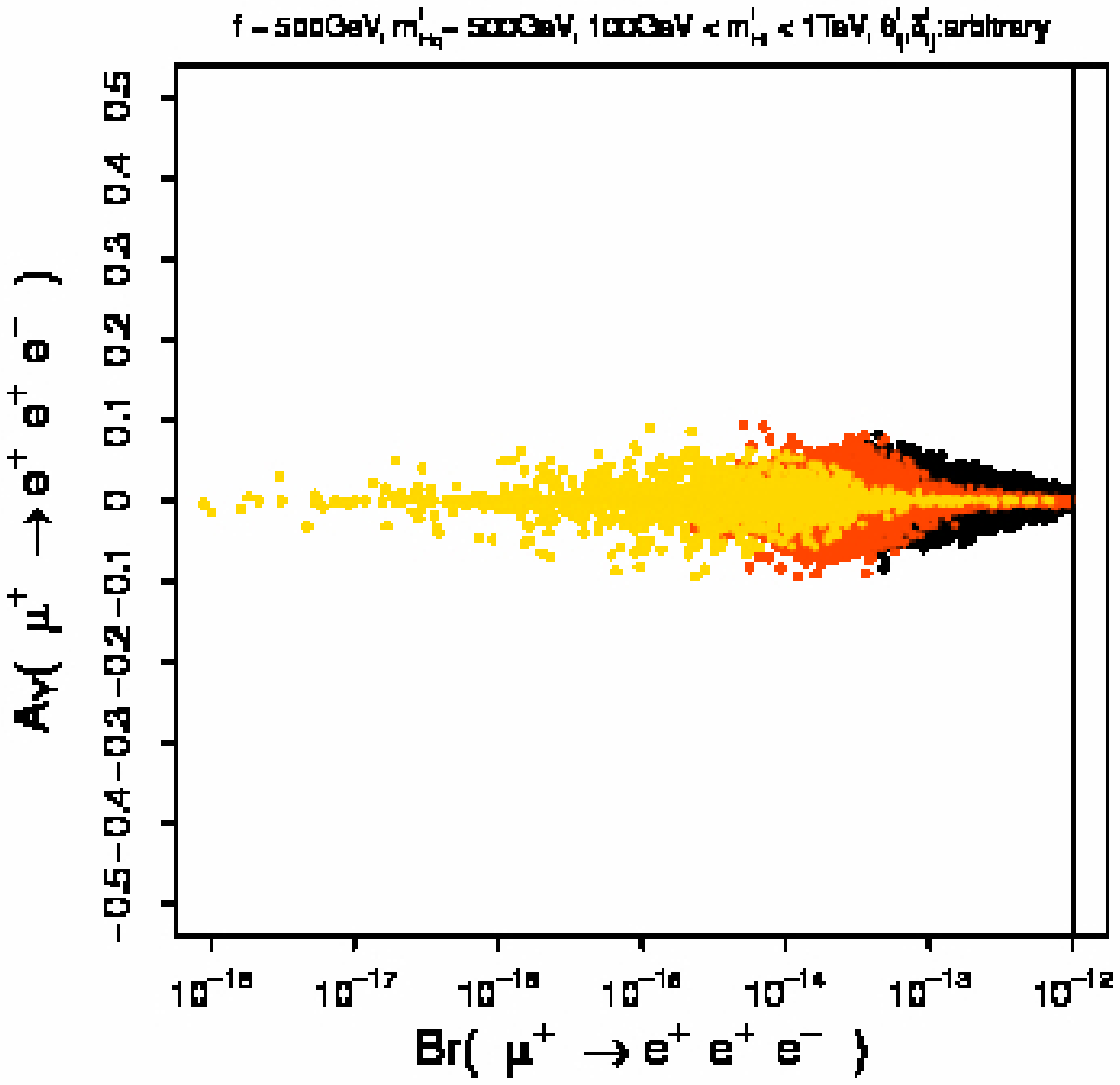} \\
  (c) & (d) \\
\end{tabular}
\caption{
Angular asymmetries of $\mu^+ \to e^+ e^+ e^-$ as functions of the 
branching ratio in the LHT for the same parameter set as in
Fig.~\ref{FigMucor}.
The correlation between $A_Z$ and $A_X$ is also shown in (c).
The vertical solid lines in (a), (b), and (d) is the experimental
upper limit of  $\text{Br}(\mu^+ \to e^+ e^+ e^-)$.
The color code is the same as in Fig.~\ref{FigMucor}.
}
\label{FigMasym}
\end{figure} 

We show the angular asymmetries of $\mu^+ \to e^+ e^+ e^-$ in
Fig.~\ref{FigMasym}.
As explained in Sec.~\ref{sec:3leptontypeI}, $A_Z$ and $A_X$ are parity
odd asymmetries and $A_Y$ is a time-reversal asymmetry.
We can see that $A_Z$ is within the range from about $-45$\% to $+30$\%, while
$A_X$ is within the range from about $-10$\% to $+20$\%.
There is a positive correlation between $A_Z$ and $A_X$ as shown in
Fig.~\ref{FigMasym}(c).
Both asymmetries do not have correlations with the branching ratio.
Possible value of the time-reversal asymmetry $A_Y$ is in the range from
about $-10\%$ to $+10\%$.

\begin{figure}[htb] 
  \includegraphics[width=30em]{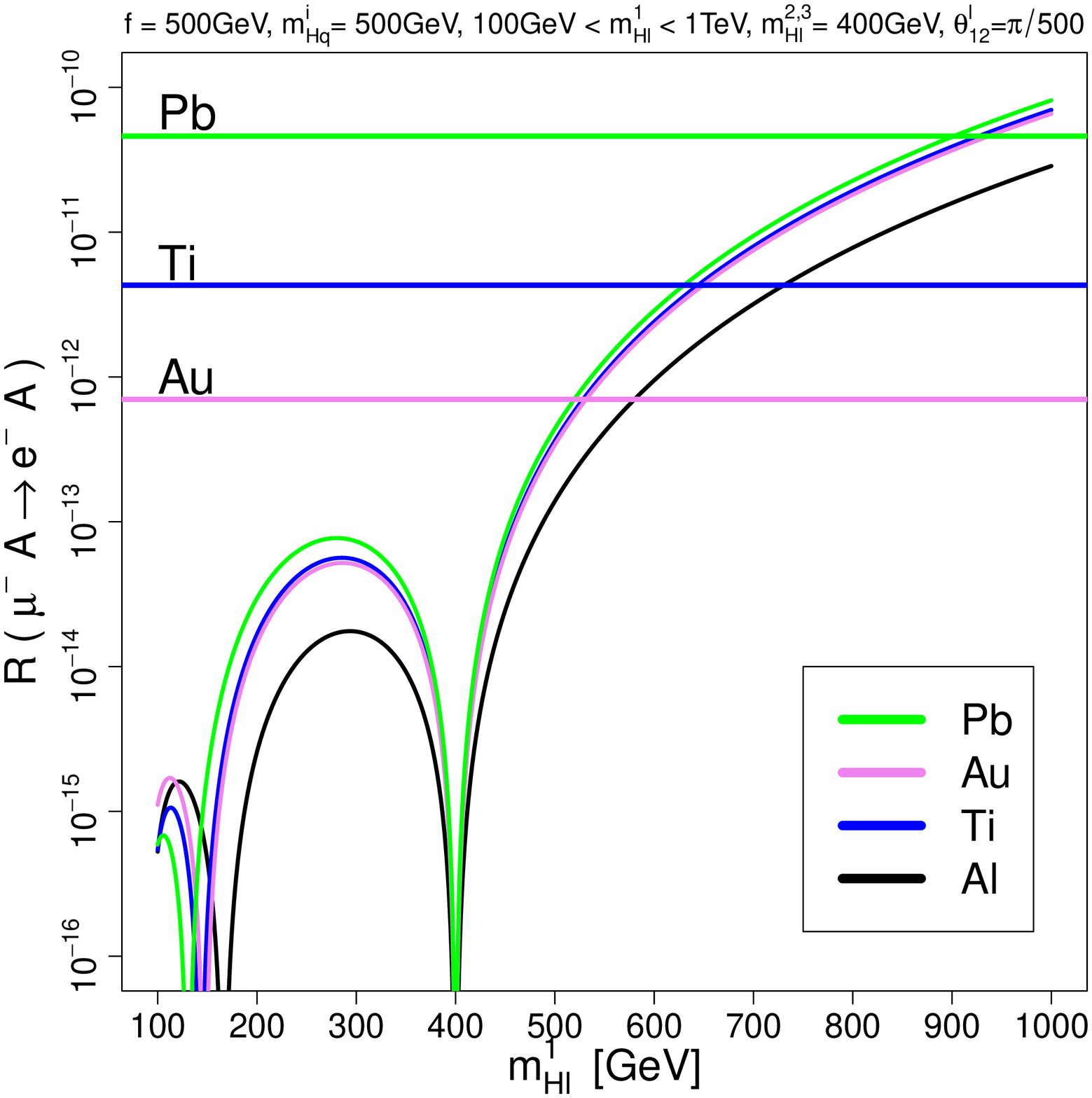}
\caption{
$\mu-e$ conversion rates for lead, gold, titanium and aluminum as functions
of the first generation T-odd lepton mass $m_{H\ell}^{1}$ for
$f=500~\text{GeV}$.
Other parameters in the T-odd lepton sector are fixed as
$m_{H\ell}^{2}=m_{H\ell}^{3}=400~\text{GeV}$,
$\theta^\ell_{12}=\pi/500$, $\theta^\ell_{23}=\theta^\ell_{13}=0$ and
$\delta^\ell_{ij}=0$.
The T-odd quark masses are also fixed as 500 GeV.
The horizontal lines are the experimental upper bounds for Pb, Au and Ti.
}
\label{FigMeccan}
\end{figure} 

In Fig.~\ref{FigMeccan}, the $\mu-e$ conversion rates for Al, Ti, Au and Pb are
plotted as functions of the mass of the first generation T-odd lepton
$m_{H\ell}^{1}$ for $f=500~\text{GeV}$.
Here, we fix the parameters in the T-odd lepton sector as
$m_{H\ell}^{2}=m_{H\ell}^{3}=400~\text{GeV}$,
$\theta^\ell_{12}=\pi/500$, $\theta^\ell_{23}=\theta^\ell_{13}=0$ and
$\delta^\ell_{ij}=0$.
T-odd quark masses are also fixed to 500~GeV, as in other scatter plots
in this paper.
The region $m_{H\ell}^{1}\gtrsim 500~\text{GeV}$ is excluded since the
branching ratio for Au exceeds the experimental upper limit.
At $m_{H\ell}^{1}=400~\text{GeV}$, all the three T-odd leptons are
degenerate in mass, so that all the LFV amplitudes vanish.
We can see that the conversion rates are suppressed also in a
region between $m_{H\ell}^{1}=100~\text{GeV}$ and $200~\text{GeV}$.
In this region, cancellation among transition amplitudes occurs
at different points for different nuclide.

\begin{figure}[htbp] 
\begin{tabular}{cc}
  \includegraphics[width=20em,clip]{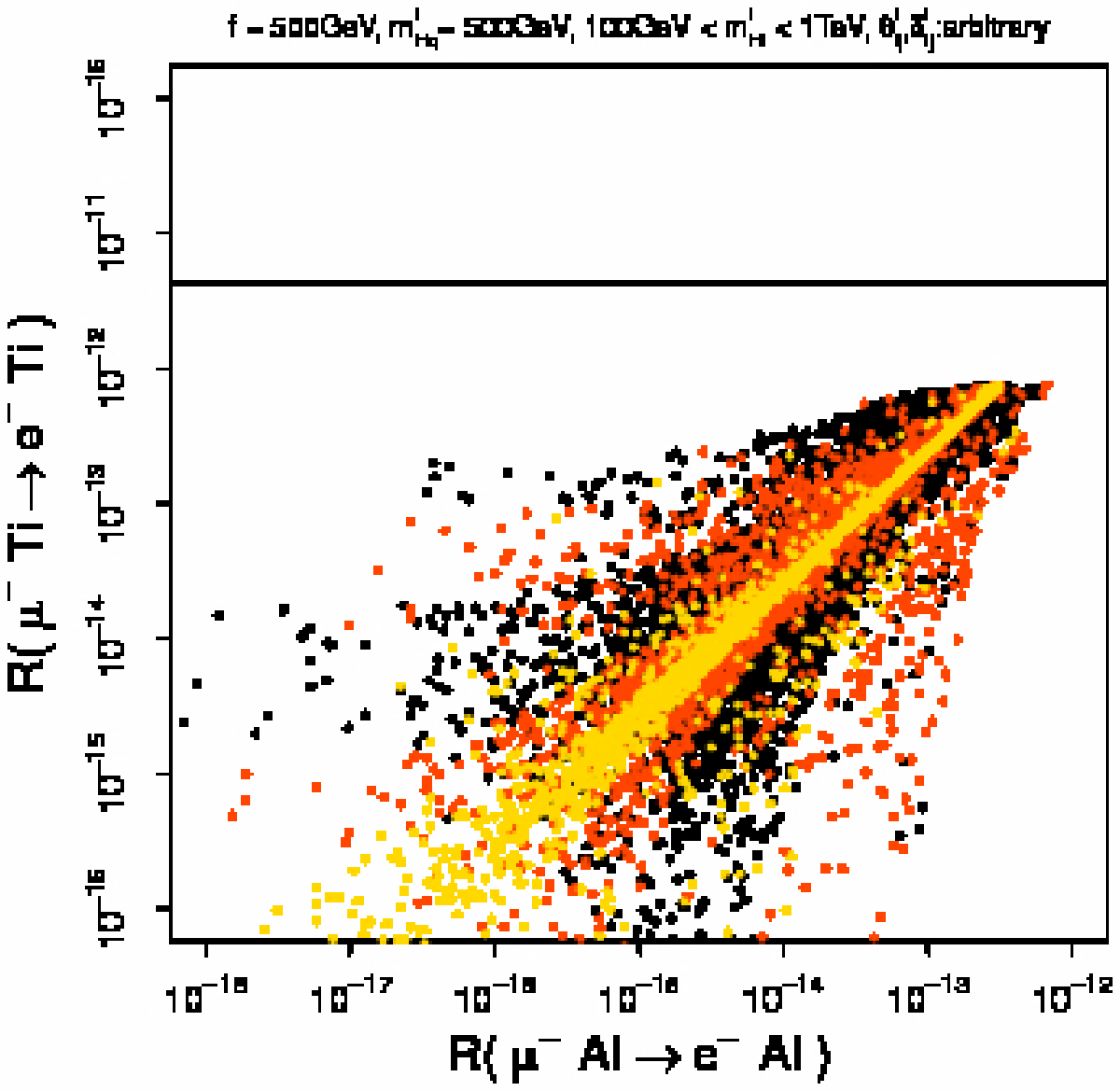} &
  \includegraphics[width=20em,clip]{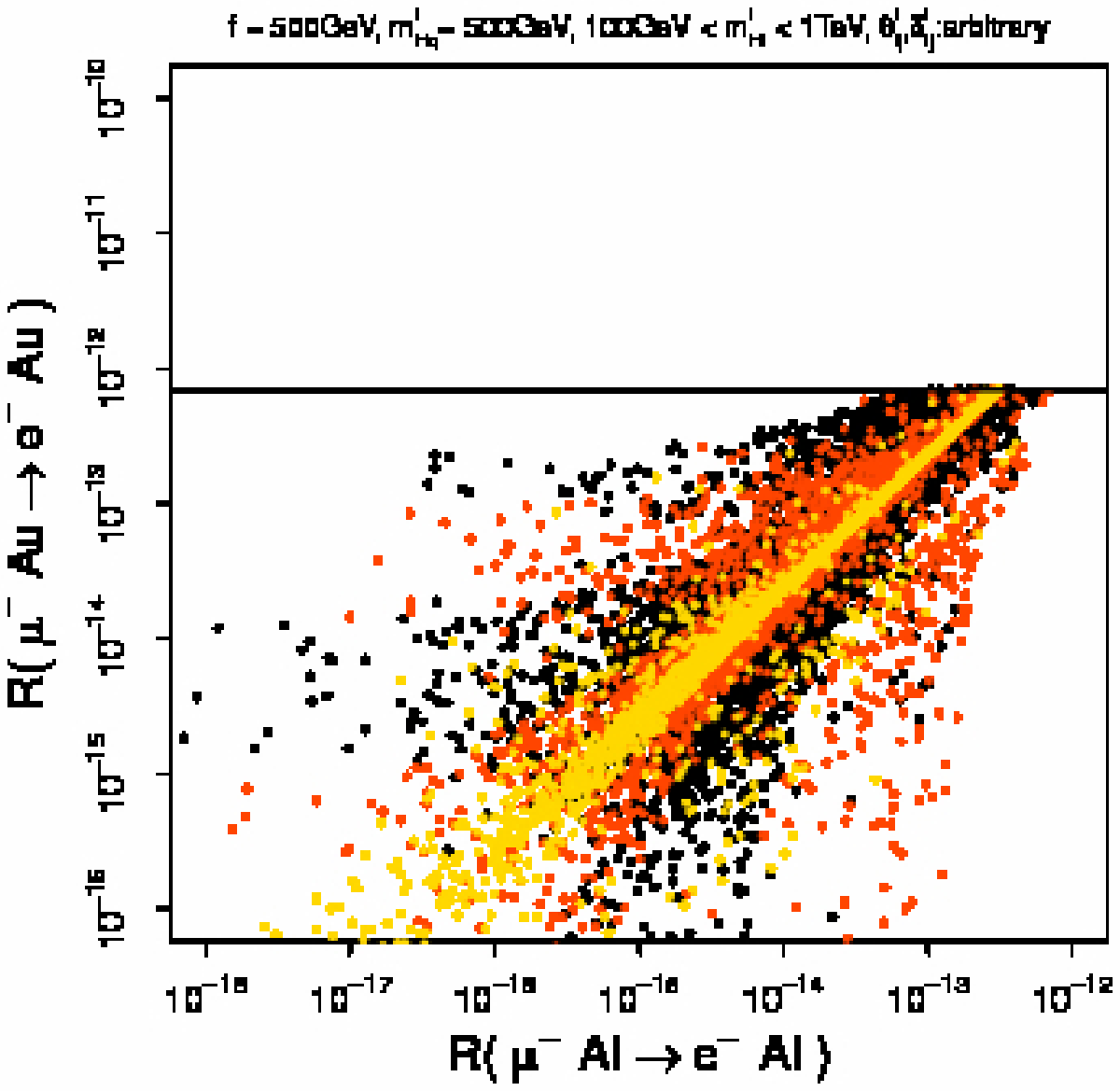} \\
  (a) & (b)
\end{tabular}
\begin{tabular}{c}
  \includegraphics[width=20em,clip]{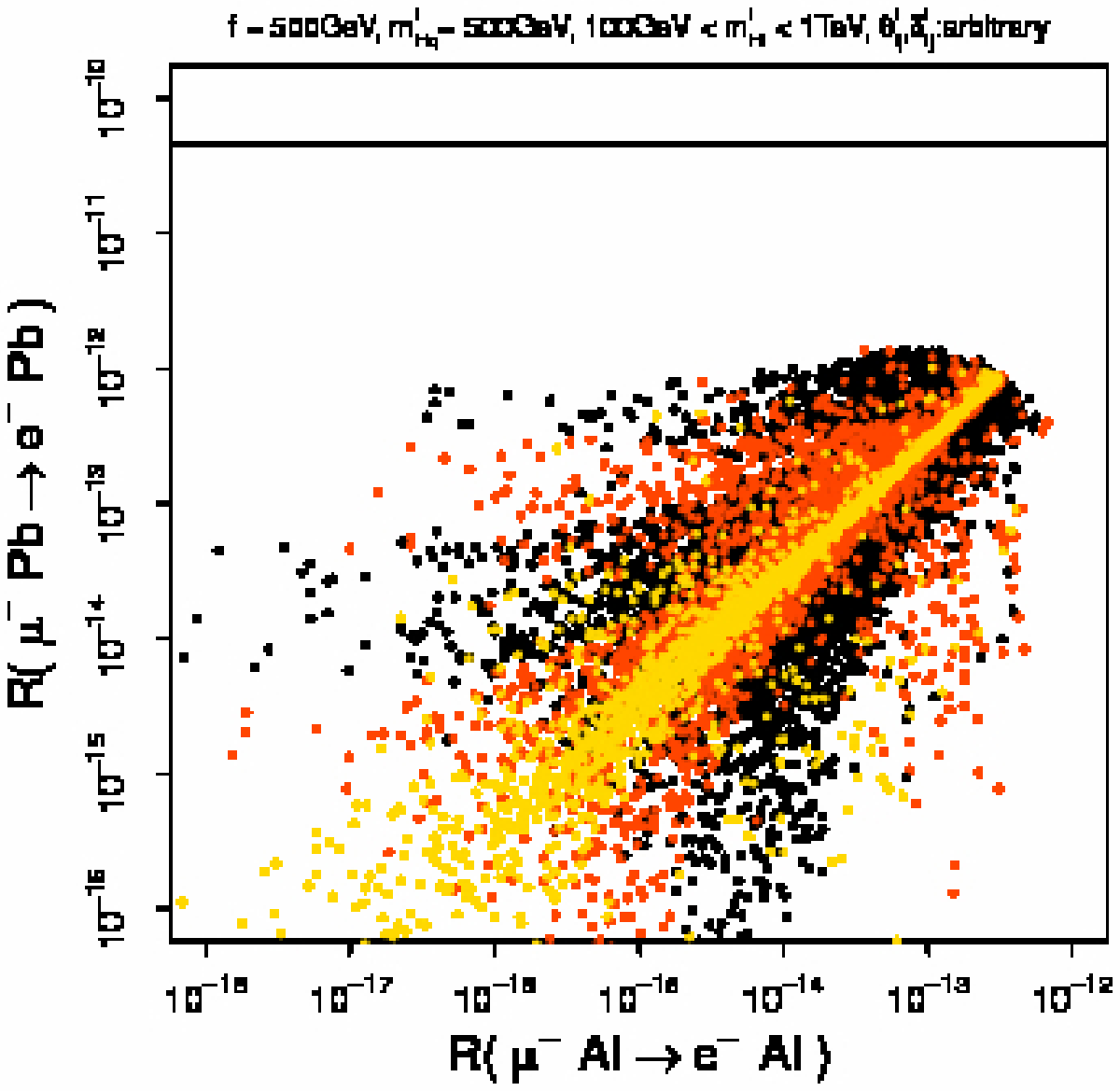} \\
  (c)
\end{tabular}
\caption{
Correlations among the $\mu-e$ conversion rates for Ti, Au, or Pb vs Al.
The horizontal lines are experimental bounds.
The input parameter set and the color code of the plots are the same as
those in Fig.~\ref{FigMucor}.
}
\label{FigMec}
\end{figure} 

We show correlation between the $\mu-e$ conversion rates 
for Al vs Ti (Au, Pb) in Fig.~\ref{FigMec}.
We can see the ratios of the conversion rates vary 
within 1 order of magnitude in most of the parameter space.
We also notice that, in some cases, the conversion rate for Ti, Au and Pb
can be close to the experimental bounds even if the rate for Al is suppressed.

\subsection{$\tau$ LFV} 
The current upper bounds for $\tau$-LFV decays are 
listed in Table~\ref{TabBound}.
These bounds are set by either the Belle and the Babar experiments.
Improvements by 1 or 2 orders of magnitude are expected 
at future B-factories at KEK and in Italy~\cite{RefBfactory}.

In this subsection, we mainly present the results on the observables in
$\tau\to \mu$ decays discussed in Sec.~\ref{SecLhtlfv}.
Quantities in $\tau\to e$ decay modes behave similarly to corresponding
ones in $\tau\to \mu$ modes.
Correlations between the observable quantities in $\tau\to \mu$ and
$\tau\to e$ modes are discussed in Sec.~\ref{sec:correlations-me-tm-te}.
\subsubsection{$\tau\to\mu\gamma$ and trilepton decay modes} 
\begin{figure}[htbp] 
\begin{tabular}{cc}
  \includegraphics[width=20em,clip]{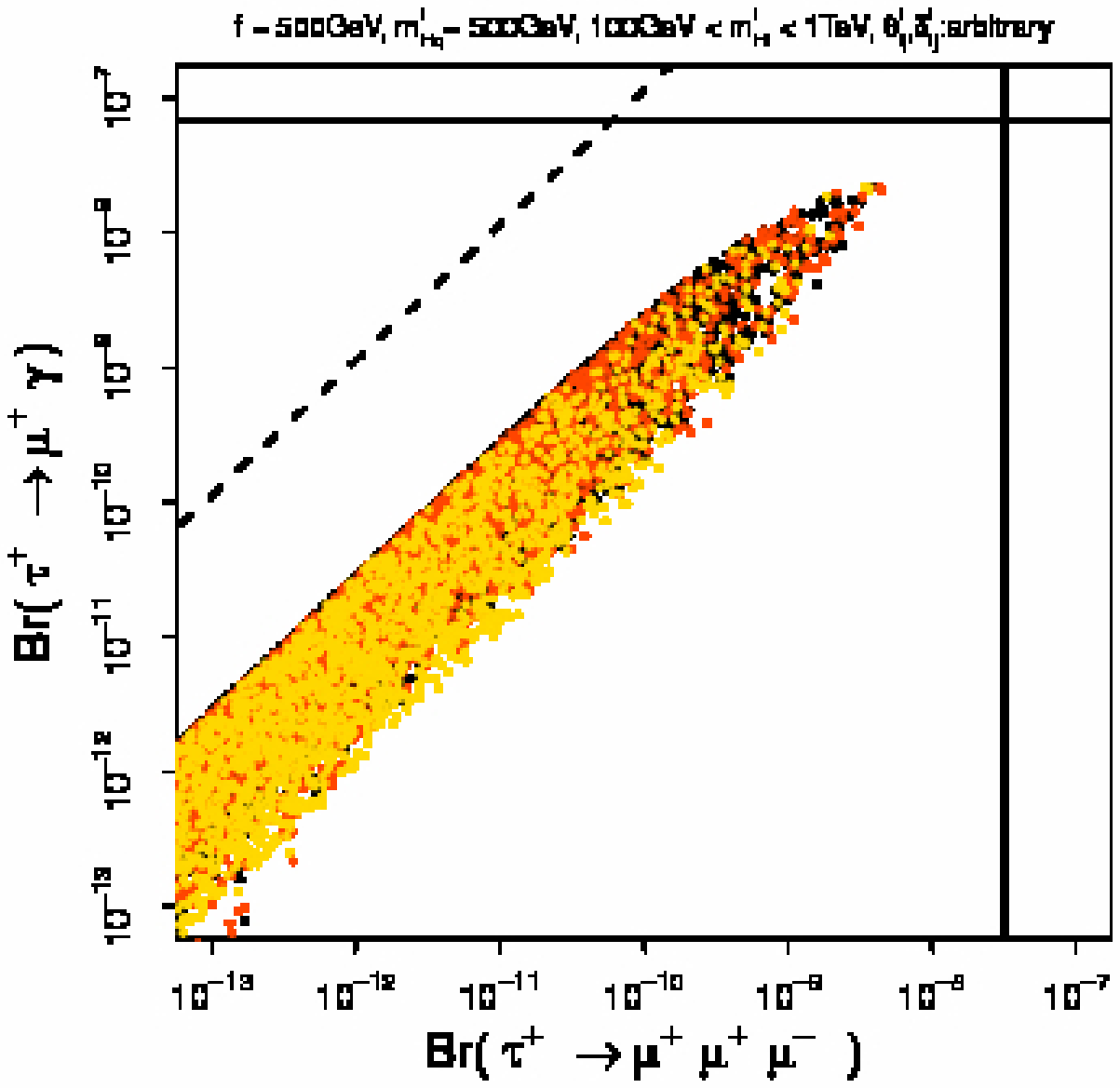} &
  \includegraphics[width=20em,clip]{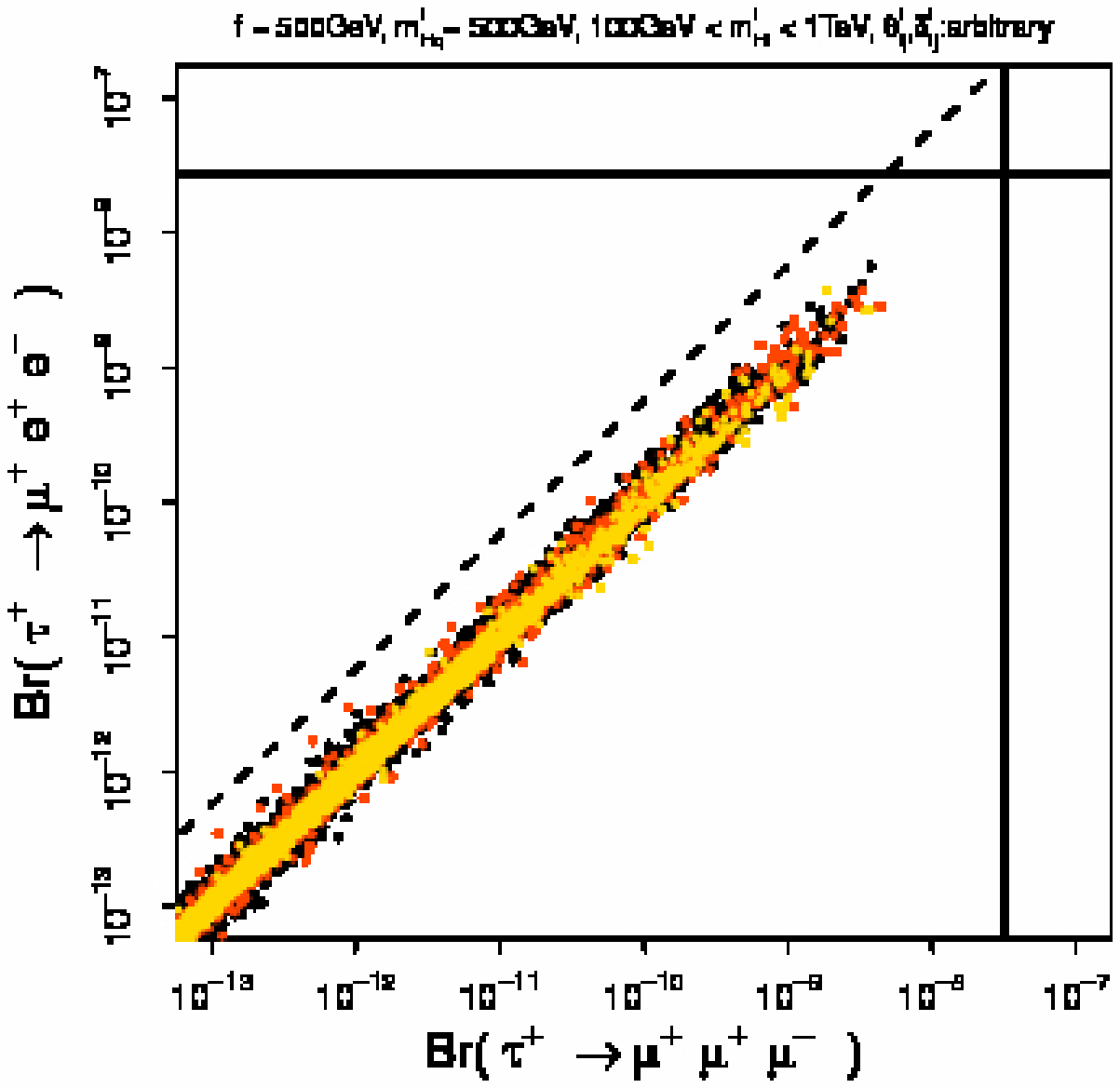} \\
  (a) & (b)
\end{tabular}
\begin{tabular}{c}
  \includegraphics[width=20em,clip]{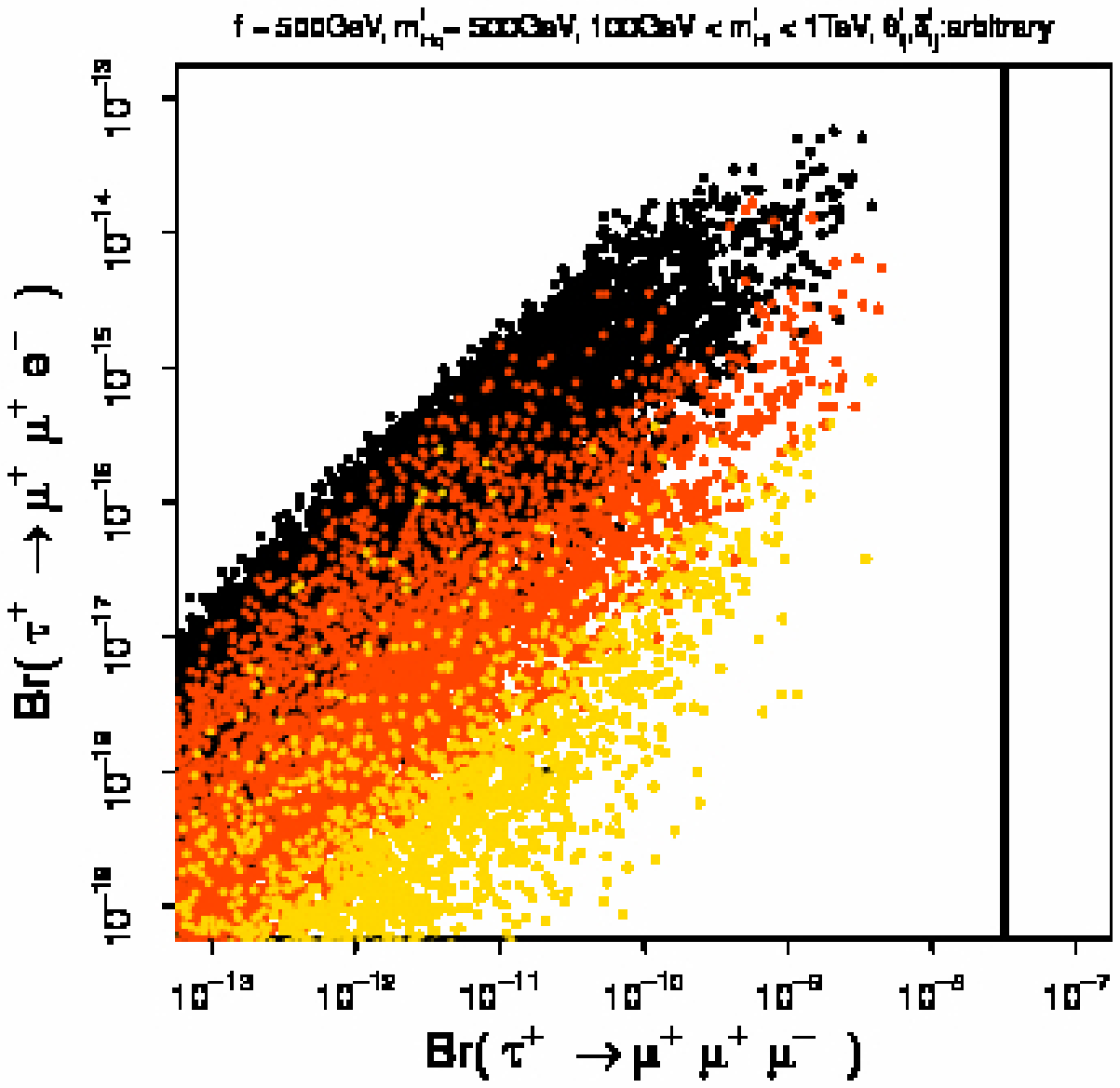} \\
  (c)
\end{tabular}
\caption{
Branching ratios of $\tau^+ \to \mu^+ \gamma$, $\tau^+ \to \mu^+ e^+ e^-$
and $\tau^+ \to \mu^+ \mu^+ e^-$ as functions of 
$\text{Br}(\tau^+ \to \mu^+ \mu^+ \mu^-)$ for the same input parameter
set as in Fig.~\ref{FigMucor}.
The horizontal and the vertical lines are the experimental bounds.
The dashed lines show the branching ratios calculated with the dipole
contributions only.
The color code is the same as in Fig.~\ref{FigMucor}.
}
\label{FigTlfv}
\end{figure} 

In Fig.~\ref{FigTlfv}, we show correlations among the branching ratios
of $\tau^+ \to \mu^+ \gamma$, $\tau^+ \to \mu^+ \mu^+ \mu^-$,
$\tau^+ \to \mu^+ e^+ e^-$ and $\tau^+ \to \mu^+ \mu^+ e^-$.
We see that the branching ratio of $\tau^+ \to \mu^+ \gamma$ can be
as large as $10^{-8}$, which is close to current experimental upper
limit.
For $\tau^+ \to \mu^+ \mu^+ \mu^-$ and $\tau^+ \to \mu^+ e^+ e^-$,
the possible maximal values of the branching ratios are about 1 order of
magnitude below the corresponding experimental limits.
The behavior of the correlation between
$\text{Br}(\tau^+ \to \mu^+ \gamma)$ and
$\text{Br}(\tau^+ \to \mu^+ \mu^+ \mu^-)$ shown in Fig.~\ref{FigTlfv}(a)
is similar to the $\mu\to e$ case given in
Fig.~\ref{FigMucor}(a):
$\text{Br}(\tau^+ \to \mu^+ \mu^+ \mu^-)$ is larger than the prediction
in the dipole-dominant case (dashed line) by 1 or
2 orders of magnitude.
Since we take all the masses and the mixing angles/phases in the T-odd
lepton sector as free parameters, there is no direct correlation between
the $\tau\to \mu$ and $\mu\to e$ transition amplitudes.
Therefore, the correlation plots among $\tau\to \mu$ processes do not
change much even if the upper limit of $\text{Br}(\mu\to e\gamma)$ is
lowered.
As shown in Fig.~\ref{FigTlfv}(c), the branching ratio of
$\tau^+ \to \mu^+ \mu^+ e^-$ is highly suppressed because this process
involves the $\mu\to e$ transition, whose magnitude is constrained by
other $\mu\to e$ processes discussed in the previous subsection.
We see that the suppression of $\text{Br}(\tau^+ \to \mu^+ \mu^+ e^-)$
is stronger for a smaller value of $\text{Br}(\mu^+ \to e^+ \gamma)$.
As for the other type III decay mode $\tau^+ \to e^+ e^+ \mu^-$, the
correlation plot with $\text{Br}(\tau^+ \to e^+ e^+ e^-)$ is almost the
same as Fig.~\ref{FigTlfv}(c).

\begin{figure}[htbp] 
\begin{tabular}{cc}
  \includegraphics[width=20em,clip]{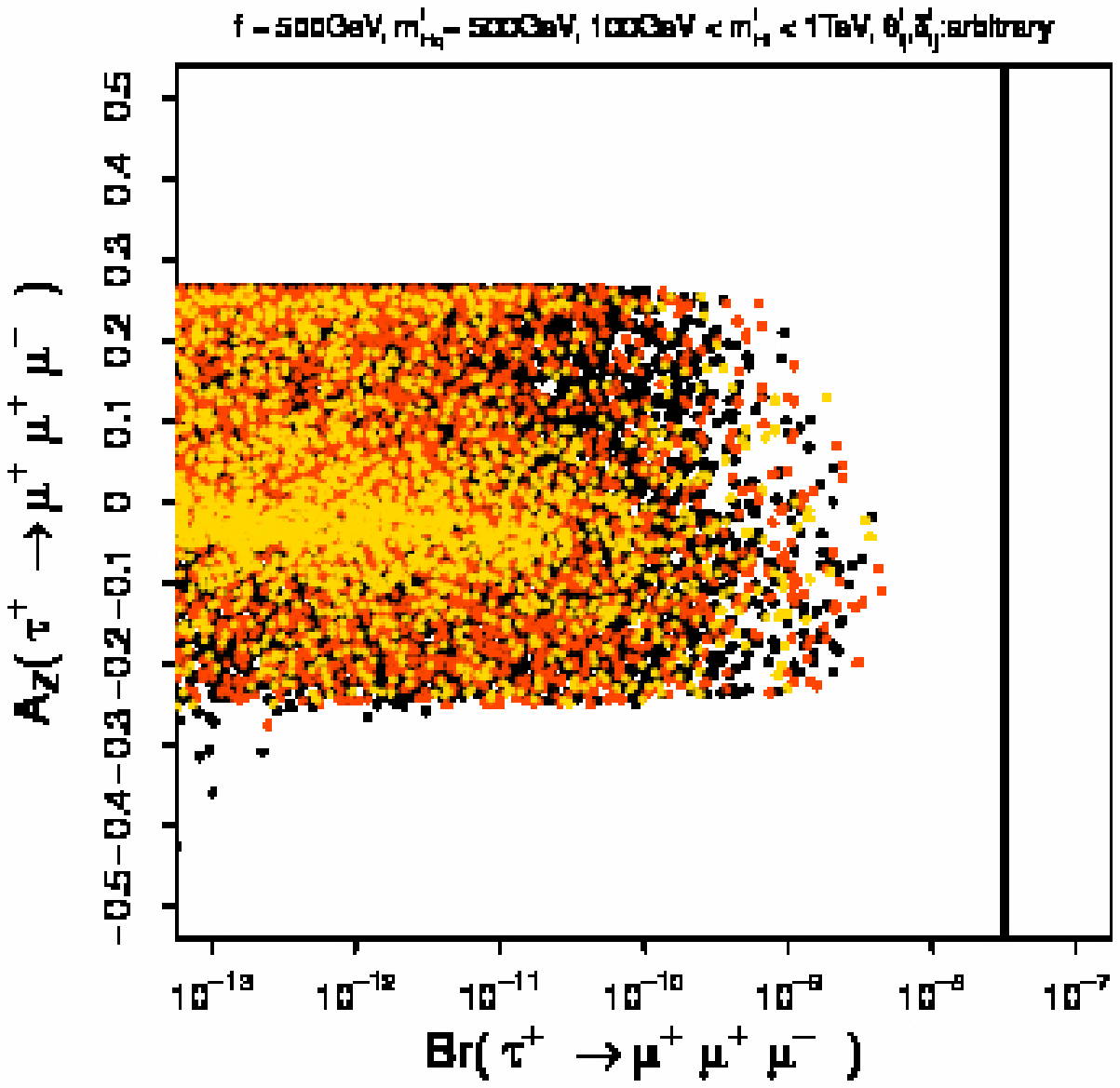} &
  \includegraphics[width=20em,clip]{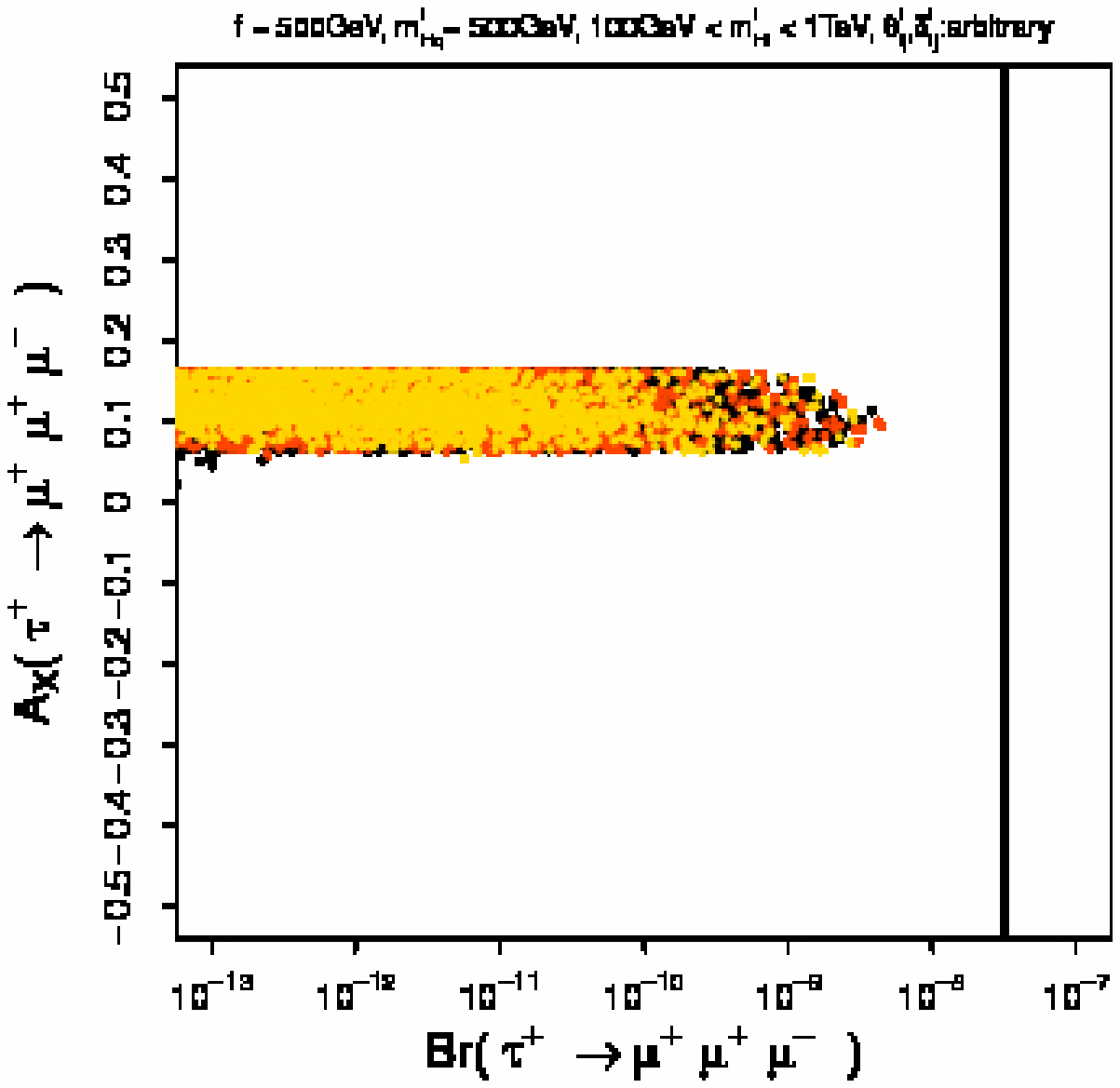} \\
  (a) & (b) \\
  \includegraphics[width=20em,clip]{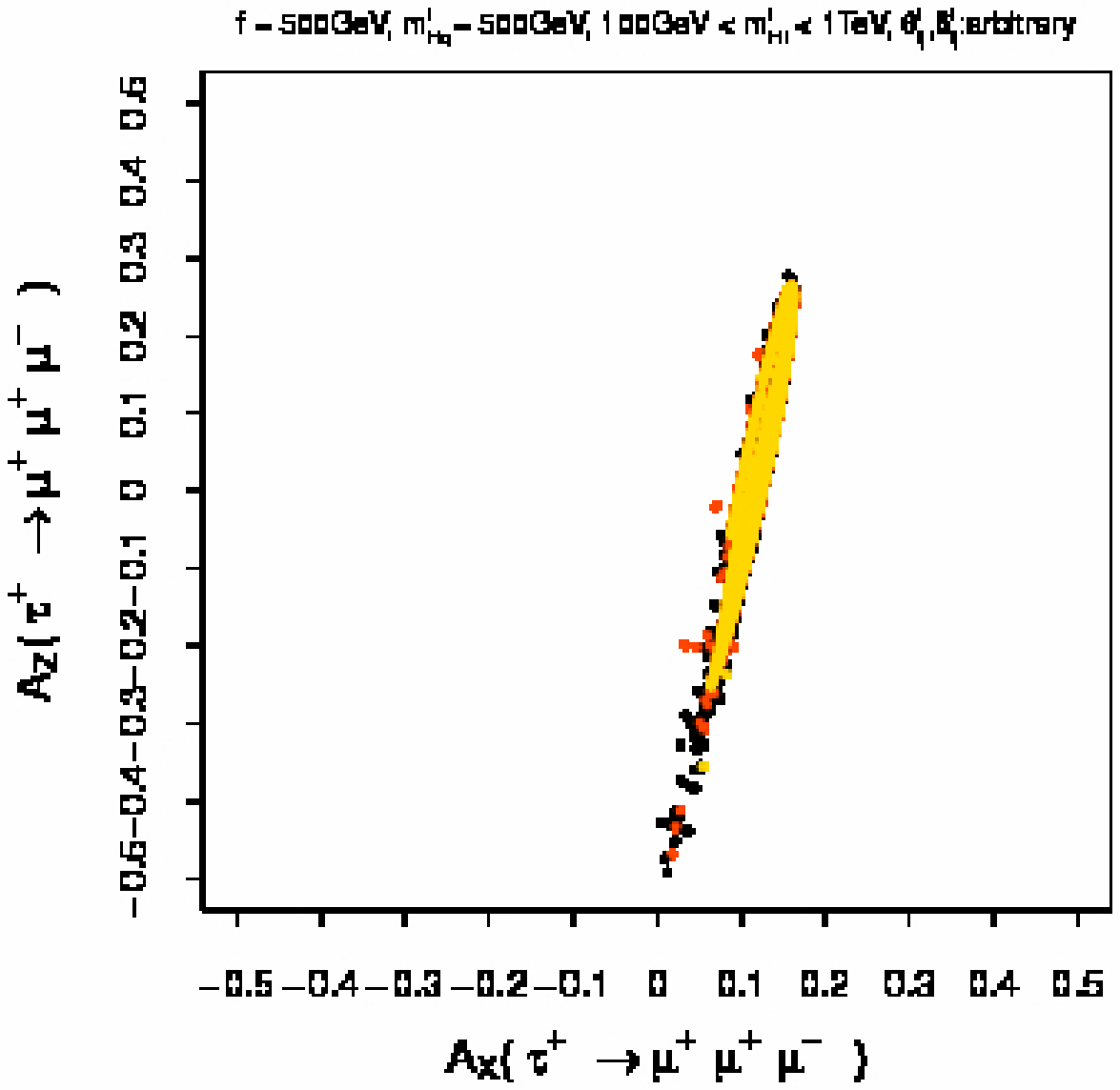} &
  \includegraphics[width=20em,clip]{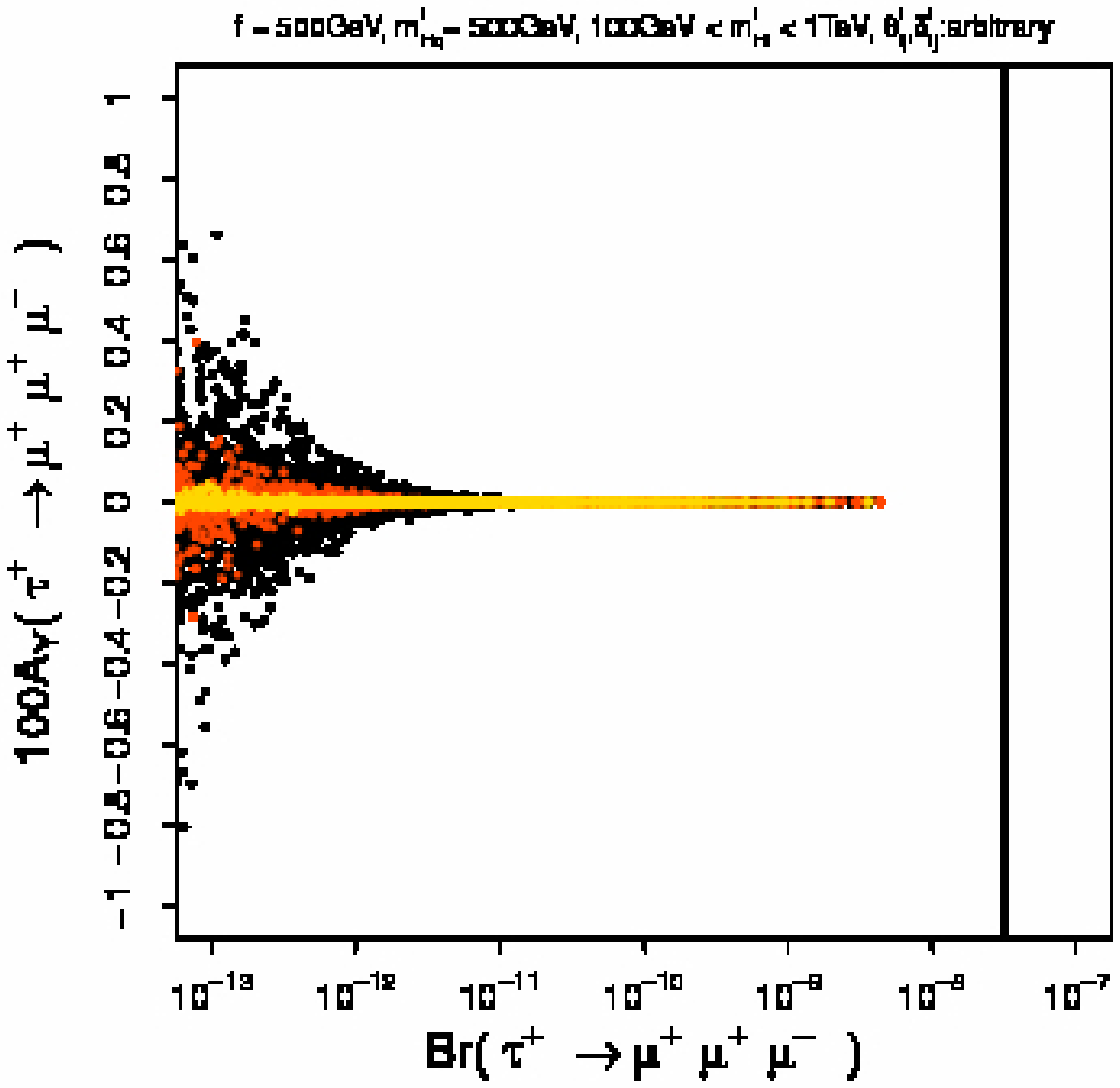} \\
  (c) & (d)
\end{tabular}
\caption{
Angular asymmetries of $\tau^+ \to \mu^+ \mu^+ \mu^-$ as functions of
the branching ratio for the same parameter set as in
Fig.~\ref{FigMucor}.
The correlation between $A_Z$ and $A_X$ is also shown in (c).
$A_Y$ is magnified by 100 in (d).
The vertical solid lines in (a), (b) and (d) are the experimental
upper limit of  $\text{Br}(\tau^+ \to \mu^+ \mu^+ \mu^-)$.
The color code is the same as in Fig.~\ref{FigMucor}.
}
\label{FigTasym}
\end{figure} 

We show angular asymmetries of $\tau^+ \to \mu^+ \mu^+ \mu^-$ in
Fig.~\ref{FigTasym}.
We find that the parity asymmetries are within the ranges
$-25\% \lesssim A_Z \lesssim +25\%$
and
$+5\% \lesssim A_X \lesssim +15\%$ irrespective of the branching ratio
for $\text{Br}(\tau^+ \to \mu^+ \mu^+ \mu^-)\gtrsim 10^{-13}$.
The time-reversal asymmetry $A_Y$ is very small for the same range of
the branching ratio.
Compared with Fig.~\ref{FigMasym}, we can see that the possible ranges of the
asymmetries for $\tau^+ \to \mu^+ \mu^+ \mu^-$ are narrower than those
for $\mu^+ \to e^+ e^+ e^-$.
This quantitative difference between the $\tau^+ \to \mu^+ \mu^+ \mu^-$
and $\mu^+ \to e^+ e^+ e^-$ cases is caused by the difference in the
allowed ranges of the branching ratios.
In the general scan in the parameter space of the T-odd lepton sector,
the decay amplitudes for $\mu^+ \to e^+ e^+ e^-$,
$\tau^+ \to \mu^+ \mu^+ \mu^-$ and $\tau^+ \to e^+ e^+ e^-$ behave in
the same way.
Consequently the distribution patterns of the asymmetries in the scatter
plots such as Figs.~\ref{FigMasym} and \ref{FigTasym} are similar if
the experimental limits and differences of cutoff parameters are neglected.
In fact, we have checked that the patterns of the scatter plots in
Figs.~\ref{FigMasym} and \ref{FigTasym} become the same if we draw all the
sample points, including those excluded by the experimental limits
as long as we take same cutoff parameters.

\begin{figure}[htb] 
\begin{tabular}{cc}
  \includegraphics[width=20em,clip]{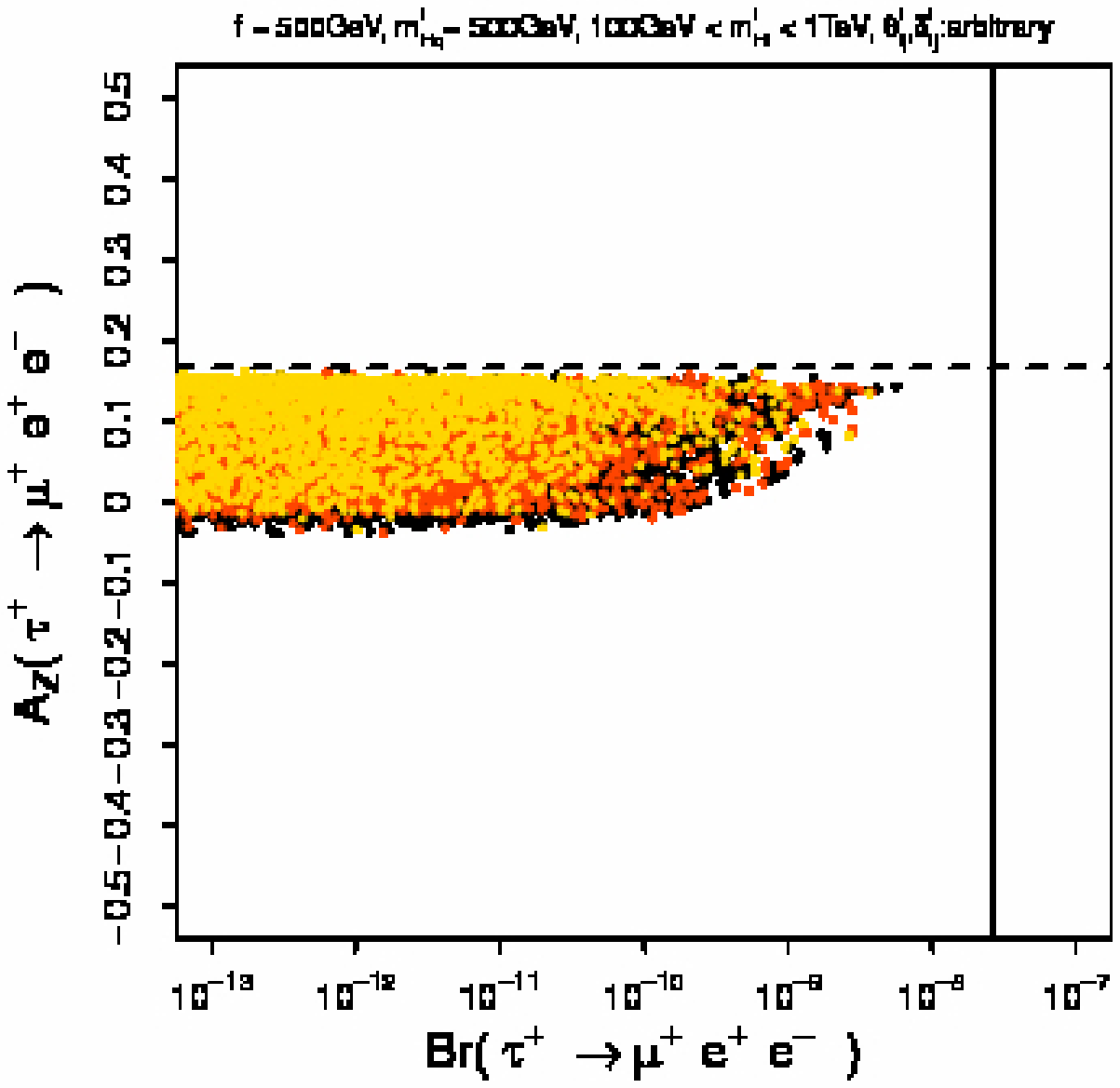} &
  \includegraphics[width=20em,clip]{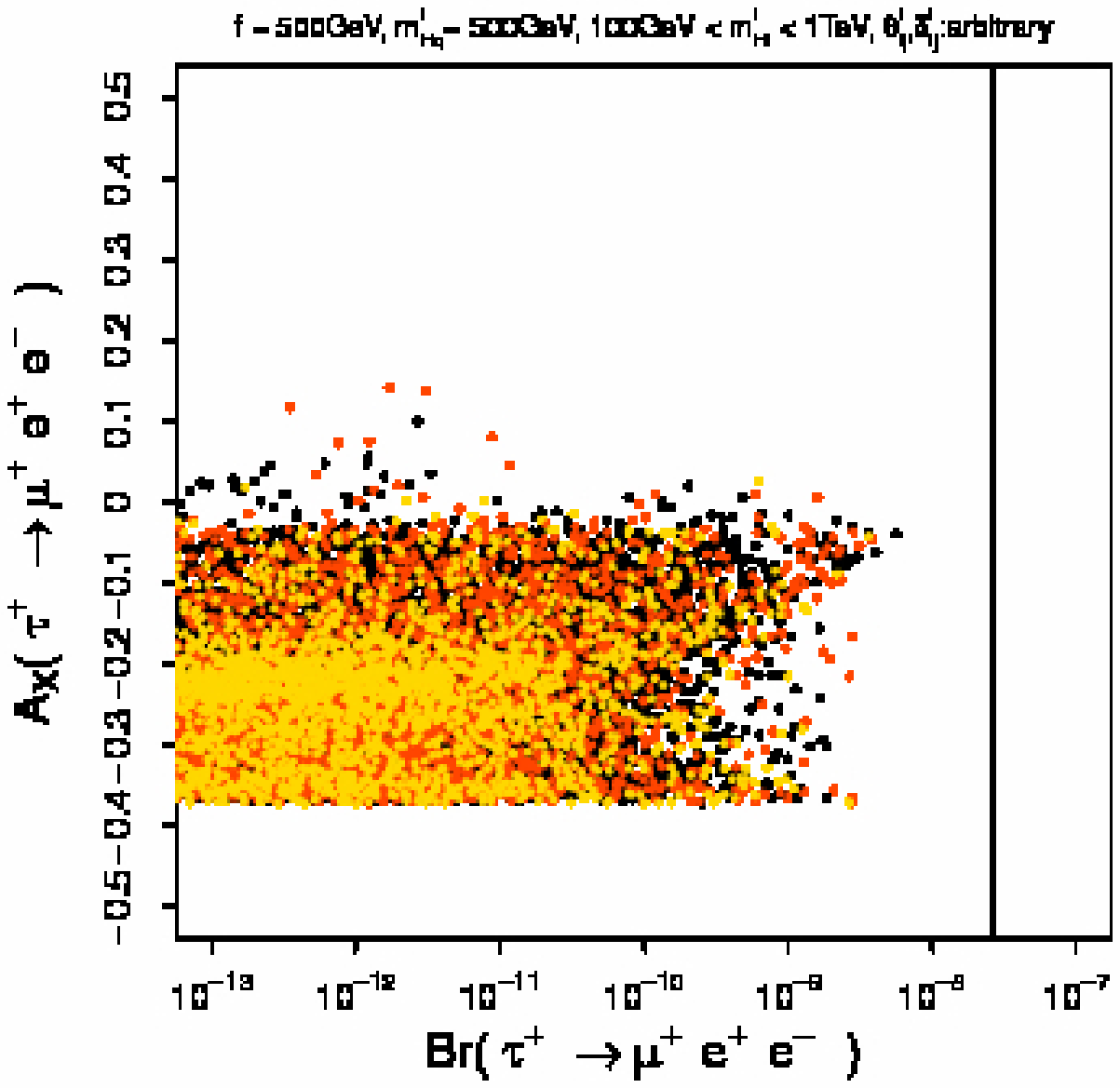} \\
  (a) & (b) \\
  \includegraphics[width=20em,clip]{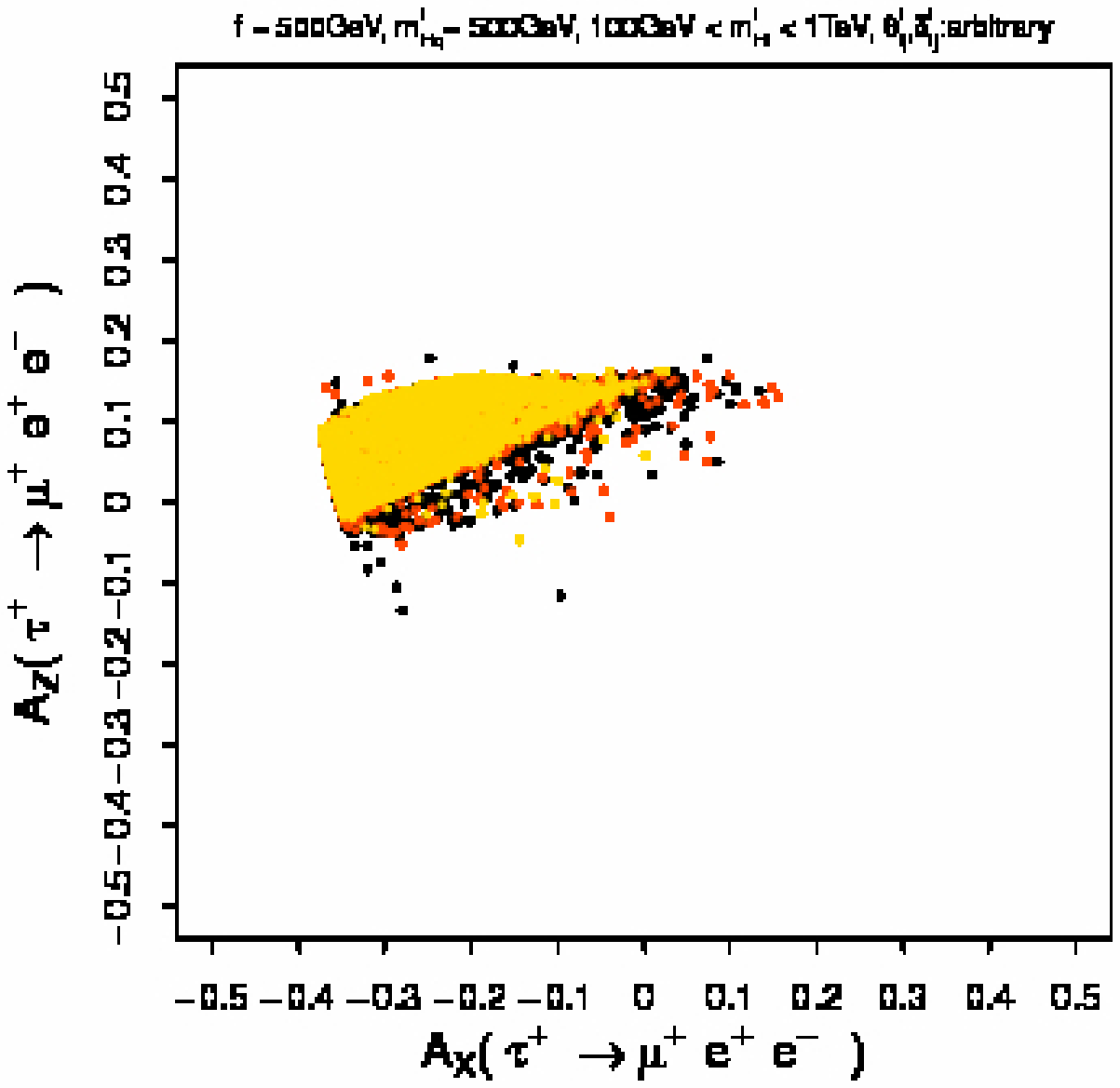} &
  \includegraphics[width=20em,clip]{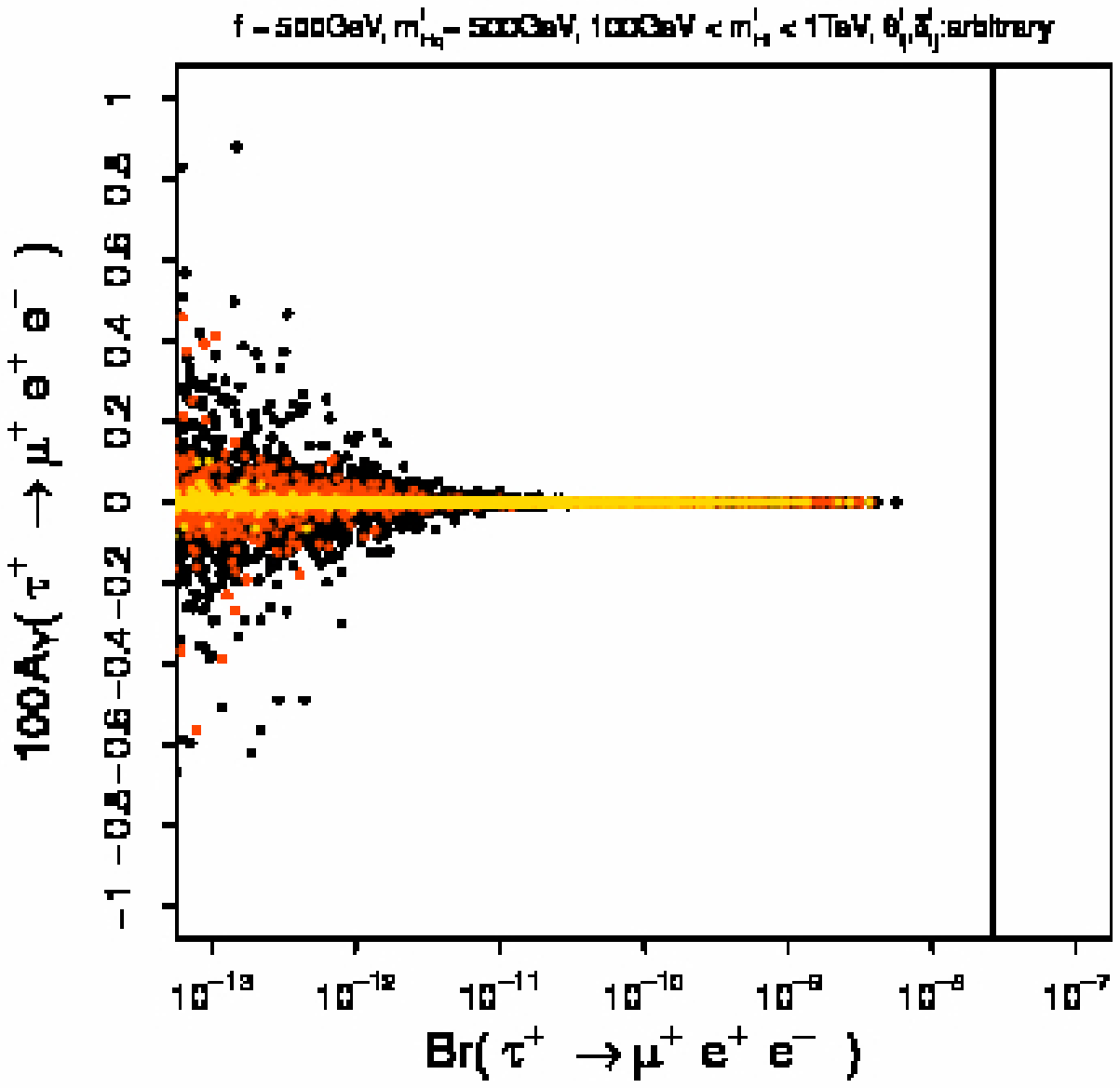} \\
  (c) & (d)
\end{tabular}
\caption{
Angular asymmetries of $\tau^+ \to \mu^+ e^+ e^-$ as functions of 
the branching ratio for the same parameter set as in
Fig.~\ref{FigMucor}.
The correlation between $A_Z$ and $A_X$ is also shown in (c).
$A_Y$ is magnified by 100 in (d).
The vertical solid lines in (a), (b), and (d) is the experimental
upper limit of  $\text{Br}(\tau^+ \to \mu^+ e^+ e^-)$.
The horizontal dashed line in (a) is the value 
in $A_{R}^{\text{LHT}} \to 0$ limit.
The color code is the same as in Fig.~\ref{FigMucor}.
}
\label{FigTang}
\end{figure}

\begin{figure}[htbp]
\begin{tabular}{cc}
  \includegraphics[width=20em,clip]{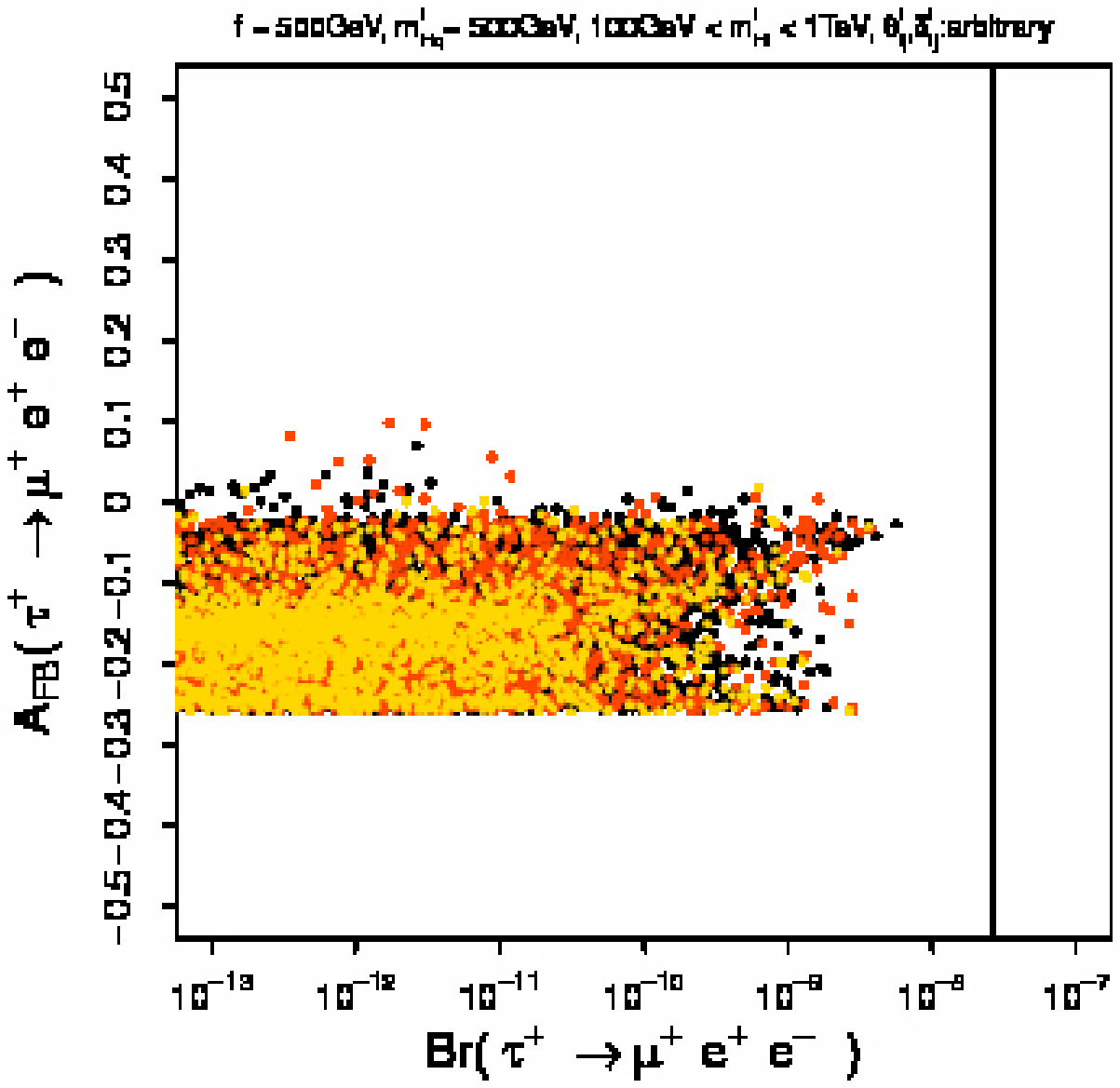} &
  \includegraphics[width=20em,clip]{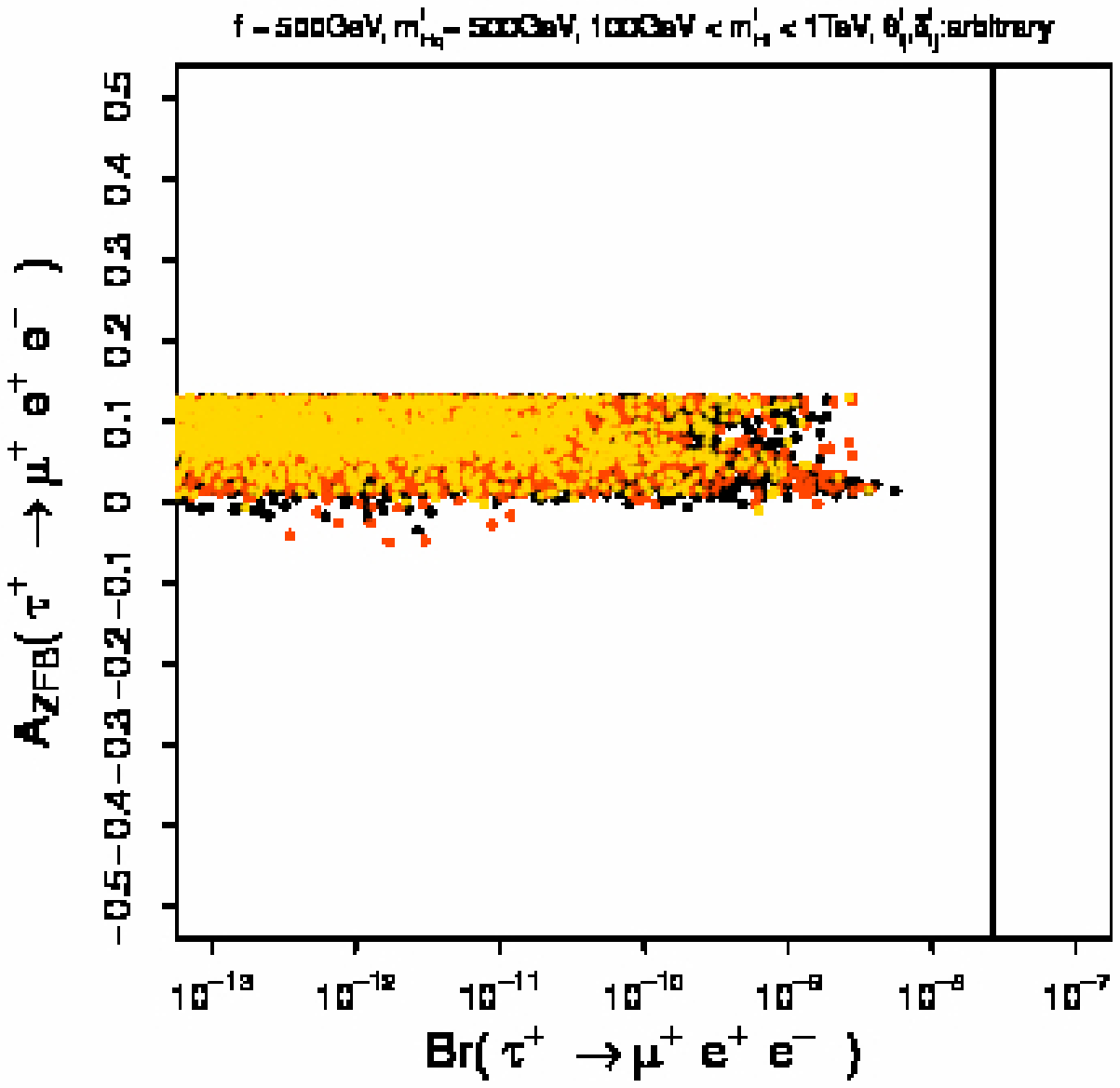} \\
  (a) & (b) \\
  \includegraphics[width=20em,clip]{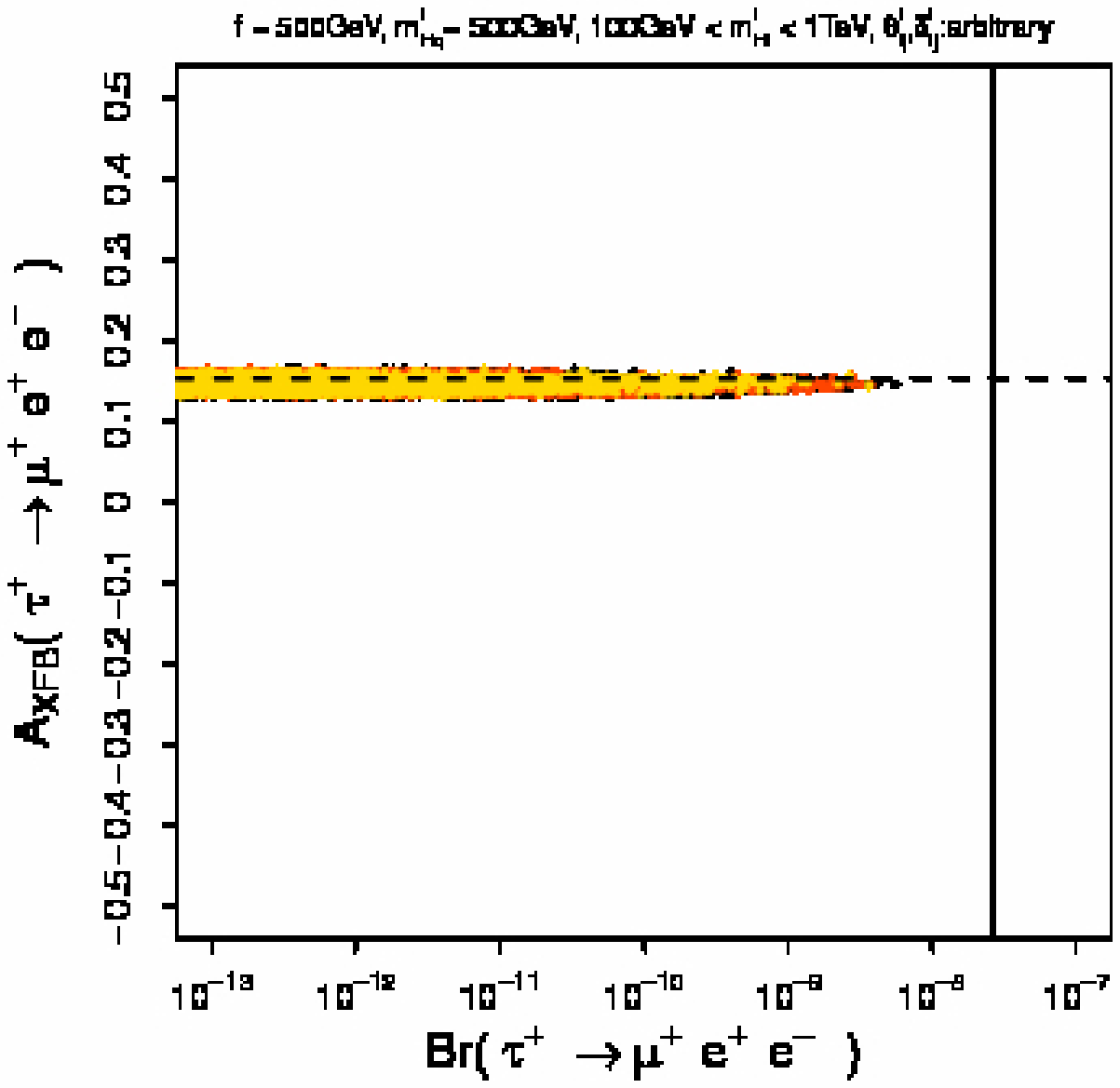} &
  \includegraphics[width=20em,clip]{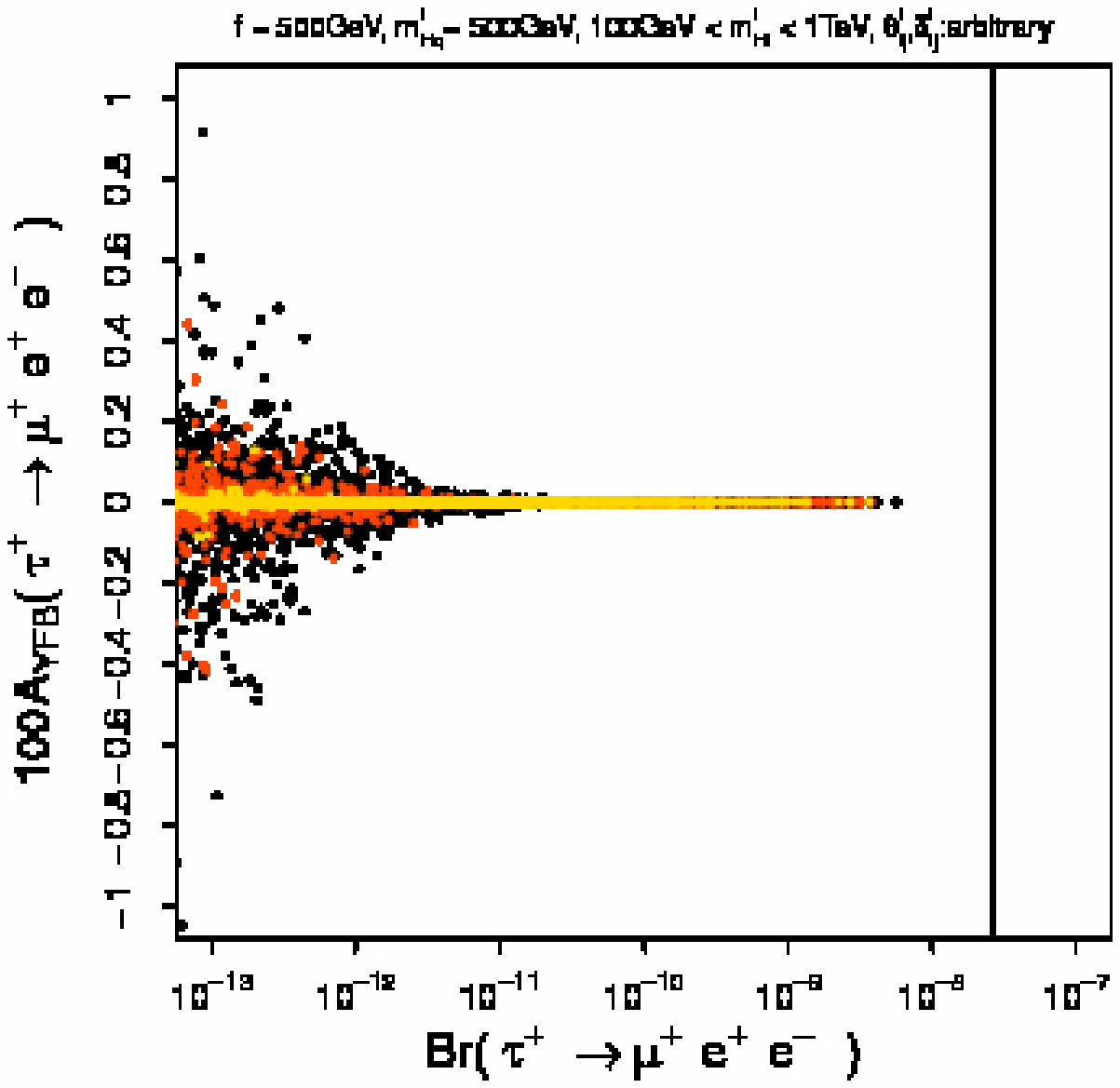} \\
  (c) & (d)
\end{tabular}
\caption{
Forward-backward asymmetry and forward-backward type angular
asymmetries of $\tau^+ \to \mu^+ e^+ e^-$ as functions of 
the branching ratio for the same parameter set as in Fig.~\ref{FigMucor}.
$A_{YFB}$ is magnified by 100 in (d).
The vertical solid line in each plot is the experimental
upper limit of  $\text{Br}(\tau^+ \to \mu^+ e^+ e^-)$.
The horizontal dashed line in (c) is the value 
in $A_{R}^{\text{LHT}} \to 0$ limit.
The color code is the same as in Fig.~\ref{FigMucor}.
}
\label{FigTfb}
\end{figure} 

Figures.~\ref{FigTang} and \ref{FigTfb} show the asymmetries of the type II
decay $\tau^+\to \mu^+ e^+ e^-$.
As given in Sec.~\ref{sec:tmeeII}, we define the seven asymmetries.
We can see that the allowed ranges of the parity asymmetries
$A_{Z}$, $A_X$, $A_{ZFB}$ and $A_{XFB}$ are almost 
independent of the branching ratio.

Taking the $A_R^{\text{LHT}}\to 0$ limit is informative to understand the
behavior of the asymmetries, because the contribution of the dipole term in
the LHT is relatively small.
In this limit, values of $A_Z$ and $A_{XFB}$ are constants:
\begin{align}
  \left.
    A_Z^{\text{II,LHT}}(\delta\to 0)
  \right|_{A_R^{\text{LHT}}\to 0} &=
  \frac{1}{6},
\\
  \left.
    A_{XFB}^{\text{II,LHT}}(\delta\to 0)
  \right|_{A_R^{\text{LHT}}\to 0} &=
  \frac{16}{105}.
\end{align}
Deviations from these values seen in Figs.~\ref{FigTang}(a) and
\ref{FigTfb}(c) are identified as effects of the dipole term.
This effect is larger in $A_Z$ because the function $D_5$ in
Eq.~(\ref{eq:AZ-II}) becomes logarithmically large for $\delta\to 0$
[see Eq.~(\ref{EqD5})], while the function $E_4$ in
Eq.~(\ref{eq:AXFB-II}) does not have such an enhancement.
On the other hand, $A_X^{\text{II,LHT}}$ and $A_{ZFB}^{\text{II,LHT}}$
depend on the relative magnitude of $g_{Ll}^{\text{II,LHT}}$ and
$g_{Lr}^{\text{II,LHT}}$, so that these asymmetries depend on input
parameters even if the dipole term is neglected.
That is why $A_X^{\text{II,LHT}}$ distributes within a wider range
compared with $A_Z^{\text{II,LHT}}$.

Since the $\tau^+\to \mu^+ e^+ e^-$ decay in the LHT is described by only
three Wilson coefficients, we can derive the following proportionality 
relation for $\delta\to 0$:
\begin{align}
  A_X : A_{FB} : A_{ZFB} = -\frac{4\pi}{35} : -\frac{1}{4} : \frac{1}{8}.
\label{EqProprel}
\end{align}
This relation holds for any values of $g_{Ll}^{\text{II,LHT}}$,
$g_{Lr}^{\text{II,LHT}}$ and $A_R^{\text{LHT}}$.
Although $A_{FB}$ is not a parity asymmetry, it is related to the parity
asymmetries $A_{X}$ and $A_{ZFB}$ because of the restricted chirality 
structure of the LHT.
We can see that Eq.~\eqref{EqProprel} holds in a good approximation in
Figs.~\ref{FigTang}(b), \ref{FigTfb}(a) and \ref{FigTfb}(b).

More relations characteristic to the LHT are discussed 
in Appendix~\ref{AppRel}.
When the LFV processes are precisely measured, 
these relations might be useful to
determine whether or not the new physics is the LHT.

The time-reversal asymmetries $A_{Y}$ and $A_{YFB}$ are very small for
$\text{Br}(\tau^+\to \mu^+ e^+ e^-)\gtrsim 10^{-13}$.

\subsubsection{Semileptonic decay modes} 
\begin{figure}[htb] 
\begin{tabular}{cc}
  \includegraphics[width=20em,clip]{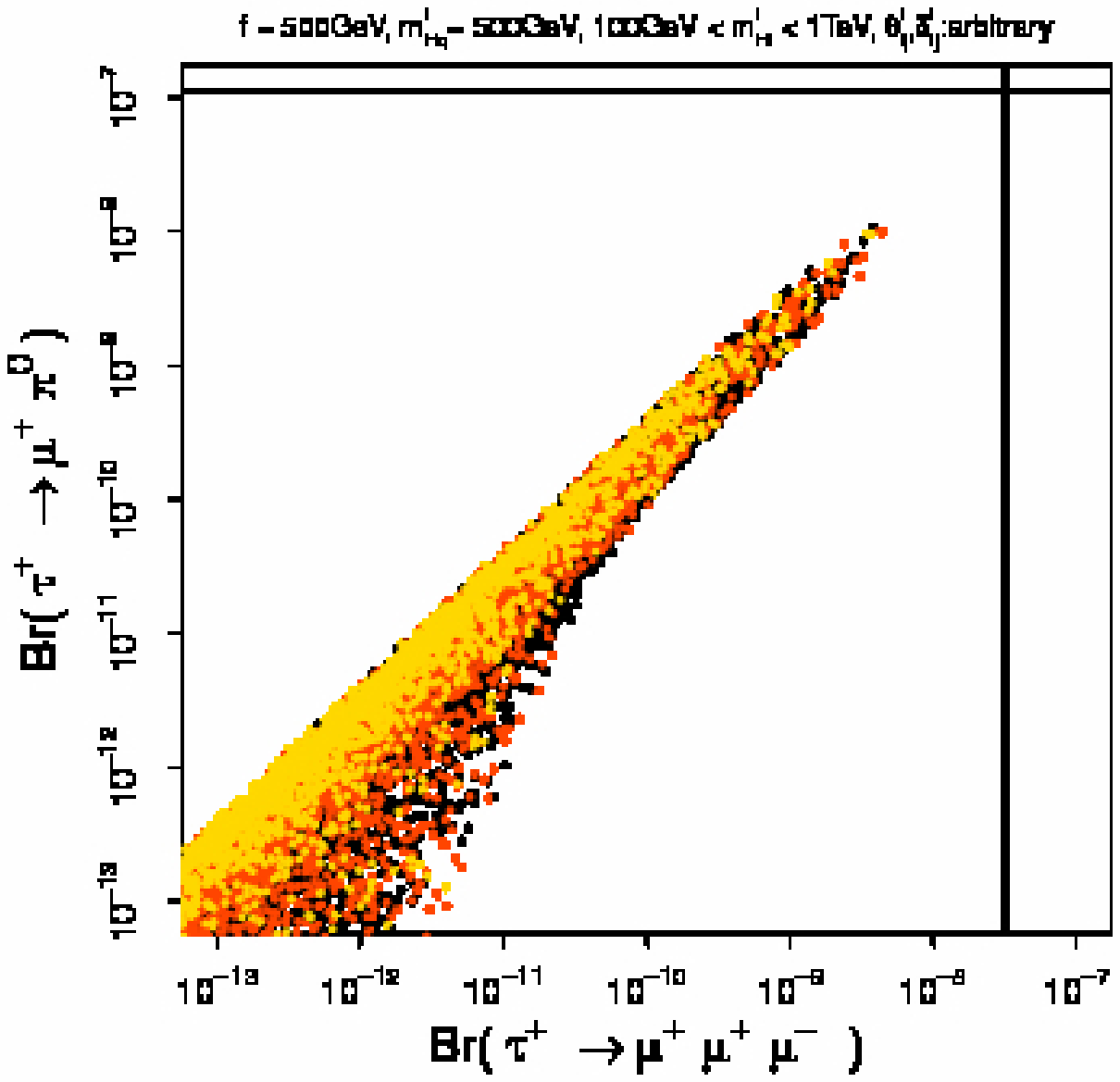} &
  \includegraphics[width=20em,clip]{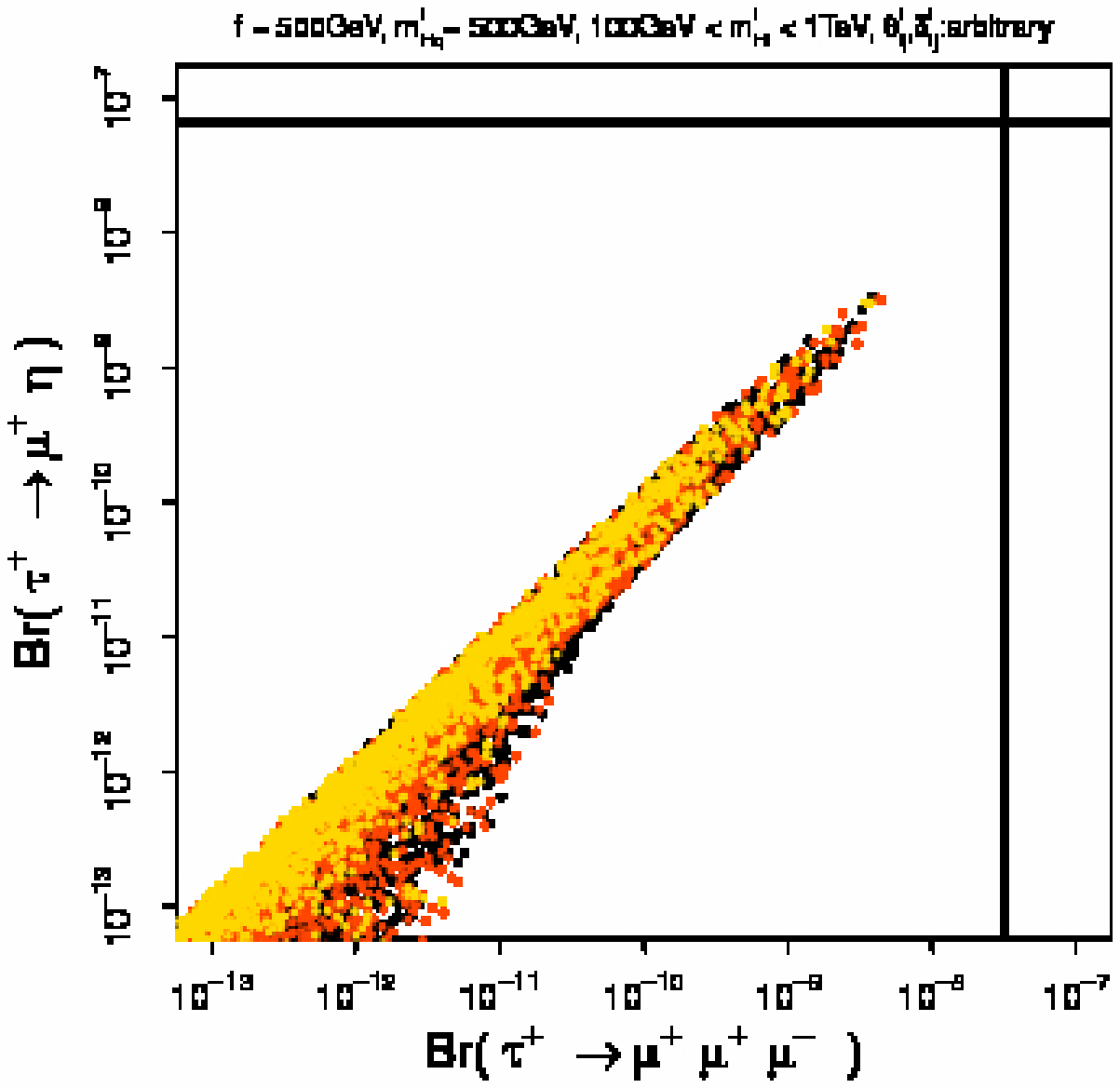} \\
  (a) & (b)
\end{tabular}
\begin{tabular}{c}
  \includegraphics[width=20em,clip]{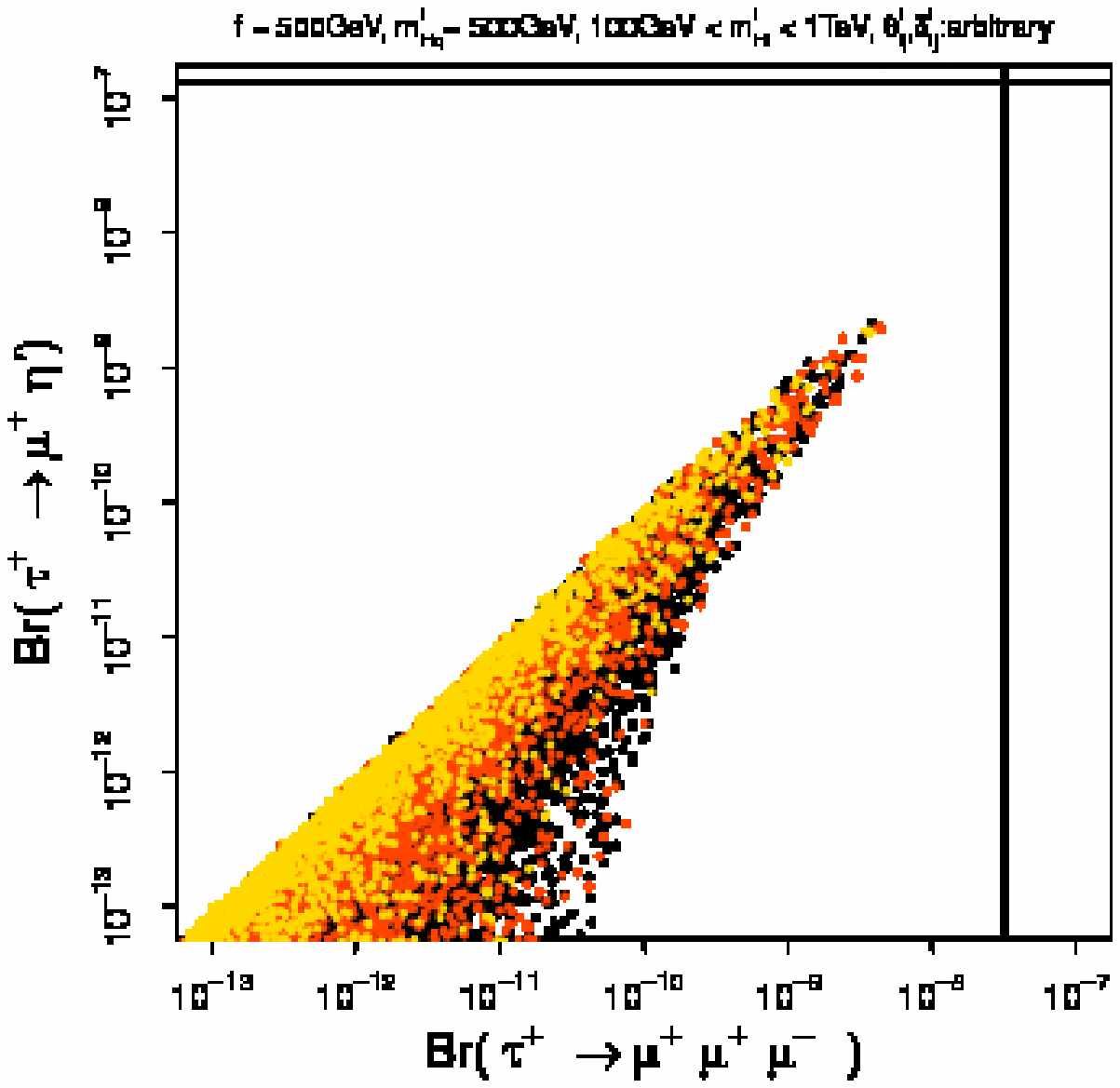} \\
  (c)
\end{tabular}
\caption{
Correlations between $\text{Br}(\tau^+\to\mu^+ P)$ and
$\text{Br}(\tau^+\to\mu^+ \mu^+ \mu^-)$ for $P=\pi^0$, $\eta$ and $\eta'$.
Input parameters are the same as those in Fig.~\ref{FigMucor}.
The vertical and the horizontal solid lines are the experimental upper limits.
The color code is the same as in Fig.~\ref{FigMucor}.
}
\label{FigTmp}
\end{figure} 

\begin{figure}[htb]
\begin{tabular}{cc}
  \includegraphics[width=20em,clip]{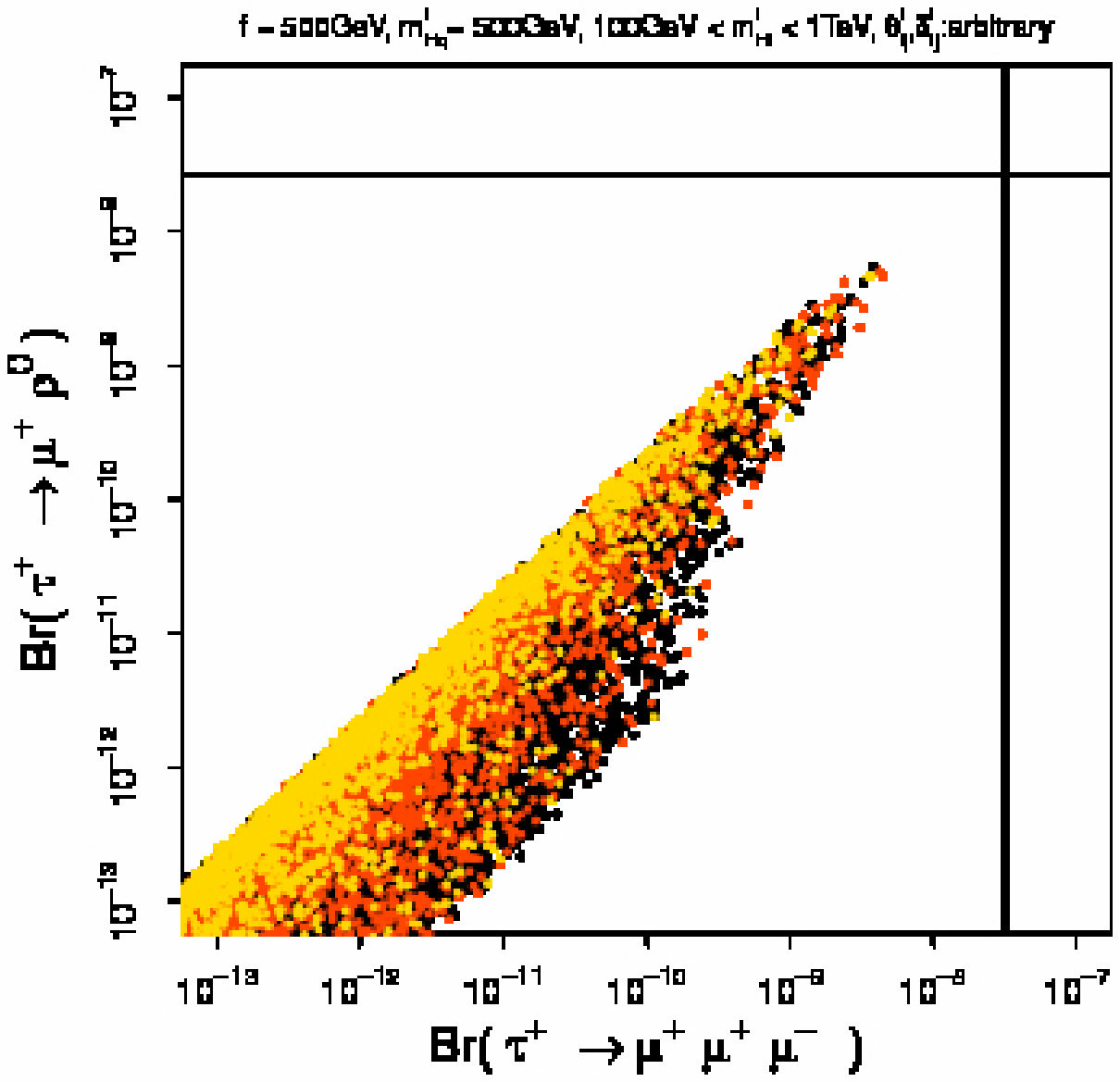} &
  \includegraphics[width=20em,clip]{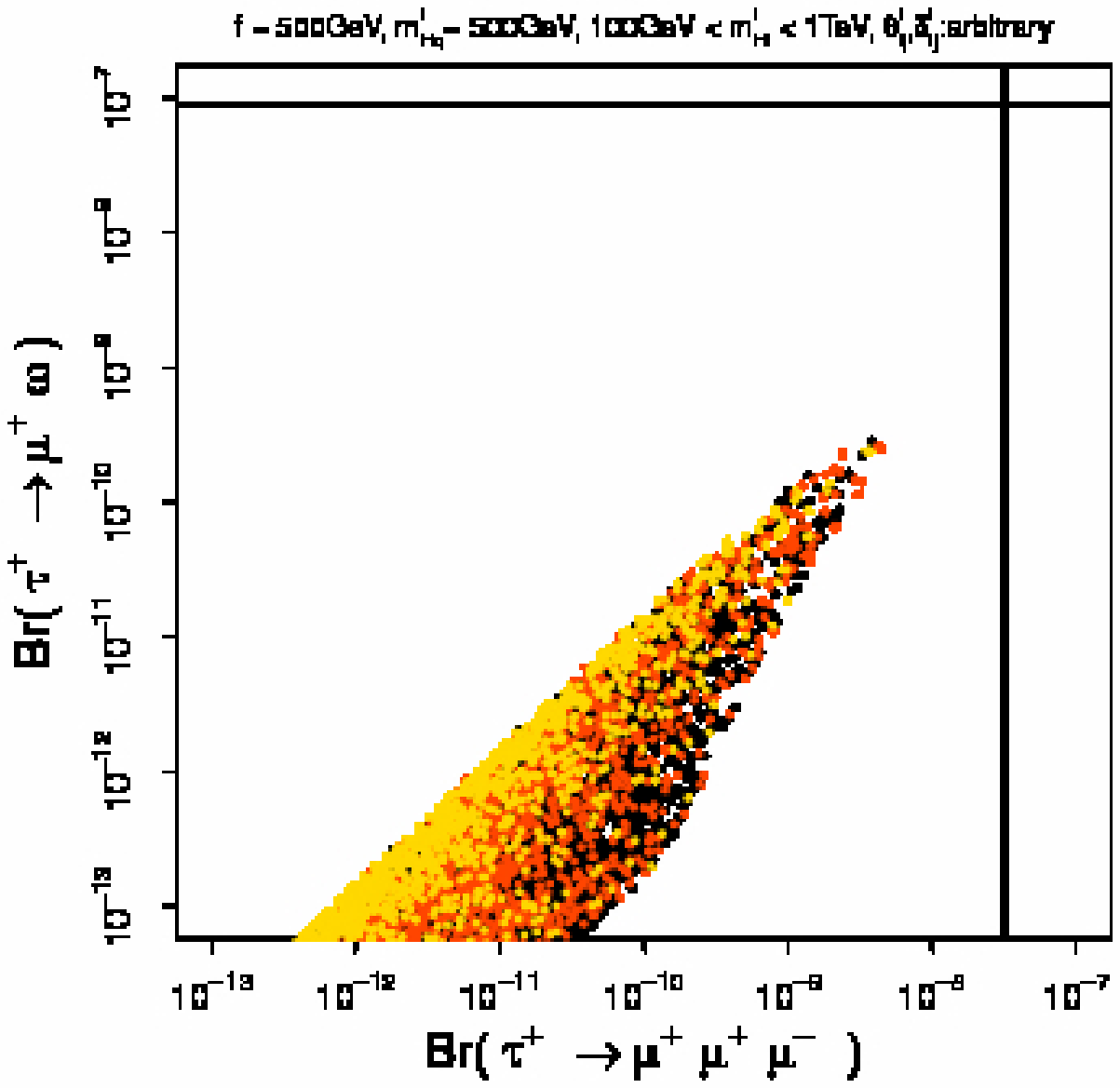} \\
  (a) & (b)
\end{tabular}
\begin{tabular}{c}
  \includegraphics[width=20em,clip]{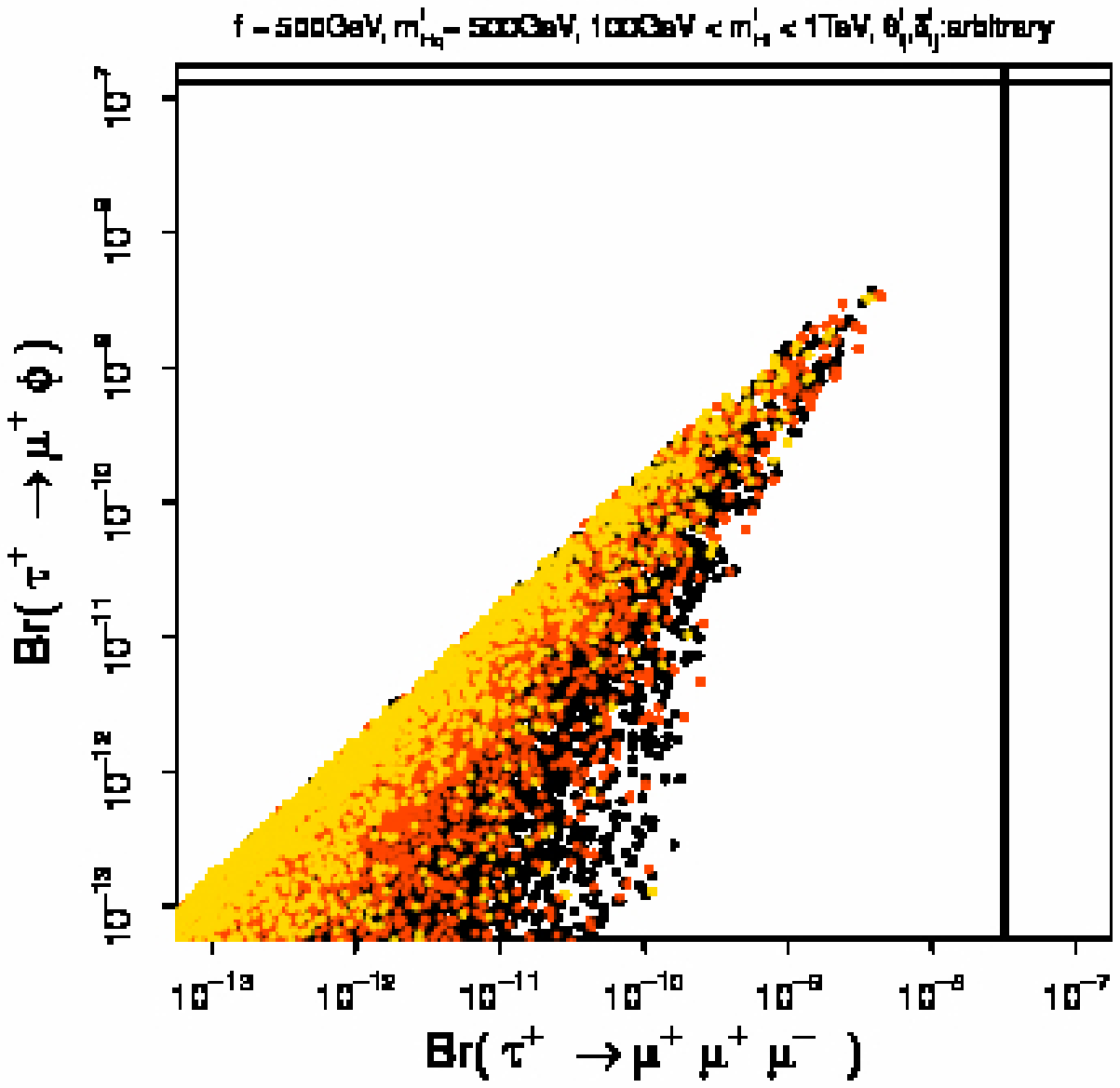} \\
  (c)
\end{tabular}
\caption{
Correlations between $\text{Br}(\tau^+\to\mu^+ V)$ and
$\text{Br}(\tau^+\to\mu^+ \mu^+ \mu^-)$ for $V=\rho^0$, $\omega$ and
$\phi$.
Input parameters are the same as those in Fig.~\ref{FigMucor}.
The vertical and the horizontal solid lines are the experimental upper limits.
The color code is the same as in Fig.~\ref{FigMucor}.
}
\label{FigTmv}
\end{figure} 
\begin{figure}[htb] 
\begin{tabular}{cc}
  \includegraphics[width=20em,clip]{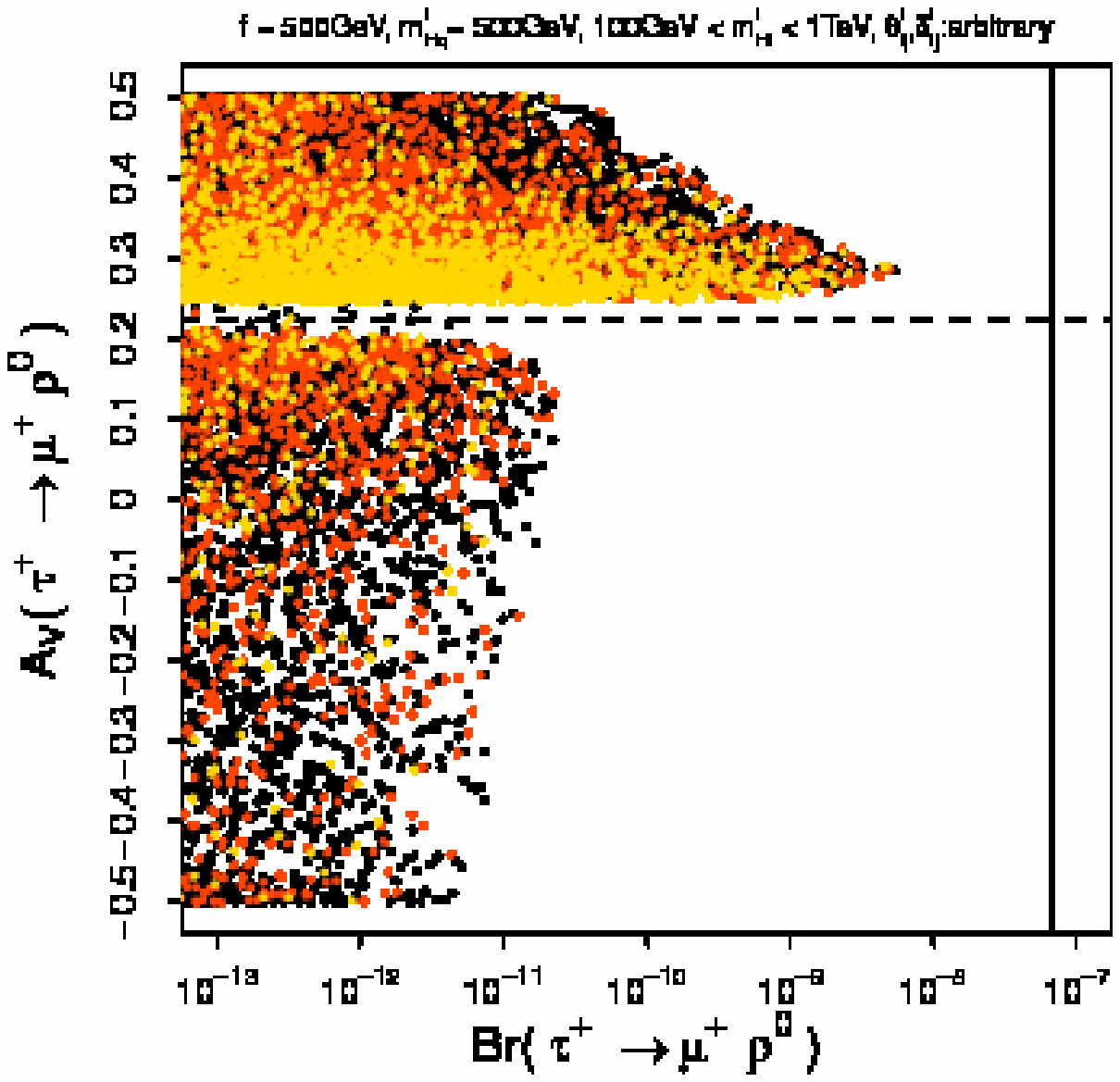} &
  \includegraphics[width=20em,clip]{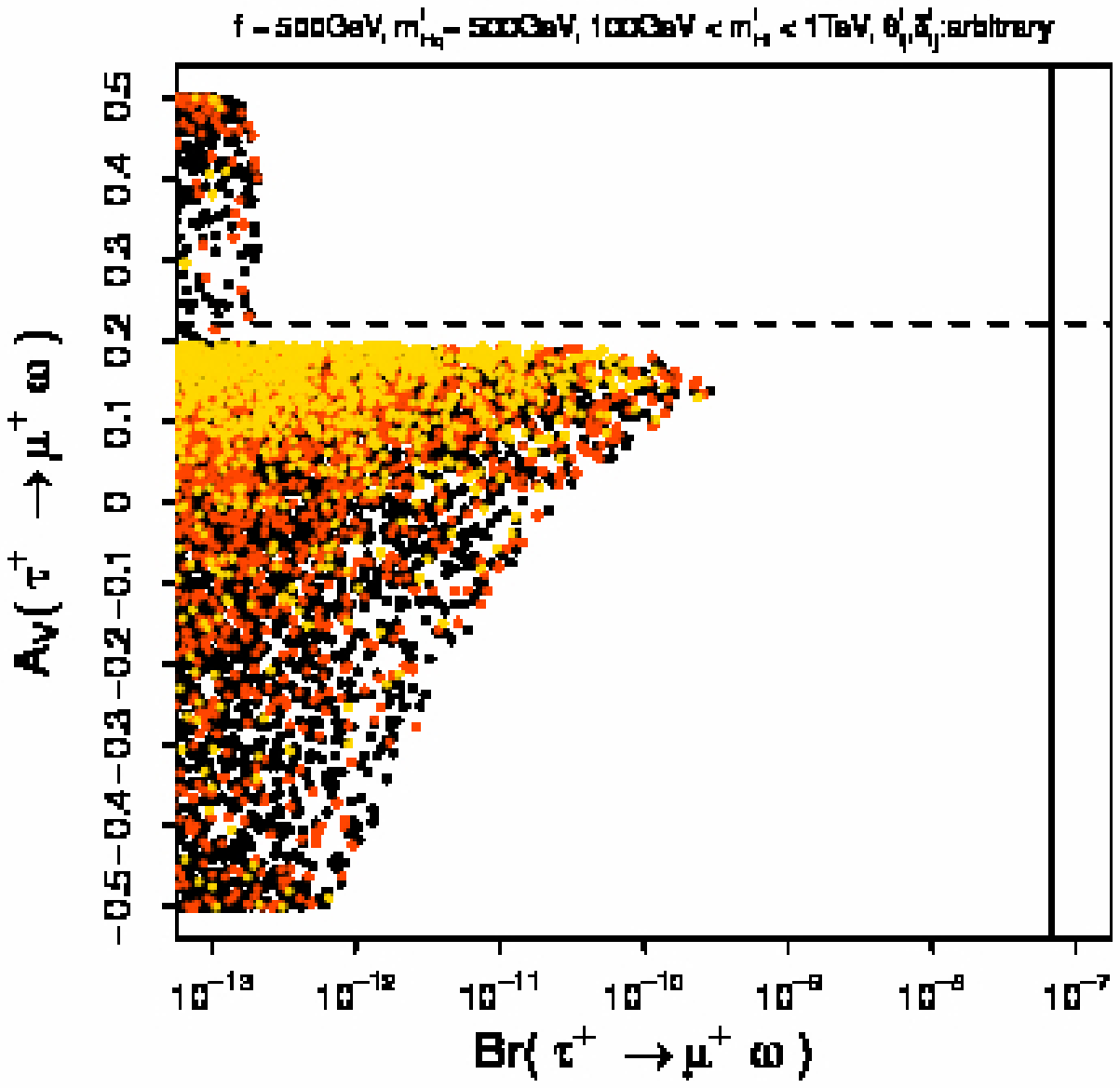} \\
  (a) & (b)
\end{tabular}
\begin{tabular}{c}
  \includegraphics[width=20em,clip]{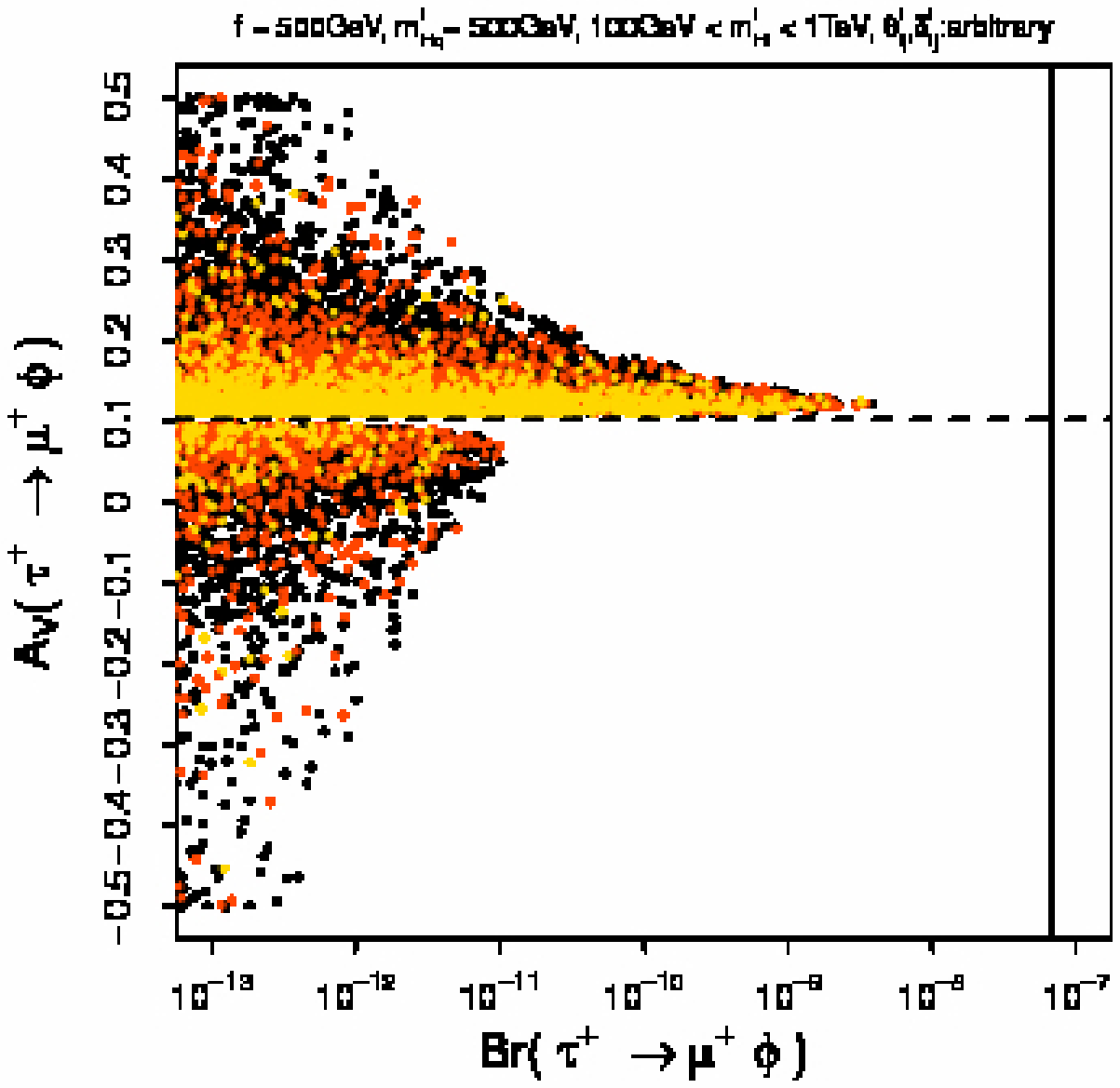}
  (c)
\end{tabular}
\caption{
Correlations between the polarization asymmetries and the branching ratios
for $\tau^+ \to \mu^+ V$ ($V=\rho^0$, $\omega$ and $\phi$) decay modes.
Input parameters and the color code are the same as those in
Fig.~\ref{FigMucor}.
The vertical solid lines are the experimental upper limits of the branching
ratios.
The dashed lines show asymmetries in $A_R^{\text{LHT}}\to 0$ limit.
}
\label{FigVasy}
\end{figure} 

The correlations between branching ratios of $\tau^+\to \mu^+ P$
($P=\pi^0,\,\eta,\,\eta'$) and $\tau^+\to \mu^+ \mu^+ \mu^-$ are shown
in Fig.~\ref{FigTmp}.
These branching ratios are roughly of the same order of magnitude for
each set of input parameters, because relevant Wilson
coefficients $g_{Ll,r}^{\text{I,LHT}}$ and $g_{Ll,r(q)}^{\text{LHT}}$ in
Eqs.~(\ref{EqVeclr1}), (\ref{EqVeclr2}), 
(\ref{eq:gLlq-tm}) and (\ref{eq:gLrq-tm}) are
similar in magnitude and the effect of the dipole term
(in $\tau^+\to \mu^+ \mu^+ \mu^-$) on the branching ratio is small.

In Fig.~\ref{FigTmv} we show the branching ratios of $\tau^+\to \mu^+ V$
($V=\rho^0,\,\omega,\,\phi$) decays.
The qualitative behavior of the correlations is the same as in the  
$\tau^+\to \mu^+ P$ case.
Polarization asymmetries are shown in Fig.~\ref{FigVasy}.
The contribution of the dipole term in $\tau^+\to \mu^+ V$ 
affects the polarization asymmetry
though the effect on the branching ratio is small.
In the $A_R^{\text{LHT}}\to 0$ limit, the asymmetry is determined as
\begin{align}
  \left.
    A_V(\tau^+\to\mu^+ V)_{\text{LHT}}
  \right|_{A_R^{\text{LHT}}\to 0}
 = \frac{1}{2} \frac{m_\tau^2 - 2 m_V^2}{m_\tau^2 + 2 m_V^2}
 \approx
\left\{
  \begin{array}{lcl}
    0.22 &\text{for}& V=\rho^0,\,\omega, \\
    0.10 &\text{for}& V=\phi.
  \end{array}
\right.
\label{eq:AVnodipole}
\end{align}
The deviations from these values seen in Fig.~\ref{FigVasy} can be
understood as effects of interferences between the dipole and the
four-Fermi terms.
In particular, in the parameter region where the branching ratios are 
larger than
$10^{-10}$, contributions of the four-Fermi terms dominate the
decay amplitude, so that values of $A_V$ are not much different from those
given in (\ref{eq:AVnodipole}).

As in the case of trilepton modes, we can derive several relations 
among various branching ratios  asymmetries of
semileptonic modes, which is discussed in Appendix~\ref{AppRel}.
\subsection{Correlations among $\mu\to e$, $\tau\to \mu$, and $\tau\to e$ 
  transitions}
\label{sec:correlations-me-tm-te}
\begin{figure}[htb] 
\begin{tabular}{cc}
  \includegraphics[width=20em,clip]{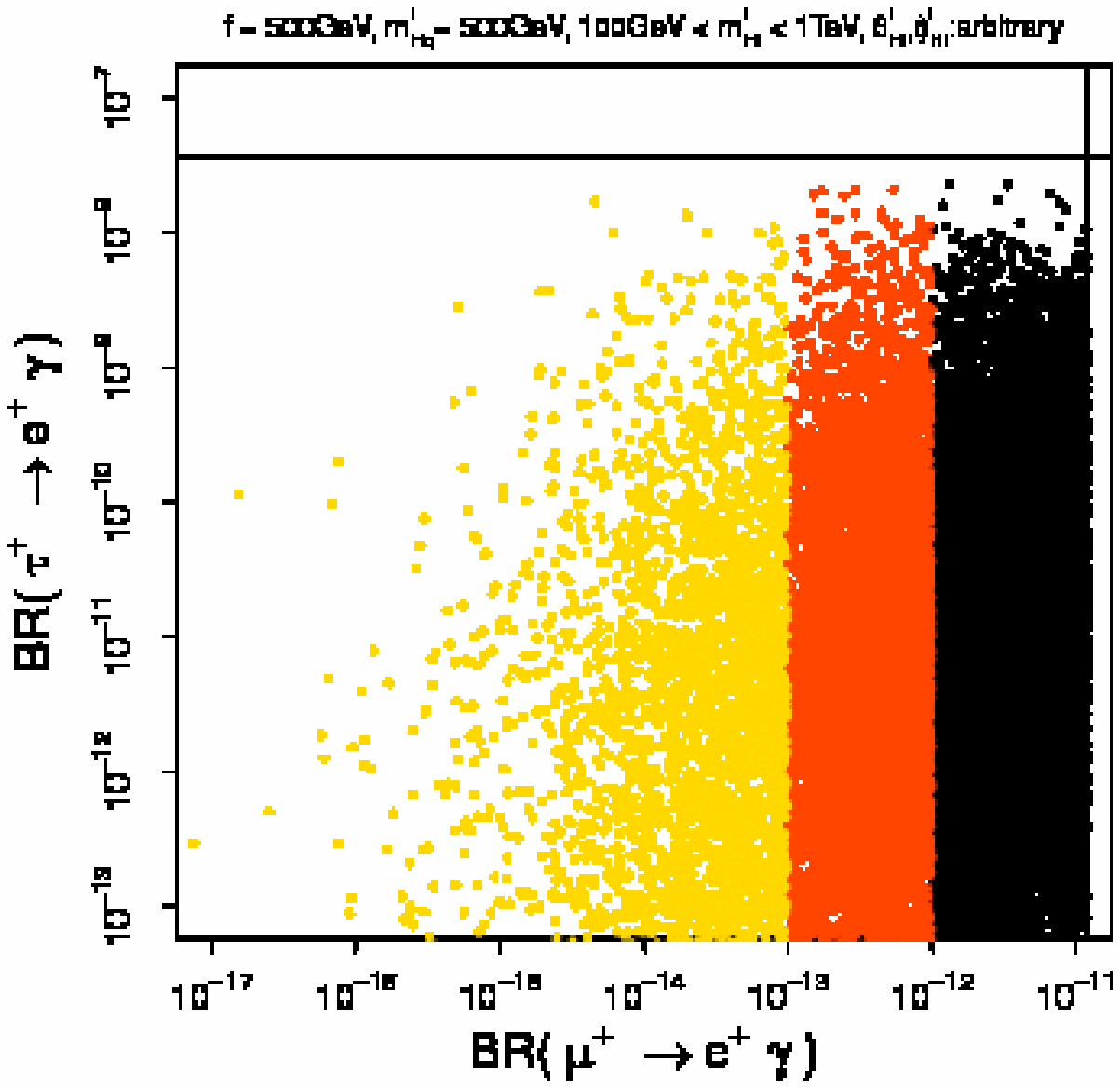} &
  \includegraphics[width=20em,clip]{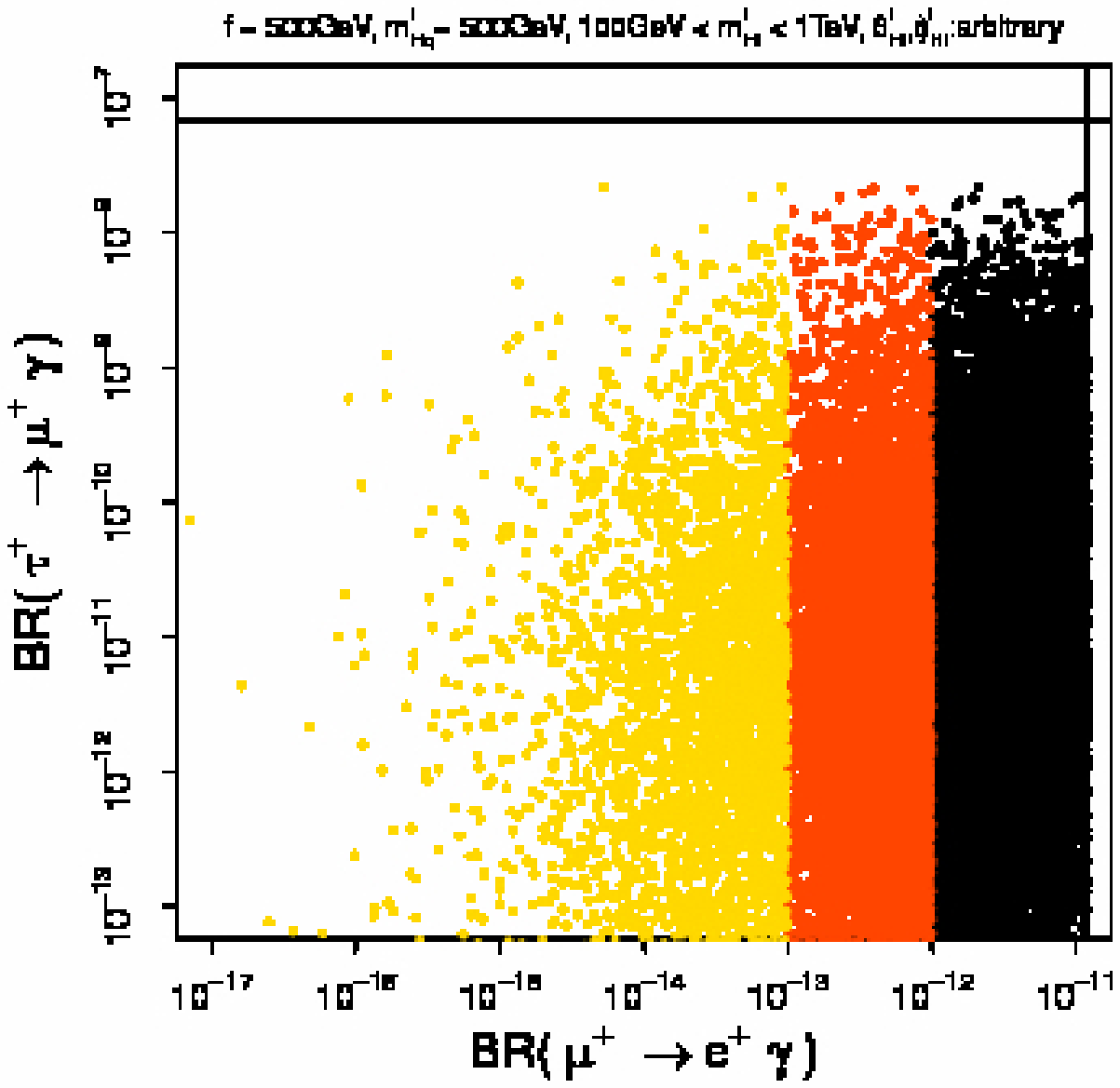} \\
  (a) & (b)
\end{tabular}
\begin{tabular}{c}
  \includegraphics[width=20em,clip]{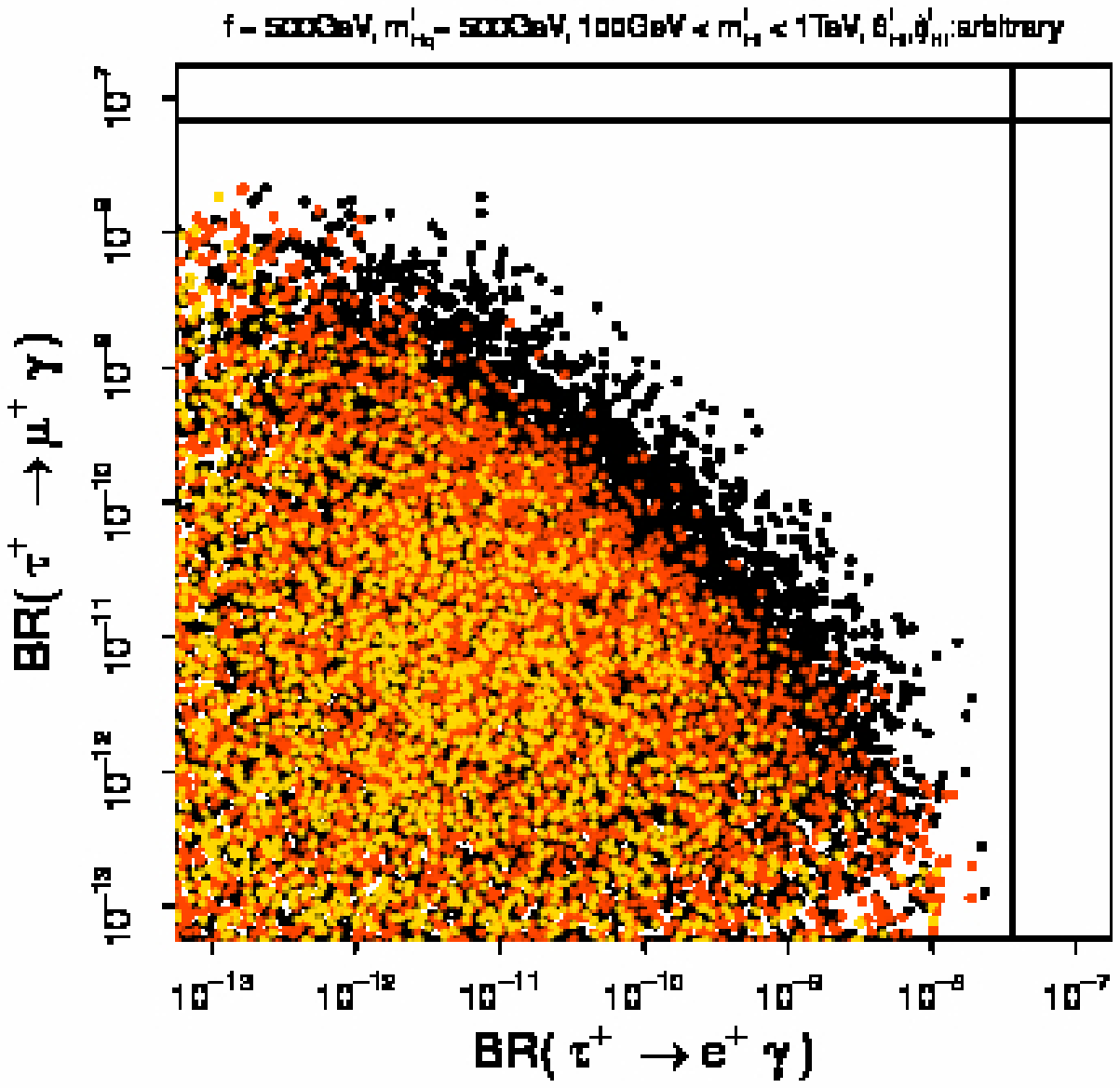} \\
  (c)
\end{tabular}
\caption{
Correlations among branching ratios of $\mu^+ \to e^+ \gamma$,
$\tau^+ \to \mu^+ \gamma$ and $\tau^+ \to e^+ \gamma$.
Input parameters and the color code are the same as those in
Fig.~\ref{FigMucor}.
The horizontal and the vertical lines are experimental upper bounds.
}
\label{fig:lfvg}
\end{figure}

\begin{figure}[htb]
\begin{tabular}{cc}
  \includegraphics[width=20em,clip]{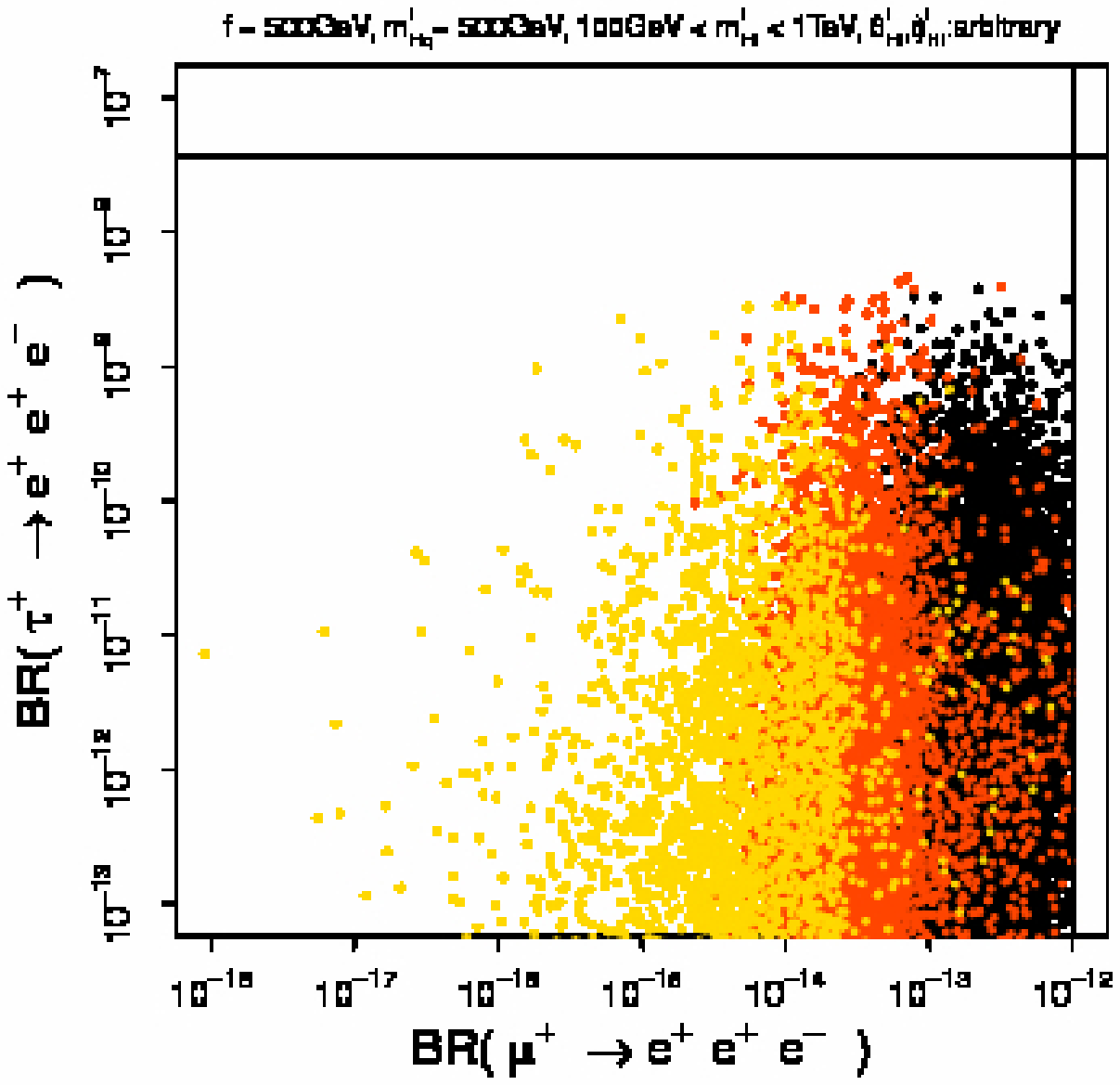} &
  \includegraphics[width=20em,clip]{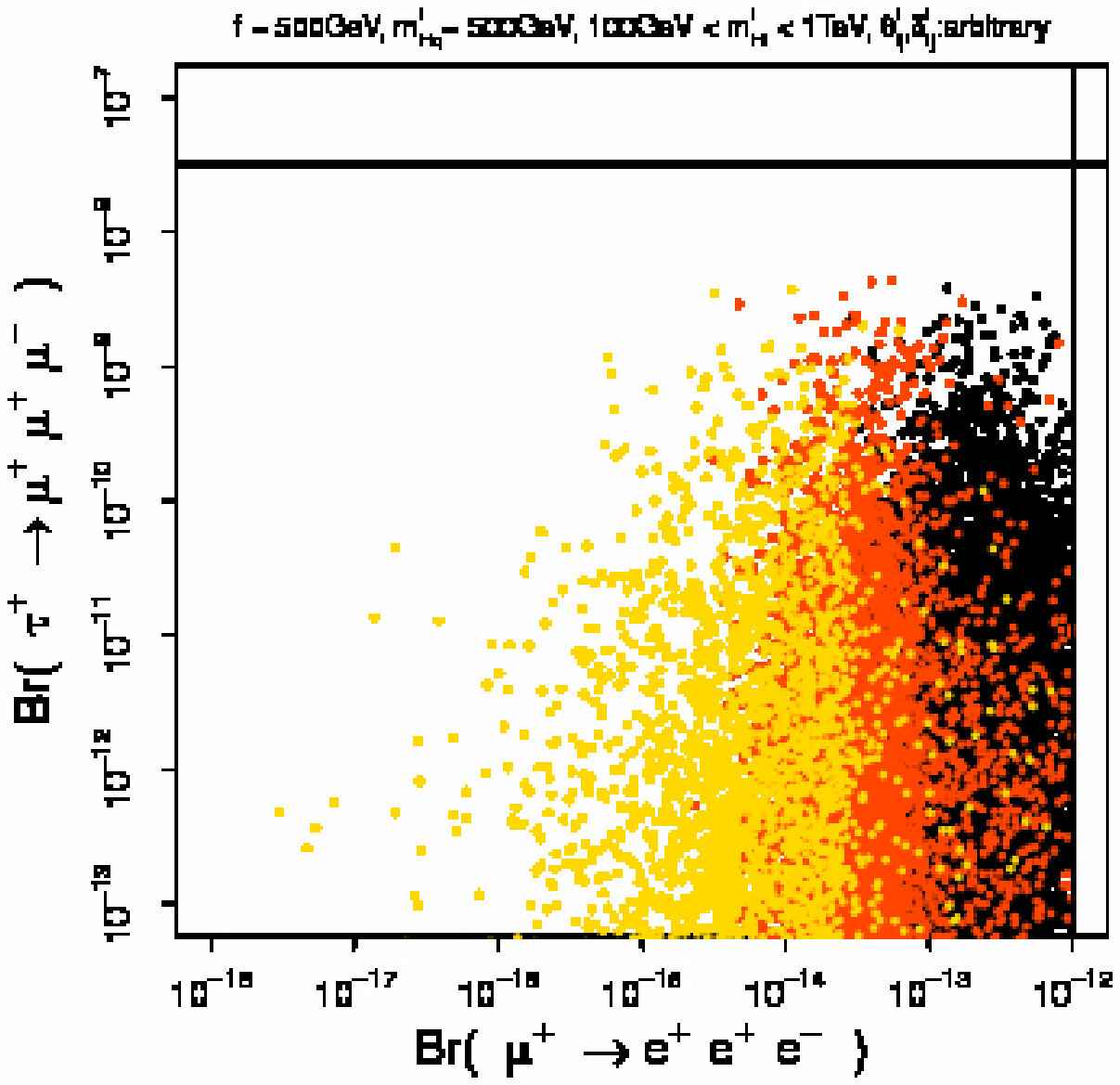} \\
  (a) & (b)
\end{tabular}
\begin{tabular}{c}
  \includegraphics[width=20em,clip]{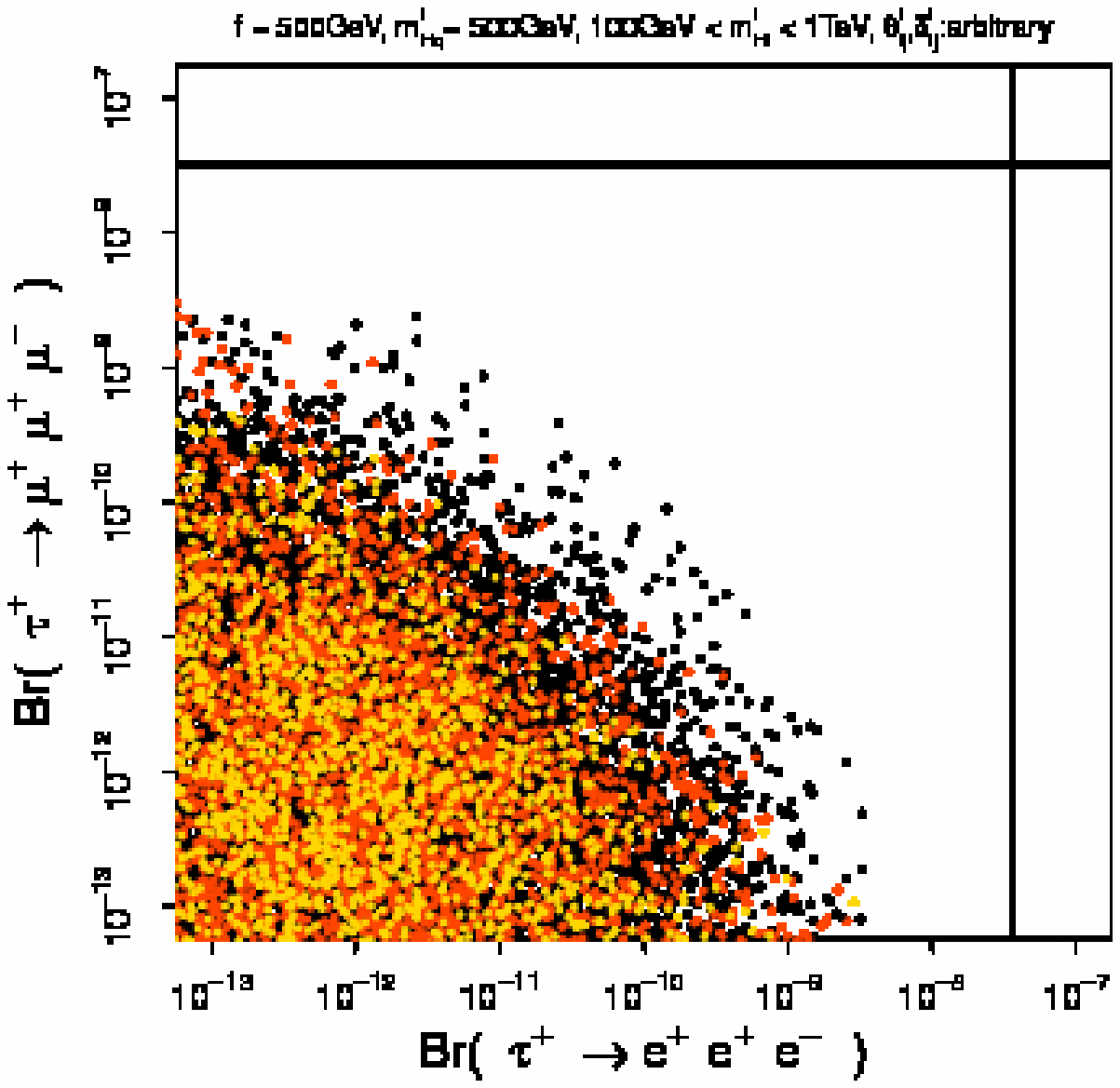} \\
  (c)
\end{tabular}
\caption{
Correlations among branching ratios of $\mu^+ \to e^+ e^+ e^-$,
$\tau^+ \to \mu^+ \mu^+ \mu^-$ and $\tau^+ \to e^+ e^+ e^-$.
Input parameters and the color code are the same as those in
Fig.~\ref{FigMucor}.
The horizontal and the vertical lines are experimental upper bounds.
}
\label{fig:lfvl}
\end{figure} 

There are three classes of processes which change the lepton flavor by one: 
$\mu\to e$, $\tau\to \mu$, and $\tau\to e$ transitions.
Here, we discuss correlations among different lepton flavor
transitions.

We show correlations among branching ratios of $\mu^+\to e^+ \gamma$,
$\tau^+\to \mu^+ \gamma$, and $\tau^+\to e^+ \gamma$ in Fig.~\ref{fig:lfvg}.
As can be seen in Fig.~\ref{fig:lfvg}(a) and (b), there are no direct
correlations between the branching ratios of $\mu^+\to e^+\gamma$ and $\tau$
LFV decays, because we take the T-odd lepton masses $m_{H\ell}^i$ and
parameters in the mixing matrix $V_{H\ell}$ as free parameters and vary
them independently.
On the other hand, in Fig.~\ref{fig:lfvg}(c), we notice that a parameter
region where both $\text{Br}(\tau^+\to \mu^+ \gamma)$ and
$\text{Br}(\tau^+\to e^+ \gamma)$ are larger than $10^{-9}$ is not allowed.
When both $\tau\to \mu$ and $\tau\to e$ transition amplitudes are
large, the corresponding $\mu\to e$ amplitudes also become large
so that the branching ratios of the $\mu \to e$ transition processes exceed 
the experimental bounds. 
In fact, we have checked that the branching ratios
$\text{Br}(\ell_1\to \ell_2 \gamma)
/\text{Br}(\ell_1\to\ell_2 \nu \bar{\nu})$
[$(\ell_1,\,\ell_2)=(\mu,\,e)$, $(\tau,\,\mu)$ and $(\tau,\,e)$]
distribute within the same range ($\lesssim 10^{-7}$) when we ignore
experimental constraints on $\mu\to e$ processes.

The correlations among 
the type I leptonic three-body decay branching ratios are
shown in Fig.~\ref{fig:lfvl}.
The behavior of the correlations is the same as the case of
Fig.~\ref{fig:lfvg}.

\section{Conclusion}
%
%
We have calculated branching ratios and angular and 
forward-backward asymmetries of $\mu$ and $\tau$-LFV processes in the LHT.
We have obtained the following results:
\begin{itemize}
 \item 
   The branching ratios of three $\mu$-LFV processes, $\mu^+ \to e^+ \gamma$,
   $\mu^+ \to e^+ e^+ e^-$ and the $\mu-e$ conversion rates,
	can be close to the present experimental bounds.
	There is a rather strong correlation between the branching ratios of 
	$\mu^+ \to e^+ \gamma$ and $\mu^+ \to e^+ e^+ e^-$.
	This is in contrast to the case of $\mu^+ \to e^+ \gamma$ and 
	the $\mu-e$ conversion where no correlation is observed.
	These features are noted in Ref.~\cite{Blanke:2007db}
 \item
   The parity asymmetry of $\mu^+ \to e^+ \gamma$ is -1/2.
   For $\mu^+ \to e^+ e^+ e^-$, the parity asymmetries can be large, 
	whereas the time-reversal asymmetry is within 10\%.
 \item 
   We have calculated the $\mu-e$ conversion rates for various muonic atoms:
   Al, Ti, Au and Pb.
	In most of the parameter space, 
	ratios of the conversion rates are found within 1 order of magnitude. 
	In some cases, however, the conversion rates for Ti, Au and Pb can be 
	close to the experimental bounds even if the rate for Al is suppressed.
 \item 
   The maximal values of the branching ratios for various $\tau$-LFV processes 
	are $10^{-9}$--$10^{-8}$ except for $O(10^{-10})$ for 
	$\tau^+ \to \mu^+ \omega$ and $\tau^+ \to e^+ \omega$, and 
	$O(10^{-13})$ for $\tau^+ \to \mu^+ \mu^+ e^-$ 
	and $\tau^+ \to e^+ e^+ \mu^-$.
 \item 
   The parity asymmetries of $\tau^+ \to \mu^+ \gamma$ and 
	$\tau^+ \to e^+ \gamma$ are -1/2 and 
	$\tau^+ \to \mu^+ P$ and $\tau^+ \to e^+ P$ are 1/2 reflecting 
	the chirality structure.
	For $\tau^+ \to \mu^+ V$ and $\tau^+ \to e^+ V$, 
	if branching ratios are larger than $10^{-10}$,
	the asymmetries are about 0.3, 0.15, and 0.1 
	for $\rho^0$, $\omega$, and $\phi$, respectively.
 \item 
	There are sizable parity asymmetries in
   $\tau^+ \to \mu^+ \mu^+ \mu^-$, $\tau^+ \to e^+ e^+ e^-$,
	$\tau^+ \to \mu^+ e^+ e^-$, and $\tau^+ \to e^+ \mu^+ \mu^-$.
	For $\tau^+ \to \mu^+ e^+ e^-$ and $\tau^+ \to e^+ \mu^+ \mu^-$,
	forward-backward asymmetries can be defined, and
	there is a relation among asymmetries as Eq.~\eqref{EqProprel}.
	The time-reversal asymmetries are found to be very suppressed.
\end{itemize}

The search for LFV in $\mu$ and $\tau$ decays has a complementary role to 
the new particle search at the LHC experiment to explore the TeV-scale physics.
We have seen that branching ratios of these processes in the LHT can be within 
the reach of ongoing and planned experiments such as
MEG, COMET and Mu2e for $\mu$-LFV processes and
$\tau$ rare decay searches at future B factories and the LHC.
We have found that various asymmetries defined with 
the help of $\mu$ and $\tau$ 
polarizations reflect the characteristic chirality structure of the LHT.
Experimental searches for LFV processes using polarization of initial leptons 
are important to identify this model among various candidates of 
TeV-scale physics models.

\begin{acknowledgments}
We thank S.Mihara for a helpful conversation.
The work of T.G. and Y.O. is supported in part by Grant-in-Aid for
Science Research, Japan Society for the Promotion of Science, No.20244037.
The work of  Y.O. is supported in part by Grant-in-Aid for
Ministry of Education, Culture, Sports, Science and Technology Japan, 
No.16081211 and No.22244031.
The work of Y.Y. is supported in part by Grant-in-Aid for
JSPS Fellows, Japan Society for the Promotion of Science,
No.22.3824.
\end{acknowledgments}

\appendix 
\section{General formulae of branching ratios and asymmetries}
\label{AppFormulae}
We present general formulae of the branching ratios and asymmetries in
the LFV processes based on the general low energy Lagrangians.
Some basic formulae can be found 
in Refs.~\cite{Okada:1999zk,Kitano:2000fg,RefLfvmeson}.

\subsection{$\mu^+ \to e^+ \gamma$, $\tau^+ \to \mu^+ \gamma$ and 
            $\tau^+ \to e^+ \gamma$}
The Lagrangian of the radiative two-body decay is 
\begin{align}
 \mathcal{L}_{\gamma} = 
  -\frac{4G_F}{\sqrt{2}}[
	  m_{\tau} A_R  \bar{\tau}_{R} \sigma^{\mu \nu} \mu_{L} F_{\mu \nu}
	+ m_{\tau} A_L  \bar{\tau}_{L} \sigma^{\mu \nu} \mu_{R} F_{\mu \nu}
	+ \text{H.c.} ].
\end{align}
The differential decay width of $\tau^+ \to \mu^+ \gamma$ is
\begin{align}
 \frac{d\text{Br}(\tau^+ \to \mu^+ \gamma)}{d \cos\theta} =
   \tau_\tau \frac{G_F^2 m_\tau^5}{\pi} 
	( |A_L|^2 +|A_R|^2 +(|A_L|^2 -|A_R|^2) \cos\theta ).
\end{align}
The branching ratio and the asymmetry defined in Eq.~\eqref{EqAsymgamma} are
\begin{align}
 \text{Br}(\tau^+ \to \mu^+ \gamma) =&
   \tau_\tau \frac{2G_F^2 m_\tau^5}{\pi} 
   (|A_R|^2 +|A_L|^2),\\
 A_\gamma (\tau^+ \to \mu^+ \gamma) =&
   \frac{1}{2}\frac{|A_L|^2 -|A_R|^2}{|A_L|^2 +|A_R|^2}.
\end{align}
\subsection{$\mu^+ \to e^+ e^+ e^-$, $\tau^+ \to \mu^+ \mu^+ \mu^-$ and 
            $\tau^+ \to e^+ e^+ e^-$}
\label{AppFormulaeT1}
The Lagrangian of the type I leptonic three-body decay is
\begin{align}
	\mathcal{L}_\text{I} =
	-\frac{4G_F}{\sqrt{2}}& \bigl[
	  m_{\tau} A_R  \bar{\tau}_{R} \sigma^{\mu \nu} \mu_{L} F_{\mu \nu}
	+ m_{\tau} A_L  \bar{\tau}_{L} \sigma^{\mu \nu} \mu_{R} F_{\mu \nu} \n &
	+ g_{Rs}^\text{I} 
	  ( \bar{\tau}_{R} \mu_{L} ) ( \bar{\mu}_{R} \mu_{L} )
	+ g_{Ls}^\text{I} 
	  ( \bar{\tau}_{L} \mu_{R} ) ( \bar{\mu}_{L} \mu_{R} ) \n &
	+ g_{Rr}^\text{I} 
	  ( \bar{\tau}_{R} \gamma^{\mu} \mu_{R}) 
	  ( \bar{\mu }_{R} \gamma_{\mu} \mu_{R})
	+ g_{Ll}^\text{I} 
	  ( \bar{\tau}_{L} \gamma^{\mu} \mu_{L}) 
	  ( \bar{\mu }_{L} \gamma_{\mu} \mu_{L}) \n &
	+ g_{Rl}^\text{I} 
	  ( \bar{\tau}_{R} \gamma^{\mu} \mu_{R}) 
	  ( \bar{\mu }_{L} \gamma_{\mu} \mu_{L})
	+ g_{Lr}^\text{I} 
	  ( \bar{\tau}_{L} \gamma^{\mu} \mu_{L}) 
	  ( \bar{\mu }_{R} \gamma_{\mu} \mu_{R})
	+ \text{H.c.}
	\bigr].
\end{align}
The differential partial decay width of $\tau^+\to\mu^+\mu^+\mu^-$ can be
written as
\begin{align}
  \frac{d^4 \text{Br}(\tau^+\to\mu^+\mu^+\mu^-)}
       {dx_b dx_c \,d\phi \,d\!\cos \theta} 
  =\tau_\tau \frac{G_F^2 m_{\tau}^5 }{128\pi^4}& (
    M_O^\text{I} (x_b,x_c) 
  + M_Z^\text{I} (x_b,x_c) \cos\theta \n &
  + M_X^\text{I} (x_b,x_c) \sin\theta \cos\phi
  + M_Y^\text{I} (x_b,x_c) \sin\theta \sin\phi ),
\label{eq:dGamma-tmmm}
\end{align}
where
\begin{subequations}
\begin{align}
  M_O^{\text{I}} (x_b,x_c) =& 
    ( C_{R1}^\text{I} +C_{L1}^\text{I}) a_1 (x_b, x_c)
   +( C_{R2}^\text{I} +C_{L2}^\text{I}) a_2 (x_b, x_c) \n &
   +( C_{R3}^\text{I} +C_{L3}^\text{I}) a_3 (x_b, x_c) \n &
   +( C_{J1}^\text{I} +C_{J2}^\text{I}) a_4 (x_b, x_c)
   +( C_{J3}^\text{I} +C_{J4}^\text{I}) a_5 (x_b, x_c) , \\
  M_Z^{\text{I}} (x_b,x_c) = &
    ( C_{R1}^\text{I} -C_{L1}^\text{I}) b_1 (x_b, x_c)
   +( C_{R2}^\text{I} -C_{L2}^\text{I}) b_2 (x_b, x_c) \n &
   +( C_{R3}^\text{I} -C_{L3}^\text{I}) a_3 (x_b, x_c) \n &
   -( C_{J1}^\text{I} -C_{J2}^\text{I}) a_4 (x_b, x_c)
   +( C_{J3}^\text{I} -C_{J4}^\text{I}) a_5 (x_b, x_c) , \\
  M_X^{\text{I}} (x_b,x_c) = &
    ( C_{R1}^\text{I} -C_{L1}^\text{I}) c_1 (x_b, x_c)
   +( C_{R2}^\text{I} -C_{L2}^\text{I}) c_2 (x_b, x_c) \n &
   +( C_{J1}^\text{I} -C_{J2}^\text{I}) c_3 (x_b, x_c)
   +( C_{J3}^\text{I} -C_{J4}^\text{I}) c_4 (x_b, x_c) , \\
  M_Y^{\text{I}} (x_b,x_c) = &
    C_{J5}^\text{I} c_3 (x_b, x_c)
   +C_{J6}^\text{I} c_4 (x_b, x_c) ,
\end{align}
\end{subequations}
where $C^{\text{I}}_{R1,\cdots}$ are defined as
\begin{subequations}
\begin{align}
  C^\text{I}_{R1} &= \abs{ e A_R }^2 , &
  C^\text{I}_{L1} &= \abs{ e A_L }^2 , 
\label{EqTypeIwilson0} \\
  C^\text{I}_{R2} &= \abs{ g^\text{I}_{Rl} }^2 , &
  C^\text{I}_{L2} &= \abs{ g^\text{I}_{Lr} }^2 , \\
  C^\text{I}_{R3} &= 
    \frac{ \abs{ g^\text{I}_{Rs} }^2 }{16} + \abs{ g^\text{I}_{Rr} }^2 ,&
  C^\text{I}_{L3} &= 
    \frac{ \abs{ g^\text{I}_{Ls} }^2 }{16} + \abs{ g^\text{I}_{Ll} }^2 , \\
  C^\text{I}_{J1} &= \text{Re}[ eA_R g_{Ll}^{\text{I}\ast} ] , &
  C^\text{I}_{J2} &= \text{Re}[ eA_L g_{Rr}^{\text{I}\ast} ] , \\
  C^\text{I}_{J3} &= \text{Re}[ eA_R g_{Lr}^{\text{I}\ast} ] , &
  C^\text{I}_{J4} &= \text{Re}[ eA_L g_{Rl}^{\text{I}\ast} ] , \\
  C^\text{I}_{J5} &= 
    \text{Im}[ eA_R g_{Ll}^{\text{I}\ast} + eA_L g_{Rr}^{\text{I}\ast} ],&
  C^\text{I}_{J6} &= 
    \text{Im}[ eA_R g_{Lr}^{\text{I}\ast} + eA_L g_{Rl}^{\text{I}\ast} ].
\label{EqTypeIwilson1}
\end{align}
\end{subequations}
The functions are defined as follows:
\begin{subequations}
\begin{align}
  a_1 (x_b, x_c) &= 
    8 \frac{ ( 1 -x_b) (2x_b^2 -2x_b +1) +( 1 -x_c) ( 2x_c^2 -2x_c +1)}
	         { ( 1-x_b) ( 1 -x_c)} ,\\
  a_2 (x_b, x_c) &= 
    2( x_b (1 -x_b) +x_c ( 1 -x_c) ) ,\\
  a_3 (x_b, x_c) &= 
    8 ( 2 -x_a -x_b) ( x_b +x_c -1) ,\\
  a_4 (x_b, x_c) &= 
    32 ( x_b + x_c -1),\\
  a_5 (x_b, x_c) &= 
    8 ( 2 -x_b -x_c),
\end{align}
\end{subequations}
\begin{subequations}
\begin{align}
  b_1 (x_b, x_c) &=
    8 \Bigl( 
	    \frac{1 -2x_b (1 -x_b)}{1 -x_c} +\frac{1 -2x_c (1 -x_c)}{1 -x_b}
	   -\frac{ 8(x_b +x_c -1)}{ 2 -x_b -x_c}
	 \Bigr),\\
  b_2 (x_b, x_c) &= 
    2 \frac{ ( x_b +x_c) ( x_b^2 +x_c^2 -3( x_b + x_c) +6 ) -4}
	        {2 -x_b -x_c},
\end{align}
\end{subequations}
\begin{subequations}
\begin{align}
  c_1 (x_b, x_c) &= 
    -32 \frac{ ( x_b -x_c) ( x_b +x_c -1)}{ 2 -x_b -x_c}
	     \sqrt{ \frac{ x_b +x_c -1}{ ( 1 -x_b) ( 1 -x_c)}} ,\\
  c_2 (x_b, x_c) &= 
    -4 \frac{ x_b -x_c}{ 2 -x_b -x_c} 
	     \sqrt{ ( 1 -x_b) ( 1 -x_c) ( x_b +x_c -1)} ,\\
  c_3 (x_b, x_c) &= 
    -16 ( x_b -x_c) ( x_b +x_c -1) 
	     \sqrt{ \frac{ x_b +x_c -1}{ ( 1 -x_b) ( 1 -x_c)}} ,\\
  c_4 (x_b, x_c) &= 
    -8 ( x_b -x_c) ( 2 -x_b -x_c) 
	     \sqrt{ \frac{ x_b +x_c -1}{ ( 1 -x_b) ( 1 -x_c)}},
\end{align}
\end{subequations}
where $x_b$ and $x_c$ are defined in Eq.~\eqref{EqDefx}.
The branching ratio and angular asymmetries defined as 
Eqs.~\eqref{eq:AZ-tmmm}--\eqref{eq:AY-tmmm} are
\begin{align}
  \text{Br}( \tau^+ \to \mu^+ \mu^+ \mu^- )(\delta) =&
  \text{Br}( \tau^+ \to \bar{\nu}_{\tau} e^+ \nu_{e}) B^\text{I}(\delta) ,
\end{align}
\begin{align}
  B^\text{I} (\delta) =& 
      ( C^\text{I}_{R1} + C^\text{I}_{L1} ) A_1(\delta)
	 + ( C^\text{I}_{R2} + C^\text{I}_{L2} ) A_2(\delta) 
	 + ( C^\text{I}_{R3} + C^\text{I}_{L3} ) A_3(\delta) \n 
	&+ ( C^\text{I}_{J1} + C^\text{I}_{J2} ) A_4(\delta) 
	 + ( C^\text{I}_{J3} + C^\text{I}_{J4} ) A_5(\delta),
\end{align}
\begin{subequations}
\begin{align}
	A_Z^\text{I}(\delta) = \frac{1}{2 B^\text{I} (\delta)} &
	\bigl(
	  ( C^\text{I}_{R1} - C^\text{I}_{L1} ) B_1 (\delta) 
	- ( C^\text{I}_{R2} - C^\text{I}_{L2} ) B_2 (\delta) \n & 
	+ ( C^\text{I}_{R3} - C^\text{I}_{L3} ) A_3 (\delta) 
	- ( C^\text{I}_{J1} - C^\text{I}_{J2} ) A_4 (\delta) \n & 
	+ ( C^\text{I}_{J3} - C^\text{I}_{J4} ) A_5 (\delta) 
	\bigr),\\
	A_X^\text{I}(\delta) = \frac{1}{2 B^\text{I} (\delta)} &
	\bigl(
	- ( C^\text{I}_{R1} - C^\text{I}_{L1} ) C_1 (\delta) 
	- ( C^\text{I}_{R2} - C^\text{I}_{L2} ) C_2 (\delta) \n &
	- ( C^\text{I}_{J1} - C^\text{I}_{J2} ) C_3 (\delta) 
	+ ( C^\text{I}_{J3} - C^\text{I}_{J4} ) C_4 (\delta)
	\bigr),\\
	A_Y^\text{I}(\delta) = \frac{1}{2 B^\text{I} (\delta)}  &
	\bigl(
	-C^\text{I}_{J5} C_3 (\delta) + C^\text{I}_{J6} C_4 (\delta) 
	\bigr).
\end{align}
\end{subequations}
$\text{Br}$, $A_Z^{\text{I}}$, $A_X^{\text{I}}$ and 
$A_Y^{\text{I}}$ extract the components $M_O^{\text{I}}$, $M_Z^{\text{I}}$,
$M_X^{\text{I}}$ and $M_Y^{\text{I}}$ 
in Eq.~\eqref{eq:dGamma-tmmm}, respectively.
Since the signs of $\cos\theta$, $\sin\theta\cos\phi$ and 
$\sin\theta\sin\phi$ are equal to $\vec{s}\cdot \vec{p}_a$,
$\vec{s}\cdot ((\vec{p}_a \times \vec{p}_b)\times \vec{p}_a)$ and
$\vec{s}\cdot (\vec{p}_a \times \vec{p}_b)$, respectively 
(see Fig.~\ref{FigPl3l}),
$A_Z^{\text{I}}$ and $A_X^{\text{I}}$ are parity odd asymmetries 
and $A_Y^{\text{I}}$ is
a time-reversal asymmetry.
The following functions are introduced in the above formulae.
The cutoff parameter, $\delta$, is defined 
in Sec.\ref{SecLhtlfv} for each of the processes.
\begin{subequations}
\begin{align}
  A_{1}(\delta) &= 
    -16 ( 1 - \delta )( 2 - \delta + 2\delta^2 )
      \ln \left( \frac{\delta}{1 - \delta} \right)
  - \frac{8}{3} ( 1 - 2\delta )( 13 - 4\delta + 4\delta^2 ), \\
  A_{2}(\delta) &= 
    ( 1 + 2\delta - 2\delta^2 )( 1 - 2\delta )^2, \\
  A_{3}(\delta) &= 
    2 ( 1 + 2\delta )( 1 - 2\delta )^3, \\
  A_{4}(\delta) &= 
    16 ( 1 - 2\delta )^3, \\
  A_{5}(\delta) &= 
    8 ( 1 + \delta )( 1 - 2\delta )^2,
\end{align}
\end{subequations}
\begin{subequations}
\begin{align}
  B_1 (\delta) =& 
    - 16 ( 2 + 21\delta + 3\delta^2 - 2\delta^3 ) \ln (2\delta)
    + 16 ( 1 - \delta )( 2 - \delta + 2\delta^2 )\ln (2( 1 - \delta ))\n
   &- \frac{8}{3} ( 1 - 2\delta )( 49 + 68\delta + 4\delta^2 ) ,\\
  B_2 (\delta) =& 
    \frac{1}{3} ( 1 -2\delta) ( 1 +8\delta -38\delta^2 -12\delta^3)
	 -16\delta^3 \ln (2\delta).
\end{align}
\end{subequations}
\begin{subequations}
\begin{align}
  C_{1}(\delta) =& 
      \frac{96}{5} ( 4 + 9\delta + \delta^{2} ) \sqrt{1 - 2\delta}
    - 48\sqrt{\delta} ( 3 + 6\delta - \delta^{2} )
      \arccos\left( \frac{3\delta - 1}{1 - \delta} \right) \n &
    + 384\delta \arccos\left( \frac{\delta}{1 - \delta} \right), \\
  C_{2}(\delta) =& 
      \frac{4}{105} ( 8 + 8\delta - 93\delta^{2} - 225\delta^{3} ) 
	   \sqrt{ 1 - 2\delta } \n &
    - 2 \delta^{ \frac{3}{2} } ( 1 - 6\delta - 3\delta^{2} )
        \arccos\left( \frac{3\delta - 1}{1 - \delta} \right)
    - 16 \delta^{3} \arccos\left( \frac{\delta}{1 - \delta} \right), \\
  C_{3}(\delta) =& 
      \frac{8}{35}\sqrt{1 - 2\delta}
      ( 48 - 57\delta - 68\delta^{2} + 85\delta^{3} ) 
    - 12 \sqrt{\delta}( 1 - \delta )^{3}
      \arccos\left( \frac{3\delta - 1}{1 - \delta} \right), \\
  C_{4}(\delta) =& 
    \frac{4}{35} \sqrt{1 - 2\delta}
      ( 64 - 41\delta + 26\delta^{2} - 85\delta^{3} ) \n &
    - 6 \sqrt{\delta}( 1 - \delta - \delta^{2} + \delta^{3} )
      \arccos\left( \frac{3\delta - 1}{1 - \delta} \right).
\end{align}
\end{subequations}
\subsection{$\tau^+ \to \mu^+ e^+ e^-$ and $\tau^+ \to e^+ \mu^+ \mu^-$} 
\label{AppFormulaeT2}
The Lagrangian of type II leptonic decay is
\begin{align}
	\mathcal{L}_{\text{II}} =
	-\frac{4G_F}{\sqrt{2}}& \bigl[
	 	m_{\tau} A_R  \bar{\tau}_{R} \sigma^{\mu \nu} \mu_{L} F_{\mu \nu}
	+  m_{\tau} A_L  \bar{\tau}_{L} \sigma^{\mu \nu} \mu_{R} F_{\mu \nu} \n &
	+  g_{Rs}^{II} ( \bar{\tau}_{R } \mu_L )
	               ( \bar{e}_{R} e_{L} )
	+  g_{Ls}^{II} ( \bar{\tau}_{ L} \mu_{R} )
	               ( \bar{e}_{L} e_{R} ) \n &
	+  g_{Rt}^{II} ( \bar{\tau}_{ R} e_{L} )
	               ( \bar{e}_{R} \mu_{L} )
	+  g_{Lt}^{II} ( \bar{\tau}_{ L} e_{R} )
	               ( \bar{e}_{L} \mu_{R} ) \n &
	+  g_{Rr}^{II} ( \bar{\tau}_{ R} \gamma^{\mu} \mu_{R} )
				      ( \bar{e}_{R} \gamma_{\mu} e_{R} )
	+  g_{Ll}^{II} ( \bar{\tau}_{ L} \gamma^{\mu} \mu_{L} )
				      ( \bar{e}_{L} \gamma_{\mu} e_{L} ) \n &
	+  g_{Rl}^{II} ( \bar{\tau}_{ R} \gamma^{\mu} \mu_{R} )
				      ( \bar{e}_{L} \gamma_{\mu} e_{L} )
	+  g_{Lr}^{II} ( \bar{\tau}_{ L} \gamma^{\mu} \mu_{L} )
				      ( \bar{e}_{R} \gamma_{\mu} e_{R} ) \n &
	+  g_{Rx}^{II} ( \bar{\tau}_{ R} \gamma^{\mu} e_{R} ) 
				      ( \bar{e}_{L} \gamma_{\mu} \mu_{L} )
	+  g_{Lx}^{II} ( \bar{\tau}_{ L} \gamma^{\mu} e_{L} )
				      ( \bar{e}_{R} \gamma_{\mu} \mu_{R} )
	+ \text{H.c.}
	\bigr].
\end{align}
The differential partial decay width of $\tau^+\to\mu^+e^+e^-$ can be given as
\begin{align}
  \frac{d^4\text{Br}(\tau^+\to\mu^+e^+e^-)}
       {dx_b dx_c \,d\phi \,d\!\cos \theta} 
  =\tau_\tau \frac{G_F^2 m_{\tau}^5 }{128\pi^4}& (
    M_O^{\text{II}} (x_b,x_c) 
  + M_Z^{\text{II}} (x_b,x_c) \cos\theta \n &
  + M_X^{\text{II}} (x_b,x_c) \sin\theta \cos\phi
  + M_Y^{\text{II}} (x_b,x_c) \sin\theta \sin\phi ),
\end{align}
where
\begin{subequations}
\begin{align}
  M_O^{\text{II}} (x_b,x_c) = & 
    ( C_{R1}^{\text{II}} +C_{L1}^{\text{II}}) d_1 (x_b, x_c)
   +( C_{R2}^{\text{II}} +C_{L2}^{\text{II}}) d_2 (x_b, x_c) \n & 
   +( C_{R3}^{\text{II}} +C_{L3}^{\text{II}}) d_3 (x_b, x_c)
   +( C_{R4}^{\text{II}} +C_{L4}^{\text{II}}) d_4 (x_b, x_c) \n & 
   +( C_{J1}^{\text{II}} +C_{J2}^{\text{II}}) d_5 (x_b, x_c)
   +( C_{J3}^{\text{II}} +C_{J4}^{\text{II}}) d_6 (x_b, x_c) \n & 
   +( C_{J5}^{\text{II}} +C_{J6}^{\text{II}}) d_7 (x_b, x_c) , \\
  M_Z^{\text{II}} (x_b,x_c) = &
    ( C_{R1}^{\text{II}} -C_{L1}^{\text{II}}) e_1 (x_b, x_c)
   +( C_{R2}^{\text{II}} -C_{L2}^{\text{II}}) d_2 (x_b, x_c) \n & 
   +( C_{R3}^{\text{II}} -C_{L3}^{\text{II}}) e_2 (x_b, x_c) 
   +( C_{R4}^{\text{II}} -C_{L4}^{\text{II}}) e_3 (x_b, x_c) \n &
   +( C_{J1}^{\text{II}} -C_{J2}^{\text{II}}) e_4 (x_b, x_c)
   +( C_{J3}^{\text{II}} -C_{J4}^{\text{II}}) e_5 (x_b, x_c) \n & 
   +( C_{J5}^{\text{II}} -C_{J6}^{\text{II}}) d_7 (x_b, x_c) , \\
  M_X^{\text{II}} (x_b,x_c) = &
    ( C_{R1}^{\text{II}} -C_{L1}^{\text{II}}) f_1 (x_b, x_c)
   +( C_{R3}^{\text{II}} -C_{L3}^{\text{II}}) f_2 (x_b, x_c) \n & 
   +( C_{R4}^{\text{II}} -C_{L4}^{\text{II}}) f_3 (x_b, x_c) \n &
   +( C_{J1}^{\text{II}} -C_{J2}^{\text{II}}) f_4 (x_b, x_c)
   +( C_{J3}^{\text{II}} -C_{J4}^{\text{II}}) f_5 (x_b, x_c) \n & 
   +( C_{J5}^{\text{II}} -C_{J6}^{\text{II}}) f_6 (x_b, x_c) , \\
  M_Y^{\text{II}} (x_b,x_c) = &
    C_{J7}^{\text{II}} g_1 (x_b, x_c)
	+C_{J8}^{\text{II}} g_2 (x_b, x_c)
   +C_{J9}^{\text{II}} f_6 (x_b, x_c).
\end{align}
\end{subequations}
$C^{\text{II}}_{R1,\cdots}$ are defined as
\begin{subequations}
\begin{align}
 C_{R1}^{\text{II}} &= \abs{eA_R}^2 , &&
 C_{L1}^{\text{II}}  = \abs{eA_L}^2 , \\
 C_{R2}^{\text{II}} &= \frac{ \abs{g_{Rs}}^2 }{4} +\abs{g_{Rx}}^2 ,&&
 C_{L2}^{\text{II}}  = \frac{ \abs{g_{Ls}}^2 }{4} +\abs{g_{Lx}}^2 , \\
 C_{R3}^{\text{II}} &= \frac{ \abs{g_{Rt}}^2 }{4} +\abs{g_{Rl}}^2 ,&&
 C_{L3}^{\text{II}}  = \frac{ \abs{g_{Lt}}^2 }{4} +\abs{g_{Lr}}^2 , \\
 C_{R4}^{\text{II}} &= \abs{g_{Rr}}^2 ,&&
 C_{L4}^{\text{II}}  = \abs{g_{Ll}}^2 ,\\
 C_{J1}^{\text{II}} &= \text{Re}[ eA_R g_{Ll}^{\ast} ] ,&&
 C_{J2}^{\text{II}}  = \text{Re}[ eA_L g_{Rr}^{\ast} ] ,\\
 C_{J3}^{\text{II}} &= \text{Re}[ eA_R g_{Lr}^{\ast} ] ,&&
 C_{J4}^{\text{II}}  = \text{Re}[ eA_L g_{Rl}^{\ast} ] ,\\
 C_{J5}^{\text{II}} &= \text{Re}[ g_{Rs} g_{Rt}^{\ast} ] ,&&
 C_{J6}^{\text{II}}  = \text{Re}[ g_{Ls} g_{Lt}^{\ast} ] ,\\
 C_{J7}^{\text{II}} &= 
   \text{Im}[ eA_R g_{Ll}^{\ast} +  eA_L g_{Rr}^{\ast} ] ,&&
 C_{J8}^{\text{II}}  = 
   \text{Im}[ eA_R g_{Lr}^{\ast} + eA_L g_{Rl}^{\ast} ] ,\\
 C_{J9}^{\text{II}} &= 
   \text{Im}[ g_{Rs} g_{Rt}^{\ast} + g_{Ls} g_{Lt}^{\ast} ].
\end{align}
\end{subequations}
The functions, $d_i(x_b,x_c)$, $e_i(x_b,x_c)$, $f_i(x_b,x_c)$ and 
$g_i(x_b,x_c)$,
are defined as
\begin{subequations}
\begin{align}
  d_1 (x_b,x_c) =& 8 \frac{ x_b( 1-x_c) +x_c( 1-x_b)}{x_b+x_c-1},\\
  d_2 (x_b,x_c) =& 2 ( 2-x_b-x_c)( x_b+x_c-1),\\
  d_3 (x_b,x_c) =& 2 x_b( 1-x_b),\\
  d_4 (x_b,x_c) =& 2 x_c( 1-x_c),\\
  d_5 (x_b,x_c) =& 8 ( 1-x_c),\\
  d_6 (x_b,x_c) =& 8 ( 1-x_b),\\
  d_7 (x_b,x_c) =& - ( 1-x_b)( x_b+x_c-1),
\end{align}
\end{subequations}
\begin{subequations}
\begin{align}
  e_1 (x_b,x_c) =& 
    \frac{8}{x_b +x_c -1}\biggl(
	  x_b (1-x_c) +x_c (1-x_b) -2 \frac{(1-x_b)^2 +(1 -x_c)^2}{2 -x_b -x_c}
	 \biggr),\\
  e_2 (x_b,x_c) =& 
   2\Bigl(
	   x_b(1-x_b)-\frac{2(1-x_b)(1-x_c)}{2-x_b-x_c}
	 \Bigr),\\
  e_3 (x_b,x_c) =& 
   2\Bigl(
	   x_c(1-x_c)-\frac{2(1-x_b)(1-x_c)}{2-x_b-x_c}
	 \Bigr),\\
  e_4 (x_b,x_c) =& -8\frac{(1-x_c)(x_b-x_c)}{2-x_b-x_c},\\
  e_5 (x_b,x_c) =& 8\frac{(1-x_b)(x_b-x_c)}{2-x_b-x_c},
\end{align}
\end{subequations}
\begin{subequations}
\begin{align}
  f_1 (x_b,x_c) =& 
    16 \frac{x_b-x_c}{2-x_b-x_c} \sqrt{ \frac{(1-x_b)(1-x_c)}{x_b+x_c-1} },\\
  f_2 (x_b,x_c) =& 
    4 \frac{1-x_b}{2-x_b-x_c} \sqrt{ (1-x_b)(1-x_c)(x_b+x_c-1) },\\
  f_3 (x_b,x_c) =& 
    -4 \frac{1-x_c}{2-x_b-x_c} \sqrt{ (1-x_b)(1-x_c)(x_b+x_c-1) },\\
  f_4 (x_b,x_c) =& 
    8 \frac{(1-x_c)(x_b+x_c)}{2-x_b-x_c} 
	  \sqrt{ \frac{(1-x_b)(1-x_c)}{x_b+x_c-1} },\\
  f_5 (x_b,x_c) =& 
   -8 \frac{(1-x_b)(x_b+x_c)}{2-x_b-x_c}
	  \sqrt{ \frac{(1-x_b)(1-x_c)}{x_b+x_c-1} },\\
  f_6 (x_b,x_c) =& 
    - \sqrt{ (1-x_b)(1-x_c)(x_b+x_c-1) },
\end{align}
\end{subequations}
\begin{subequations}
\begin{align}
  g_1 (x_b,x_c) =& -24  (1-x_c) \sqrt{ \frac{(1-x_b)(1-x_c)}{(x_b+x_c-1)}},\\
  g_2 (x_b,x_c) =&  24  (1-x_b) \sqrt{ \frac{(1-x_b)(1-x_c)}{(x_b+x_c-1)}}.
\end{align}
\end{subequations}
The branching ratio and seven asymmetries defined 
in Eqs.~\eqref{eq:AZ-tmmm}--\eqref{eq:AY-tmmm} 
and \eqref{EqDeffbasym}--\eqref{EqDeffbasymY} are
\begin{align}
  \text{Br}(\tau^+ \to \mu^+ e^+ e^-)(\delta) =& 
   \text{Br}( \tau^+ \to \bar{\nu}_{\tau} e^+ \nu_{e} ) 
	B^{\text{II}}(\delta) ,
\end{align}
\begin{align}
  B^{\text{II}} (\delta)= &
	 ( C_{R1}^{\text{II}} +C_{L1}^{\text{II}}) D_1(\delta)
	+\biggl( 
	   C_{R2}^{\text{II}} +C_{L2}^{\text{II}} 
	  -\frac{1}{4}( C_{J5}^{\text{II}} +C_{J6}^{\text{II}})
	  \biggr) D_2(\delta) \n &
	+( C_{R3}^{\text{II}} +C_{L3}^{\text{II}} 
	  +C_{R4}^{\text{II}} +C_{L4}^{\text{II}}) D_3(\delta)
	+( C_{J1}^{\text{II}} +C_{J2}^{\text{II}} 
	  +C_{J3}^{\text{II}} +C_{J4}^{\text{II}}) D_4(\delta),
\end{align}
\begin{subequations}
\begin{align}
  A_Z^{\text{II}} (\delta)= \frac{1}{2 B^{\text{II}} (\delta)} &
  \biggl(
   -( C_{R1}^{\text{II}} -C_{L1}^{\text{II}}) D_5(\delta) \n &
   +\Bigl( C_{R2}^{\text{II}} -C_{L2}^{\text{II}} 
	  -\frac{1}{4}( C_{J5}^{\text{II}} -C_{J6}^{\text{II}})
	 \Bigr) D_2(\delta) \n &
	-( C_{R3}^{\text{II}} -C_{L3}^{\text{II}} 
	  +C_{R4}^{\text{II}} -C_{L4}^{\text{II}}) D_6(\delta) \n &
	-\frac{1}{3} ( C_{J1}^{\text{II}} -C_{J2}^{\text{II}} 
	              +C_{J3}^{\text{II}} -C_{J4}^{\text{II}}) D_4(\delta)
  \biggr), \\
  A_X^{\text{II}}(\delta)= \frac{\pi}{2 B^{\text{II}}(\delta)} &
  \biggl(
    \Bigl( C_{R3}^{\text{II}} -C_{L3}^{\text{II}} 
	  -C_{R4}^{\text{II}} +C_{L4}^{\text{II}} 
	  -\frac{1}{2}( C_{J5}^{\text{II}} - C_{J6}^{\text{II}})
	 \Bigr) E_1(\delta) \n &
   +( C_{J1}^{\text{II}} -C_{J2}^{\text{II}} 
	  -C_{J3}^{\text{II}} +C_{J4}^{\text{II}}) E_2(\delta)
  \biggr), \\
  A_Y^{\text{II}}(\delta)= \frac{\pi}{2 B^{\text{II}}(\delta)} &
  \biggl(
   -( C_{J7}^{\text{II}} -C_{J8}^{\text{II}}) E_3(\delta)
   -\frac{1}{2} C_{J9}^{\text{II}} E_1(\delta) 
  \biggr),
\end{align}
\end{subequations}
\begin{subequations}
\begin{align}
  A_{FB}^{\text{II}}(\delta)= \frac{1}{ B^{\text{II}}(\delta)} &
  \biggl(
	-\frac{1}{4}
	 \Bigl( C_{R3}^{\text{II}} +C_{L3}^{\text{II}} 
	   -C_{R4}^{\text{II}} -C_{L4}^{\text{II}} 
	   -\frac{1}{2} ( C_{J5}^{\text{II}} +C_{J6}^{\text{II}} )
	 \Bigr) D_2(\delta) \n &
	+\frac{1}{2}( C_{J1}^{\text{II}} +C_{J2}^{\text{II}} 
	             -C_{J3}^{\text{II}} -C_{J4}^{\text{II}}) D_4(\delta)
  \biggr), \\
  A_{ZFB}^{\text{II}}(\delta)= \frac{1}{2 B^{\text{II}}(\delta)} &
  \biggl(
   -\frac{1}{4}
	  \Bigl( C_{R3}^{\text{II}} -C_{L3}^{\text{II}} 
	   -C_{R4}^{\text{II}} +C_{L4}^{\text{II}}
	   -\frac{1}{2}( C_{J5}^{\text{II}} -C_{J6}^{\text{II}})
	  \Bigr) D_2(\delta) \n &
	-\frac{1}{2}( C_{J1}^{\text{II}} -C_{J2}^{\text{II}} 
	             -C_{J3}^{\text{II}} +C_{J4}^{\text{II}}) D_4(\delta)
  \biggr), \\
  A_{XFB}^{\text{II}}(\delta)= \frac{1}{2 B^{\text{II}}(\delta)} &
  \biggl(
    ( C_{R1}^{\text{II}} - C_{L1}^{\text{II}}) E_4(\delta) \n &
   -\frac{4}{3}( C_{R3}^{\text{II}} -C_{L3}^{\text{II}} 
	             +C_{R4}^{\text{II}} -C_{L4}^{\text{II}}) E_1(\delta) \n &
   +\frac{4}{3}( C_{J1}^{\text{II}} -C_{J2}^{\text{II}} 
	             +C_{J3}^{\text{II}} -C_{J4}^{\text{II}}) E_2(\delta)
  \biggr),
\label{eq:AXFB-II}
\\
  A_{YFB}^{\text{II}}(\delta)= \frac{1}{2 B^{\text{II}}(\delta)} &
  \biggl(
   -\frac{4}{3}( C_{J7}^{\text{II}} + C_{J8}^{\text{II}}) E_3(\delta)
  \biggr),
\end{align}
\end{subequations}
with
\begin{subequations}
\begin{align}
  D_1 (\delta)&= 
    \frac{16}{3} ( -( 1 -\delta) ( 8 -\delta -\delta^2) -6\ln \delta ),\\
  D_2 (\delta)&= ( 1 -\delta)^3 (1 +3\delta),\\
  D_3 (\delta)&= ( 1 -\delta)^3 (1 +\delta),\\
  D_4 (\delta)&= 8( 1 -\delta)^3,\\
  D_5 (\delta)&= 
	 \frac{16}{3} ( -( 1 -\delta) ( 10 -5\delta +\delta^2) -6\ln \delta),
  \label{EqD5}\\
  D_6 (\delta)&= \frac{1}{3}( 1 -\delta)^3 (1 -3\delta),
\end{align}
\end{subequations}
\begin{subequations}
\begin{align}
  E_1 (\delta)&= 
    \frac{1}{35} ( 1 -\delta^{1/2})^3 
	  ( 8 +24\delta^{1/2} +48\delta +45\delta^{3/2} +15\delta^2) ,\\
  E_2 (\delta)&= 
    \frac{2}{35} ( 1 -\delta^{1/2})^3 
	   ( 64 +87\delta^{1/2} +69\delta +45\delta^{3/2} +15\delta^2) ,\\
  E_3 (\delta)&= 
    \frac{6}{35} ( 1 -\delta^{1/2})^4 
	   ( 16 +29\delta^{1/2} +20\delta +5\delta^{3/2}),\\
  E_4 (\delta)&= 
    \frac{32}{15} ( 1 -\delta^{1/2})^3
	   ( 8 +9\delta^{1/2} +3\delta).
\end{align}
\end{subequations}

\subsection{$\tau^+ \to \mu^+ \mu^+ e^-$ and $\tau^+ \to e^+ e^+ \mu^-$} 
\label{AppFormulaeT3}
The effective Lagrangian of type III leptonic decay is
\begin{align}
	\mathcal{L}_{\text{III}} =
	-\frac{4G_F}{\sqrt{2}}& \bigl[
	   g_{Rs}^{III} ( \bar{\tau}_R \mu_{L} )
		             ( \bar{e   }_R \mu_{L} )
	+  g_{Ls}^{III} ( \bar{\tau}_{L} \mu_{R} )
	                ( \bar{e   }_{L} \mu_{R} ) \n &
	+  g_{Rr}^{III} ( \bar{\tau}_{R} \gamma^{\mu} \mu_{R} )
				       ( \bar{e   }_{R} \gamma_{\mu} \mu_{R} )
	+  g_{Ll}^{III} ( \bar{\tau}_{L} \gamma^{\mu} \mu_{L} )
				       ( \bar{e   }_{L} \gamma_{\mu} \mu_{L} ) \n &
	+  g_{Rl}^{III} ( \bar{\tau}_{R} \gamma^{\mu} \mu_{R} )
				       ( \bar{e   }_{L} \gamma_{\mu} \mu_{L} )
	+  g_{Lr}^{III} ( \bar{\tau}_{L} \gamma^{\mu} \mu_{L} )
				       ( \bar{e   }_{R} \gamma_{\mu} \mu_{R} )
	+ \text{H.c.}
	\bigr],
\end{align}
The differential partial decay width of $\tau^+\to\mu^+\mu^+e^-$ is
written as
\begin{align}
  \frac{d^4\text{Br}(\tau^+\to\mu^+\mu^+e^-)}
       {dx_b dx_c \,d\phi \,d\!\cos \theta} 
  =\tau_\tau \frac{G_F^2 m_{\tau}^5 }{128\pi^4}& (
    M_O^{\text{III}} (x_b,x_c) 
  + M_Z^{\text{III}} (x_b,x_c) \cos\theta \n &
  + M_X^{\text{III}} (x_b,x_c) \sin\theta \cos\phi
  + M_Y^{\text{III}} (x_b,x_c) \sin\theta \sin\phi ),
\end{align}
where
\begin{subequations}
\begin{align}
  M_O^{\text{III}} (x_b,x_c) = &
    ( C_{R2}^{\text{III}} +C_{L2}^{\text{III}}) a_2 (x_b, x_c)
   +( C_{R3}^{\text{III}} +C_{L3}^{\text{III}}) a_3 (x_b, x_c),\\
  M_Z^{\text{III}} (x_b,x_c) = &
    ( C_{R2}^{\text{III}} -C_{L2}^{\text{III}}) b_2 (x_b, x_c)
   +( C_{R3}^{\text{III}} -C_{L3}^{\text{III}}) a_3 (x_b, x_c),\\
  M_X^{\text{III}} (x_b,x_c) = &
    ( C_{R2}^{\text{III}} -C_{L2}^{\text{III}}) c_2 (x_b, x_c),\\
  M_Y^{\text{III}} (x_b,x_c) = & 0. 
\end{align}
\end{subequations}
Integrating them in $x_b$-$x_c$ plane, 
branching ratio and parity asymmetries are given as
\begin{align}
  \text{Br}( \tau \to \mu^+ \mu^+ e^- ) (\delta) = &
  \text{Br}( \tau^+ \to \bar{\nu}_{\tau} e^+ \nu_{e} ) 
  ( ( C^{\text{III}}_{R2} + C^{\text{III}}_{L2} ) A_2 (\delta) 
  + ( C^{\text{III}}_{R3} + C^{\text{III}}_{L3} ) A_3 (\delta) ),
\end{align}
\begin{align}
	A_Z^{\text{III}} (\delta) =& \frac{1}{2}\frac{
	 -( C^{\text{III}}_{R2} - C^{\text{III}}_{L2} ) B_2 (\delta) 
	 +( C^{\text{III}}_{R3} - C^{\text{III}}_{L3} ) A_3 (\delta)
	}{
     ( C^{\text{III}}_{R2} + C^{\text{III}}_{L2} ) A_2 (\delta) 
   + ( C^{\text{III}}_{R3} + C^{\text{III}}_{L3} ) A_3 (\delta) 
	}, \\
	A_X^{\text{III}} (\delta) =& \frac{1}{2} \frac{
	  ( C^{\text{III}}_{R2} - C^{\text{III}}_{L2} ) C_2 (\delta)
	}{
     ( C^{\text{III}}_{R2} + C^{\text{III}}_{L2} ) A_2 (\delta) 
   + ( C^{\text{III}}_{R3} + C^{\text{III}}_{L3} ) A_3 (\delta) 
	} ,\\
	A_Y^{\text{III}} (\delta) =& 0,
\end{align}
where coefficients $C^{\text{III}}_i$s are given by replacement of I with III
in coefficients of the type I, 
Eqs.~\eqref{EqTypeIwilson0}--\eqref{EqTypeIwilson1}.
In this type, time-reversal asymmetry, $A_Y$, does not appear.

\subsection{Semileptonic decays of $\tau$} 
\label{AppFormulaeMeson}
The effective Lagrangian of semileptonic $\tau$ decays is
\begin{align}
	\mathcal{L}_{\text{had}} =
	-\frac{4G_F}{\sqrt{2}} \Bigl[
	&  m_\tau A_R  \bar{\tau}_{R} \sigma^{\mu \nu} \mu_{L} F_{\mu \nu}
	+  m_\tau A_L  \bar{\tau}_{L} \sigma^{\mu \nu} \mu_{R} F_{\mu \nu} \n
	+\sum_{q=u,d,s} \bigl( &
		g_{Rs(q)} ( \bar{\tau}_{R} \mu_{L} )( \bar{q}_{R} q_{L} )
	+  g_{Ls(q)} ( \bar{\tau}_{L} \mu_{R} )( \bar{q}_{L} q_{R} ) \n 
	+& g_{Rt(q)} ( \bar{\tau}_{R} \mu_{L} )( \bar{q}_{L} q_{R} )
	+  g_{Lt(q)} ( \bar{\tau}_{L} \mu_{R} )( \bar{q}_{R} q_{L} ) \n 
	+& g_{Rr(q)} ( \bar{\tau}_{R} \gamma^{\mu} \mu_{R} )
				 ( \bar{q}_{R} \gamma_{\mu} q_{R} )
	+  g_{Ll(q)} ( \bar{\tau}_{L} \gamma^{\mu} \mu_{L} )
				 ( \bar{q}_{L} \gamma_{\mu} q_{L} ) \n
	+& g_{Rl(q)} ( \bar{\tau}_{R} \gamma^{\mu} \mu_{R} )
				 ( \bar{q}_{L} \gamma_{\mu} q_{L} )
	+  g_{Lr(q)} ( \bar{\tau}_{L} \gamma^{\mu} \mu_{L} )
				 ( \bar{q}_{R} \gamma_{\mu} q_{R} ) \n
	+& g_{RT(q)} ( \bar{\tau}_{R} \sigma^{\mu \nu} \mu_{L} )
				 ( \bar{q}_{R} \sigma_{\mu \nu} q_{L} )
	+  g_{LT(q)} ( \bar{\tau}_{L} \sigma^{\mu \nu} \mu_{R} )
				 ( \bar{q}_{L} \sigma_{\mu \nu} q_{R} )
	+ \text{H.c.}
	\bigl) \Bigl].
\label{EqLaghad}
\end{align}

The branching ratios and asymmetries for $\tau^+ \to \mu^+ P$ decays are
\begin{align}
	\text{Br}(\tau^+ \to \mu^+ P) &= \tau_{\tau}
	  \frac{G_F^2 m_{\tau}^3 }{4 \pi}
	   \left( 1 - \frac{m_P^2}{m_{\tau}^2} \right)^2 
		( \abs{G_{RP}}^2 + \abs{G_{LP}}^2 ), \\
	A_P (\tau^+ \to \mu^+ P) &= \frac{1}{2}
		\frac{ \abs{G_{RP}}^2 - \abs{G_{LP}}^2 }
			  { \abs{G_{RP}}^2 + \abs{G_{LP}}^2 }.
\end{align}
Here, $G_{RP}$ and $G_{LP}$ for $P=\pi$ and $\eta$ are given by
\begin{subequations}
\begin{align}
	G_{R\pi} = &
	 \frac{f_{\pi}}{2\sqrt{2}} \Bigl(
	 g_{Lr(u)} -g_{Lr(d)} -g_{Ll(u)} +g_{Ll(d)} \n & \qquad 
	-\frac{m_{\pi}^2}{2m_q m_\tau} 
	 ( g_{Rt(u)} -g_{Rt(d)} -g_{Rs(u)} +g_{Rs(d)} )
	\Bigr),\\
	G_{L\pi} = & 
	 \frac{f_{\pi}}{2\sqrt{2}} \Bigl(
	 g_{Rr(u)} -g_{Rr(d)} -g_{Rl(u)} +g_{Rl(d)} \n & \qquad
	+\frac{m_{\pi}^2}{2m_q m_\tau} 
	 ( g_{Lt(u)} -g_{Lt(d)} -g_{Ls(u)} +g_{Ls(d)} )
	\Bigr),\\
	G_{R\eta} = & 
	  \frac{1}{2} \Bigl(
	  \frac{ f_{\eta}^q }{\sqrt{2}} 
	    ( g_{Lr(u)} + g_{Lr(d)} - g_{Ll(u)} - g_{Ll(d)} )
	+ f_{\eta}^s ( g_{Lr(s)} - g_{Ll(s)} ) \n &
	- \frac{ h_{\eta}^q m_{\eta}^2 }{ 2\sqrt{2} m_q m_\tau } 
		 ( g_{Rt(u)} -g_{Rt(d)} -g_{Rs(u)} +g_{Rs}(d) )
	+ \frac{ h_{\eta}^s m_{\eta}^2 }{ 2 m_q m_\tau } ( g_{Rt(s)} - g_{Rs(s)} )
	\Bigr),\\
	G_{L\eta} = & 
	  \frac{1}{2} \Bigl(
	  \frac{ f_{\eta}^q }{\sqrt{2}} 
	    ( g_{Rr(u)} + g_{Rr(d)} - g_{Rl(u)} - g_{Rl(d)} )
	+ f_{\eta}^s ( g_{Rr(s)} - g_{Rl(s)} ) \n &
	- \frac{ h_{\eta}^q m_{\eta}^2 }{ 2\sqrt{2} m_q m_\tau } 
		 ( g_{Lt(u)} -g_{Lt(d)} -g_{Ls(u)} +g_{Ls(d)} )
	+ \frac{ h_{\eta}^s m_{\eta}^2 }{ 2 m_q m_\tau } ( g_{Lt(s)} - g_{Ls(s)} )
	\Bigr).
\end{align}
\end{subequations}
The expressions for $\eta'$ are obtained by the replacement $\eta \to \eta'$
in the last two equations. 
To describe the decay constants for $\eta$ and $\eta'$,
we used the formalism in Ref.~\cite{Feldmann:1998vh}.
We assume isospin symmetry so that $m_u = m_d = m_q$ in the following.
The decay constants are
\begin{subequations}
\begin{align}
 -if_{\pi} p^{\mu} &= 
  \bra{0} 
	\frac{ \bar{u} \gamma^{\mu} \gamma^5 u 
	      -\bar{d} \gamma^{\mu} \gamma^5 d }{\sqrt{2}}
  \ket{\pi(p)}
	, \\
 -if_q p^{\mu} &= 
  \bra{0} 
	\frac{ \bar{u} \gamma^{\mu} \gamma^5 u 
	      +\bar{d} \gamma^{\mu} \gamma^5 d }{\sqrt{2}}
  \ket{\eta_q (p)}
	, \\
 -if_s p^{\mu} &=
  \bra{0} 
	 \bar{s} \gamma^{\mu} \gamma^5 s
  \ket{\eta_s (p)} 
   ,\\
 \frac{ih_q}{2m_q} &= 
  \bra{0} 
	\frac{ \bar{u} \gamma^5 u 
	      +\bar{d} \gamma^5 d }{\sqrt{2}}
  \ket{\eta_q (p)}
	, \\
 \frac{ih_s}{2m_s} &=
  \bra{0} 
	 \bar{s} \gamma^5 s
  \ket{\eta_s (p)} 
   ,
\end{align}
\end{subequations}
For $\eta$ and $\eta'$, the decay constants,
$f_{\eta^{(')}}{^{q,s}}$ are $h_{\eta^{(')}}{^{q,s}}$ are expressed by 
$f_q$, $f_s$ and a mixing angle, $\phi_\eta$, as follows.
\begin{align}
	\begin{pmatrix}
		f_{\eta}^q  & f_{\eta}^s \\
		f_{\eta'}^q & f_{\eta'}^s
	\end{pmatrix}
	= 
	\begin{pmatrix}
		\cos \phi_\eta & -\sin \phi_\eta \\
		\sin \phi_\eta &  \cos \phi_\eta
	\end{pmatrix}
	\begin{pmatrix}
		f_q & 0 \\ 0 & f_s 
	\end{pmatrix},
\end{align}
\begin{align}
	\begin{pmatrix}
		h_{\eta}^q  & h_{\eta}^s \\
		h_{\eta'}^q & h_{\eta'}^s
	\end{pmatrix}
	= 
	\begin{pmatrix}
		\cos \phi_\eta & -\sin \phi_\eta \\
		\sin \phi_\eta &  \cos \phi_\eta
	\end{pmatrix}
	\begin{pmatrix}
		h_q & 0 \\ 0 & h_s 
	\end{pmatrix},
\end{align}
where 
\begin{subequations}
\begin{align}
	h_q &= f_q ( m_{\eta}^2 \cos^2 \phi_\eta + m_{\eta'}^2 \sin^2 \phi_\eta) 
	  -\sqrt{2} f_s( m_{\eta'}^2 - m_{\eta}^2) \cos\phi_\eta \, \sin\phi_\eta \\
	h_s &=  f_s ( m_{\eta'}^2 \cos^2 \phi_\eta + m_{\eta}^2 \sin^2 \phi_\eta ) 
		  - \frac{1}{\sqrt{2}} 
		  		f_s ( m_{\eta'}^2 - m_{\eta}^2 ) \cos \phi_\eta \, \sin \phi_\eta .
\end{align}
\end{subequations}

Similarly, for vector mesons, branching ratios and asymmetries are 
\begin{align}
	\text{Br}(\tau^+ \to \mu^+ V) =& \tau_{\tau}
		 \frac{G_F^2 m_\tau^3}{\pi} 
		 \left( 1 - \frac{m_V^2}{m_{\tau}^2} \right)^2 \n & \times
		 \biggl( 
			C_{AV+} \frac{2 m_{\tau}^2 +m_V^2}{m_{\tau}^2} 
		 + C_{V+} \frac{m_{\tau}^2 +2 m_V^2}{4 m_V^2} 
		 - 3 \text{Re}[ C_{IV+} ] 
		 \biggr), \\
	A_V (\tau^+ \to \mu^+ V) =& \frac{1}{2}
		\frac{
		  C_{AV-} ( 2 m_{\tau}^2 -m_V^2 ) / m_{\tau}^2
		+ C_{V-} ( m_{\tau}^2 -2 m_V^2 ) / (4 m_V^2 )
		+ \text{Re}[ C_{IV-} ]
		}{
		  C_{AV+} ( 2 m_{\tau}^2 +m_V^2 ) / m_{\tau}^2 
		+ C_{V+} ( m_{\tau}^2 +2 m_V^2 ) / (4 m_V^2 )
		- 3 \text{Re}[ C_{IV+} ] 
		},			
\end{align}
with
\begin{subequations}
\begin{align}
	C_{AV+} &= \abs{G_{LAV}}^2 + \abs{G_{RAV}}^2 , &
	C_{V+}  &= \abs{G_{LV}}^2  + \abs{G_{RV}}^2  ,\\
	C_{AV-} &= \abs{G_{LAV}}^2 - \abs{G_{RAV}}^2 , &
	C_{V-}  &= \abs{G_{LV}}^2  - \abs{G_{RV}}^2  , \\
	C_{IV+} &= G_{RAV} G_{LV}^{\ast} + G_{LAV} G_{RV}^{\ast} , &
	C_{IV-} &= G_{RAV} G_{LV}^{\ast} - G_{LAV} G_{RV}^{\ast} .
\end{align}
\end{subequations}
The effective couplings $G_{RAV}$, $G_{RV}$, $G_{LAV}$ and $G_{LV}$ 
for $V =\rho^0$, $\omega$ and $\phi$ are
\begin{subequations}
\begin{align}
	G_{RA\rho} &= -\frac{f_{\rho}^T }{2}
	               \frac{ g_{RT(u)} -g_{RT(d)} }{\sqrt{2}} 
					  +f_{\rho} \frac{m_\tau}{m_{\rho}} 
					   \frac{Q_u -Q_d}{\sqrt{2}} eA_R ,\\
	G_{LA\rho} &= -\frac{f_{\rho}^T}{2} 
	               \frac{ g_{LT(u)} -g_{LT(d)} }{\sqrt{2}} 
					  +f_{\rho} \frac{m_\tau}{m_{\rho}} 
					   \frac{Q_u -Q_d}{\sqrt{2}} eA_L ,\\
   G_{R\rho}  &= \frac{f_{\rho} m_{\rho}}{2\sqrt{2} m_\tau } 
						( g_{Rr(u)} -g_{Rr(d)} +g_{Rl(u)} -g_{Rl(d)} ) ,\\
   G_{L\rho}  &= \frac{f_{\rho} m_{\rho}}{2\sqrt{2} m_\tau } 
						( g_{Ll(u)} -g_{Ll(d)} +g_{Lr(u)} -g_{Lr(d)} ) ,\\
	G_{RA\omega} &= -\frac{f_{\omega}^T}{2}
	                 \frac{ g_{RT(u)} +g_{RT(d)} }{\sqrt{2}} 
					    +f_{\omega} \frac{m_\tau}{m_{\omega}} 
					     \frac{Q_u +Q_d}{\sqrt{2}} eA_R ,\\
	G_{LA\omega} &= -\frac{f_{\omega}^T}{2}
	                 \frac{ g_{LT(u)} +g_{LT(d)} }{\sqrt{2}} 
					    +f_{\omega} \frac{m_\tau}{m_{\omega}} 
					     \frac{Q_u +Q_d}{\sqrt{2}} eA_L ,\\
   G_{R\omega}  &= \frac{f_{\omega} m_{\omega}}{2\sqrt{2} m_\tau } 
						( g_{Rr(u)} +g_{Rr(d)} +g_{Rl(u)} +g_{Rl(d)} ) ,\\
   G_{L\omega}  &= \frac{f_{\omega} m_{\omega}}{2\sqrt{2} m_\tau } 
						( g_{Ll(u)} +g_{Ll(d)} +g_{Lr(u)} +g_{Lr(d)} ) ,\\
	G_{RA\phi} &= -\frac{f_{\phi}^T}{2} g_{RT(s)} 
	              +f_{\phi} \frac{m_\tau}{m_{\phi}} Q_s eA_R ,\\
	G_{LA\phi} &= -\frac{f_{\phi}^T}{2}  g_{LT(s)} 
	              +f_{\phi} \frac{m_\tau}{m_{\phi}} Q_s eA_L ,\\
   G_{R\phi}  &= \frac{f_{\phi} m_{\phi}}{2 m_\tau} 
					  ( g_{Rr(s)} +g_{Rl(s)} ) ,\\
   G_{L\phi}  &= \frac{f_{\phi} m_{\phi}}{2 m_\tau} 
					  ( g_{Ll(s)} +g_{Lr(s)} )
\end{align}
\end{subequations}
The decay constants $f_V$ are defined as
\begin{subequations}
\begin{align}
  m_{\rho} f_{\rho} \epsilon^{\mu} &= 
	\bra{0} 
		\frac{ \bar{u} \gamma^{\mu} u -\bar{d} \gamma^{\mu} d }{\sqrt{2}}
	\ket{\rho(p)}
	, \\
  m_{\omega} f_{\omega} \epsilon^{\mu} &= 
	\bra{0} 
		\frac{ \bar{u} \gamma^{\mu} u +\bar{d} \gamma^{\mu} d }{\sqrt{2}}
	\ket{\omega(p)}
	, \\
  m_{\phi} f_{\phi} \epsilon^{\mu} &=
	\bra{0} 
		 \bar{s} \gamma^{\mu} s
	\ket{\phi(p)} ,
\end{align}
\end{subequations}
where $\epsilon^\mu$ is the polarization vector of the vector mesons.
$f_V^T$ are also defined with the same flavor combinations,
\begin{align}
	\bra{0}
		\bar{q} \sigma^{\mu \nu} q 
	\ket{V(p)}
	= -i f_V^T (p^{\mu} \epsilon^{\nu} -p^{\nu} \epsilon^{\mu}).
\end{align}
$f_V$ can be extracted from partial decay widths for $e^+ e^-$ decay modes,
\begin{subequations}
\begin{align}
  f_{\rho}^2 &= 
    \frac{3 m_{\rho} \Gamma_{\rho \to e^+ e^-}}{2\pi \alpha^2},\\
  f_{\omega}^2 &=
    \frac{27 m_{\omega} \Gamma_{\omega \to e^+ e^-}}{2\pi \alpha^2}, \\
  f_{\phi}^2 &=
    \frac{27 m_{\phi} \Gamma_{\phi \to e^+ e^-}}{4 \pi \alpha^2}.
\end{align}
\end{subequations}
The values of the decay constants are listed in Table~\ref{TabPara}.

\begin{table} 
\begin{tabular}[t]{|ll|}
\hline
  Pseudoscalar & \\
\hline
  $f_{\pi}    $ & 130 MeV \\
  $f_q/f_{\pi}$ & 1.07    \\
  $f_s/f_{\pi}$ & 1.34    \\
  $\phi_{\eta}$ & 39.3$^{\circ}$ \\
\hline
\end{tabular}
\begin{tabular}[t]{|ll|}
\hline
  Vector & \\
\hline 
  $f_{\rho}$   & 221 MeV \\
  $f_{\omega}$ & 196 MeV \\
  $f_{\phi}$   & 228 MeV \\
\hline 
\end{tabular}
\caption{Decay constants of pseudoscalar and vector mesons.}
\label{TabPara}
\end{table}
\subsection{$\mu-e$ conversion} 
\label{AppFormulaeMeconv}
The effective Lagrangian for $\mu$-$e$ conversion is obtained 
by the replacement of
$\tau \to \mu$ and $\mu \to e$ in Eq.~\eqref{EqLaghad}.
According to Ref.\cite{Kitano:2002mt} the rates of 
coherent $\mu-e$ conversions are
\begin{align}
	\text{R}(\mu^-A\to e^-A)
	=& \frac{ 2 G_F^2 }{ \omega_{ \text{capt}} }  \abs{ 
		- A_R D  + 2 \sum_{N = p,n} \bigl(
		  ( \tilde{g}_{Ls}^{(N)} +\tilde{g}_{Lt}^{(N)} ) S^{(N)}
		+ ( \tilde{g}_{Ll}^{(N)} +\tilde{g}_{Lr}^{(N)} ) V^{(N)}
		\bigr) }^2 \n &
		+ ( L \leftrightarrow R ),
\label{EqMec}
\end{align}
where $\omega_{\text{capt}}$ is the muon capture rate of each atom
and the $p$ and $n$ mean proton and neutron, respectively.
$\tilde{g}_{Ls}^{(N)}$, $\cdots$ in the above equation are
written in terms of the Wilson coefficients in the effective Lagrangian:
\begin{subequations}
\begin{align}
  \tilde{g}_{Ls}^{(p)} + \tilde{g}_{Lt}^{(p)}
	&= \sum_q G_{S}^{(q,p)}
	     \bigl( g_{Ls(q)} + g_{Lt(q)} \bigr) , 
   \label{EqMuecsclrP}\\
  \tilde{g}_{Ls}^{(n)} + \tilde{g}_{Lt}^{(n)}
	&= \sum_q G_{S}^{(q,n)}
	     \bigl( g_{Ls(q)} + g_{Lt(q)} \bigr) , 
   \label{EqMuecsclrN}\\
  \tilde{g}_{Ll}^{(p)} + \tilde{g}_{Lr}^{(p)}
	&= 2 g_{Ll(u)} + 2 g_{Lr(u)} + g_{Ll(d)} + g_{Lr(d)} ,\\
  \tilde{g}_{Ll}^{(n)} + \tilde{g}_{Lr}^{(n)} 
	&= g_{Ll(u)} + g_{Lr(u)} + 2 g_{Ll(d)} + 2 g_{Lr(d)} ,
	\label{EqMueconvvector}
\end{align}
\end{subequations}
where $G_S^{(q,N)}$ are given in Ref.~\cite{Kitano:2002mt}:
\begin{subequations}
\begin{align}
	G_{S}^{(u,p)} = G_{S}^{(d,n)} = 5.1 ,\\
	G_{S}^{(d,p)} = G_{S}^{(u,n)} = 4.3 ,\\
	G_{S}^{(s,p)} = G_{S}^{(s,n)} = 2.5 .
\end{align}
\end{subequations}
$D$, $S^{(p)}$, $S^{(n)}$, $V^{(p)}$ and $V^{(n)}$,
in Eq.~\eqref{EqMec} are overlap integrals 
defined in Ref.~\cite{Kitano:2002mt}. 
They depend on nuclides. 
In this paper, we calculate the conversion rates for Al, Ti, Au and Pb using
the numerical values listed in Table~\ref{TabMec}.
\begin{table}[htb]
\begin{tabular}{lcccccc}	
	\hline \hline
	Nucleus & 
	$D$    & $S^{(p)}$ & $V^{(p)}$ & $S^{(n)}$ & $V^{(n)}$ 
	& $\omega_{\text{capt}}$ \\
	\hline
	${}_{13}^{27}$Al & 
	0.0357 & 0.0153    & 0.0159    & 0.0163    & 0.0169    & 0.7054   \\
	${}_{22}^{48}$Ti &
	0.0870 & 0.0371    & 0.0399    & 0.0462    & 0.0495    & 2.59     \\
	${}_{\ 79}^{197}$Au &
	0.167 & 0.0523     & 0.0859    & 0.0610    & 0.108     & 13.07     \\
	${}_{\ 82}^{208}$Pb &
	0.163  & 0.0493    & 0.0845    & 0.0686    & 0.120     & 13.45    \\
	\hline \hline
\end{tabular}
\caption{
Values of overlap integrals and capture rates. They are taken from 
Table II for Al and Au, IV for Ti,
VI for Pb and VIII in Ref.\cite{Kitano:2002mt}.
The integrals are in units of $m_{\mu}^{5/2}$ and 
rates are in units of $10^6$/sec.
}
\label{TabMec}
\end{table}
\section{Functions used in the Wilson coefficients} 
\label{AppWilson}
The following functions are used to describe the Wilson coefficients 
in Sec.\ref{SecLhtlfv}:
\begin{subequations}
\begin{align}
    N_{CM}(x)
  =& \frac{ x }{ 4 (1-x)^3 } \left( 
     -1 +5x +2x^2 +\frac{ 6x^2 }{ 1-x } \ln x \right), 
  \\
    N_{NM}(x)
  =& \frac{ x }{ 4 (1-x)^3 } \left( 
     2 +5x -x^2 +\frac{ 6x }{ 1-x } \ln x \right), 
  \\
    P_{\gamma}(x) 
  =& -2 \left( 
     N_{CC}(x) + \frac{1}{2} N_{NC}(x) + \frac{1}{10} N_{NC}(x) \right), 
  \\
	 N_{CC}(x)
  =& \frac{x}{12(1-x)^3} \left( 
     12 + x -7x^2 + ( 12 - 10x + x^2 )\frac{ 2x }{ 1-x }\ln x \right), 
  \\
	 N_{NC}(x)
  =& \frac{x}{12(1-x)^3} \left( 
      18 - 11x -x^2 + ( 15 - 16x + 4x^2 )\frac{ 2x }{ 1-x }\ln x \right)
		-\frac{2}{3} \ln x, 
  \\
	 P_Z(x)
  =& - \frac{x}{1-x} \left( 
     6 - x + (2 + 3x)\frac{\ln x}{1-x} \right), 
\end{align}
\end{subequations}
\begin{subequations}
\begin{align}
    B_{(e)}(x,y) 
  =& \frac{1}{4} \left( 
     B_{C(e)}(x,y) + B_{N}(x,y) + B_{NX}(x,y) \right), 
  \\
    B_{(u)}(x,y) 
  =& \frac{1}{4} \left( 
     B_{C(u)}(x,y) + B_{N}(x,y) - B_{NX}(x,y) \right), 
  \\
    B_{(d)}(x,y) 
  =& \frac{1}{4} \left( 
     B_{C(d)}(x,y) + B_{N}(x,y) + B_{NX}(x,y) \right), 
\end{align}
\end{subequations}
\begin{subequations}
\begin{align}
	 B_{C(u)}(x,y)
  =& ( 16 + xy ) B_{X}(x,y,1) - 8 xy B_{XC}(x,y) ,
  \\
    B_{C(d,e)}(x,y)
  =& -( 4 + xy ) B_{X}(x,y,1) + 8 xy B_{XC}(x,y) ,
  \\
    B_N (x,y)
  =& 3 \left( B_{X}(x,y,1) +\frac{\eta}{25} B_{X1}(x',y',1) \right) ,
  \\
    B_{NX}(x,y)
  =& \frac{6\eta}{5} B_{X} (x,y,\eta) 
  \\
    B_{X}(x,y,\eta)
  =& -\frac{ \eta^2 \ln \eta }{ ( 1 -\eta )( \eta - x )( \eta - y ) }
    + \frac{ x^2 \ln x }{ ( 1 - x )( \eta - x )( x - y ) } \n &
	 + \frac{ y^2 \ln y }{ ( 1 - y )( \eta - y )( y - x ) }, 
  \\
    B_{XC}(x,y)
  =& \frac{1}{(1-x)(1-y)}
    + \frac{1}{x-y} \left( 
	   \frac{x\ln x}{(1-x)^2} -\frac{y\ln y}{(1-y)^2} \right),
\end{align}
\end{subequations}
where,
\begin{align}
  x  = \frac{m_H^2}{m_{W_H}^2}, \quad
  x' = \frac{m_H^2}{m_{A_H}^2} = \frac{x}{\eta}, \quad
  \eta = \frac{\tan^2 \theta_W}{5}.
\end{align}
\section{Consistency test of the LHT 
         using LFV branching ratios and asymmetries} 
\label{AppRel}
We introduce useful formulae to parametrize the various branching ratios
and asymmetries in the LHT.
Since the observables are expressed by a small number of parameters, 
we can obtain implicit relations among these quantities,
which can be used for a consistency check of the model.

The nonzero elements of the Wilson coefficients are only of three types:
$A_R^{\text{LHT}}$, $g_{Lr}^{\text{LHT}}$ and $g_{Ll}^{\text{LHT}}$.
Here we consider the $\tau \to \mu$ transition. 
Corresponding formulae for the $\tau \to e$ transition are also derived.
According to Eqs.~\eqref{EqVeclr1}, \eqref{EqVeclr2}, 
\eqref{EqWilsontypeII}, \eqref{eq:gLlq-tm} and \eqref{eq:gLrq-tm}, 
we can derive the following parametrization:
\begin{subequations}
\begin{align}
  \frac{ g_{Lr}^{ \text{I,LHT} } }{ eA_R^{ \text{LHT} } } =& 
    -z, &
  \frac{ g_{Ll}^{ \text{I,LHT} } }{ eA_R^{ \text{LHT} } } =& 
    -z + \Delta^{ \text{I} },  \label{EqZ1} \\
  \frac{ g_{Lr}^{ \text{II,LHT} } }{ eA_R^{ \text{LHT} } } =& 
    -z, &
  \frac{ g_{Ll}^{ \text{II,LHT} } }{ eA_R^{ \text{LHT} } } =& 
    -z + \Delta^{ \text{II} }, \label{EqZ2} \\
  \frac{ g_{Lr(q)}^{ \text{LHT} } }{ eA_R^{ \text{LHT} } } =& 
    Q_{q}\, z ,&
  \frac{ g_{Ll(q)}^{ \text{LHT} } }{ eA_R^{ \text{LHT} } } =& 
    Q_{q}\, z + \Delta_{q}.
\end{align}
\end{subequations}
$z$ is a complex-valued function associated with 
photon penguin diagrams and a part of $Z$ penguin diagrams.

In leptonic decays, the following formulae are obtained using 
the above parameters:
\begin{subequations}
\begin{align}
  \frac{96\pi}{\alpha} 
  \frac{ \text{Br}(\tau^+ \to \mu^+ \mu^+ \mu^-) }
       { \text{Br}(\tau^+ \to \mu^+ \gamma)} =&
     -8 \left( 4 \ln(\delta) +\frac{13}{3} \right) 
	+\abs{z}^2 +2\abs{ -z +\Delta^{\text{I}} }^2 \n &
	+ 8\text{Re}[-3z + 2\Delta^{\text{I}} ], \\
  \frac{96\pi}{\alpha} 
  \frac{ \text{BA}_Z (\tau^+ \to \mu^+ \mu^+ \mu^-) }
       { \text{Br}(\tau^+ \to \mu^+ \gamma)} =&
    -4 \left( 4\ln(\delta) + \frac{49}{3} \right) 
	 +\frac{1}{6} \abs{z}^2 -\abs{ -z +\Delta^{\text{I}} }^2 \n &
	 - 4\text{Re}[-z +2\Delta^{\text{I}}],\\
  \frac{96\pi}{\alpha} 
  \frac{ \text{BA}_X (\tau^+ \to \mu^+ \mu^+ \mu^-) }
       { \text{Br}(\tau^+ \to \mu^+ \gamma)} =&
    \frac{64}{35} \left( 
	  -21 +\frac{1}{12} \abs{z}^2 
	  -\text{Re}[ -z + 3\Delta^{\text{I}} ]
	 \right), \\
  \frac{96\pi}{\alpha} 
  \frac{ \text{BA}_Y (\tau^+ \to \mu^+ \mu^+ \mu^-) }
       { \text{Br}(\tau^+ \to \mu^+ \gamma)} =&
    \frac{64}{35} \text{Im}[-z +3\Delta^{\text{I}}],
\end{align}
\end{subequations}
\begin{subequations}
\begin{align}
  \frac{96\pi}{\alpha} 
  \frac{ \text{Br}(\tau^+ \to \mu^+ e^+ e^-) }
       { \text{Br}(\tau^+ \to \mu^+ \gamma)} =& 
     -32 \biggl( \ln(\delta) +\frac{4}{3} \biggr) 
	  +\abs{z}^2 +\abs{-z +\Delta^{\text{II}} }^2 
	+ 8\text{Re}[ -2z +\Delta^{\text{II}} ] , \\
  \frac{96\pi}{\alpha} 
  \frac{ \text{BA}_Z (\tau^+ \to \mu^+ e^+ e^-) }
       { \text{Br}(\tau^+ \to \mu^+ \gamma)} =& 
   \frac{1}{6} \left( 
	  32 ( 3\ln(\delta) +5 ) +\abs{z}^2 +\abs{ -z +\Delta^{\text{II}} }^2 
	 -8\text{Re}[ -2z +\Delta^{\text{II}} ]
	\right),\\
  \frac{96\pi}{\alpha} 
  \frac{ \text{BA}_X (\tau^+ \to \mu^+ e^+ e^-)}
       { \text{Br}(\tau^+ \to \mu^+ \gamma)} =& 
    \frac{4\pi}{35} \left( 
	     -\abs{z}^2 +\abs{-z +\Delta^{\text{II}} }^2 
	   + 16\text{Re}[ \Delta^{\text{II}}]
	 \right), \label{EqPropX}\\
  \frac{96\pi}{\alpha} 
  \frac{ \text{BA}_Y (\tau^+ \to \mu^+ e^+ e^-)}
       { \text{Br}(\tau^+ \to \mu^+ \gamma)} =& 
    \frac{48\pi}{35} \text{Im}[ \Delta^{\text{II}}],
\end{align}
\end{subequations}
\begin{subequations}
\begin{align}
  \frac{96\pi}{\alpha} 
  \frac{ \text{BA}_{FB} (\tau^+ \to \mu^+ e^+ e^-) }
       { \text{Br}(\tau^+ \to \mu^+ \gamma)}
  =& 
     \frac{1}{4} ( -\abs{z}^2 +\abs{ -z +\Delta^{\text{II}} }^2  
	+ 16 \text{Re}[ \Delta^{\text{II}} ]), \label{EqPropFB}\\
  \frac{96\pi}{\alpha} 
  \frac{ \text{BA}_{ZFB} (\tau^+ \to \mu^+ e^+ e^-) }
       { \text{Br}(\tau^+ \to \mu^+ \gamma)}
  =& 
    -\frac{1}{8} \left( 
	     -\abs{z}^2 +\abs{ -z +\Delta^{\text{II}} } 
	   + 16\text{Re}[ \Delta^{\text{II}} ]
	 \right),\label{EqPropZFB}\\
  \frac{96\pi}{\alpha} 
  \frac{ \text{BA}_{XFB} (\tau^+ \to \mu^+ e^+ e^-) }
       { \text{Br}(\tau^+ \to \mu^+ \gamma)}
  =& 
    \frac{16}{105} \left( 
	     56 +\abs{z}^2 +\abs{ -z +\Delta^{\text{II}} }^2
	   + 16 \text{Re}[ -2z +\Delta^{\text{II}} ] 
	 \right), \\
  \frac{96\pi}{\alpha} 
  \frac{ \text{BA}_{YFB} (\tau^+ \to \mu^+ e^+ e^-) }
       { \text{Br}(\tau^+ \to \mu^+ \gamma)}
  =& 
    \frac{64}{35} \text{Im}[ -2z +\Delta^{\text{II}} ],
\end{align}
\end{subequations}
where BA$_{a}$ ($a = Z$, $X$, $Y$, $FB$, $ZFB$, $XFB$ and $YFB$) 
denotes asymmetry, 
$A_{a}$ multiplied by the branching ratio. 
We take the limit $\delta \to 0$ except for
logarithmically divergent terms in
$A_1(\delta)$, $B_1(\delta)$, $D_1(\delta)$ and $D_5(\delta)$.
The proportional relation \eqref{EqProprel} is found 
from Eqs.~\eqref{EqPropX}, \eqref{EqPropFB} and \eqref{EqPropZFB}.
Furthermore, ten independent observables can be expressed by 
three complex functions.

Similarly, the following formulae are derived for semileptonic decays,
\begin{subequations}
\begin{align}
  \frac{16}{\alpha} 
  \frac{\text{Br}(\tau^+ \to \mu^+ \pi^0) }
       {\text{Br}(\tau^+ \to \mu^+ \gamma)}
  =
    \frac{\pi f_{\pi}^2}{m_{\tau}^2} 
	 ( 1 -x_{\pi} )^2 &
	 \abs{ \Delta_u -\Delta_d }^2 ,\\
  \frac{1}{2\alpha} 
  \frac{\text{Br}(\tau^+ \to \mu^+ \rho^0)}
       {\text{Br}(\tau^+ \to \mu^+ \gamma)}
  =
    \frac{\pi f_{\rho}^2}{m_{\tau}^2} 
	 ( 1 -x_{\rho} )^2 &
    \biggl( 
	    \frac{ (1-x_{\rho})^2 }{ x_{\rho} (1 +2x_{\rho})} \n &
		+\frac{1 +2x_{\rho}}{32} 
		 \abs{ 2z +\Delta_u -\Delta_d -\frac{12}{1 +2x_{\rho}} }^2 
	 \biggr), \\
  \frac{1}{\alpha} 
  \frac{\text{BA}(\tau^+ \to \mu^+ \rho^0)}
       {\text{Br}(\tau^+ \to \mu^+ \gamma)}
  =
    \frac{\pi f_{\rho}^2}{m_{\tau}^2} 
	 ( 1 -x_{\rho} )^2 &
    \biggl(
	   -\frac{(1 -x_{\rho})^2 }{x_{\rho} (1 -2x_{\rho})} \n &
		+\frac{1 -2x_{\rho}}{32} 
		 \abs{ 2z +\Delta_u -\Delta_d +\frac{4}{1 -2x_{\rho}} }^2 
	 \biggr),
\end{align}
\end{subequations}
\begin{subequations}
\begin{align}
  \frac{8}{\alpha} 
  \frac{\text{Br}(\tau^+ \to \mu^+ \eta^{(')})}
       {\text{Br}(\tau^+ \to \mu^+ \gamma)}
  =
    \frac{\pi f_{\pi}^2}{m_{\tau}^2} 
	 ( 1 -x_{\eta^{(')}} &)^2 
	 \abs{ \frac{f^q_{\eta^{(')}} }{\sqrt{2}f_{\pi} } (\Delta_u +\Delta_d) 
	      +\frac{f^s_{\eta^{(')}} }{f_{\pi}} \Delta_s }^2 , \\
  \frac{1}{2\alpha} 
  \frac{\text{Br}(\tau^+ \to \mu^+ \omega)}
       {\text{Br}(\tau^+ \to \mu^+ \gamma)}
  =
    \frac{\pi f_{\omega}^2}{m_{\tau}^2} 
	 ( 1 -x_{\omega} )^2 &
    \biggl( 
	    \frac{1}{9} \frac{(1 -x_{\omega})^2}{x_{\omega}(1 +x_{\omega}) } \n &
		+\frac{1 +2x_{\omega} }{32} 
		 \abs{ \frac{2}{3}z +\Delta_u +\Delta_d -\frac{4}{1 +2x_{\omega} } }^2 
	 \biggr), \\
  \frac{1}{2\alpha} 
  \frac{\text{Br}(\tau^+ \to \mu^+ \phi)}
       {\text{Br}(\tau^+ \to \mu^+ \gamma)}
  =
    \frac{\pi f_{\phi}^2}{m_{\tau}^2} 
	 ( 1 -x_{\phi} )^2 &
    \biggl(
	    \frac{2}{9} \frac{(1 -x_{\phi})^2 }{x_{\phi} (1 +2x_{\phi}) } \n &
		+\frac{1 +2x_{\phi}}{16} 
		 \abs{ -\frac{2}{3}z +\Delta_s +\frac{4}{1 +2x_{\phi}} }^2 
	 \biggr),\\
  \frac{1}{\alpha} 
  \frac{\text{BA}(\tau^+ \to \mu^+ \omega)}
       {\text{Br}(\tau^+ \to \mu^+ \gamma)}
  =
    \frac{\pi f_{\omega}^2}{m_{\tau}^2} 
	 ( 1 -x_{\omega})^2 &
    \biggl(
	   -\frac{1}{9} \frac{(1 -x_{\omega})^2}{x_{\omega} (1 -2x_{\omega})} \n &
	   +\frac{1 -2x_{\omega} }{32} 
	    \abs{ \frac{2}{3}z +\Delta_u +\Delta_d +\frac{4}{3(1 -2x_{\omega})} }^2 
	 \biggr), \\
  \frac{1}{\alpha} 
  \frac{\text{BA}(\tau^+ \to \mu^+ \phi)}
       {\text{Br}(\tau^+ \to \mu^+ \gamma)}
  =
    \frac{\pi f_{\phi}^2}{m_{\tau}^2} 
	 ( 1 -x_{\phi} )^2 &
    \biggl(
	   -\frac{2(1 -x_{\phi})^2}{9x_{\phi} (1 -2x_{\phi}) } \n &
		+\frac{1 -2x_{\phi}}{16} 
		 \abs{ -\frac{2}{3}z +\Delta_s -\frac{4}{3(1 -2x_{\phi})} }^2 
	 \biggr),
\end{align}
\end{subequations}
where $x_{a} = m_a^2 /m_{\tau}^2$.
From the above expressions, we can see that various consistency tests 
are possible with complex functions $z$, $\Delta_u -\Delta_d$, 
$\Delta_u +\Delta_d$ and $\Delta_s$.
 
\end{document}